\documentclass[12pt]{article}
\usepackage{amssymb}
\usepackage{amsmath}

\begin{document}

\begin{center}
\textbf{PERTURBATIONAL\ NON-CANONICAL\bigskip\ THEORY\ OF\ MOLECULAR\
ORBITALS\ AND\ ITS\ APPLICATIONS\bigskip }

Viktorija Gineityte

\bigskip Institute of Theoretical Physics and Astronomy, Vilnius University,
Sauletekio al. 3, LT-10257 Vilnius, Lithuania

Email: viktorija.gineityte@tfai.vu.lt

Mach, 2018\bigskip
\end{center}

\textbf{Abstract}

The article contains a summary of fundamentals of the perturbational non-
canonical molecular orbital (PNCMO) theory formerly developed by the author.
In some respects, the PNCMO theory is a generalization of the well-known
simple PMO theory: First, the usual diagonalization problem (and/or the
eigenvalue equation) for a certain model Hamiltonian matrix ($\mathbf{H}$)
is now replaced by two interrelated non-canonical one-electron problems,
namely by the block-diagonalization problem for the matrix $\mathbf{H}$\
following from the Brillouin theorem and determining non-canonical
(localized) MOs (NCMOs) and by the commutation equation for the respective
one-electron density matrix (charge-bond order (CBO)) matrix. Second,
perturbative solutions of the above-specified alternative problems are
sought in terms of entire submatrices (blocks) of the matrix $\mathbf{H}$\
instead of usual matrix elements (e.g. of Coulomb and resonance parameters).
Third, a generalized version of the perturbation theory (PT) is used in
place of the standard Rayleigh-Schr\"{o}dinger PT (RSPT), wherein
non-commutative quantities stand for the usual (commutative) ones (cf. the
so-called non-commutative RSPT (NCRSPT)). As a result, algebraic expressions
are derived for the principal quantum-chemical characteristics (including
the CBO matrix, the NCMO representation matrix and the total energy) that
embrace definite classes of Hamiltonian matrices and thereby of molecules.
To illustrate the point, saturated and conjugated hydrocarbons are taken as
examples. Arguments are given that the PNCMO theory possibly forms the basis
of a novel way of qualitative chemical thinking.\bigskip

\textbf{Introduction\bigskip }

The perturbational molecular orbital (PMO) theory [1-3] is among the most
popular and efficient tools for evaluation of observed properties of
molecules and for qualitative chemical thinking in general. As with any
perturbative approach, the main advantage of the above-mentioned theory
consists in a possibility of a direct comparison of two (or several)
slightly different molecules or molecular systems, e.g. of a certain
hydrocarbon and its heteroatom-containing derivative(s). In other words,
application of the PMO theory allows our attention to be concentrated on
consequences of an alteration in the structure of the system(s) under
interest instead of properties of each molecule separately.

The PMO theory is commonly understood as a combination of the standard
Rayleigh-Schr\"{o}dinger perturbation theory (RSPT) [4, 5] and of the H\"{u}%
ckel type (and/or tight binding) approximation for Hamiltonian matrix
elements (see e.g. [6-9]). In a broader sense, however, any perturbative
solution of a secular (eigenvalue) equation for a H\"{u}ckel type
Hamiltonian matrix ($\mathbf{H}$) of molecule may be ascribed to the PMO
theory along with the subsequent steps, e.g. derivation of the total energy
and/or of the relevant charge-bond order (CBO) matrix using the eigenvectors
and eigenvalues obtained. It is then no surprise that the results of the
theory under discussion are usually interpreted in terms of individual
orbitals (underlying the matrix $\mathbf{H}$) and their interactions, the
latter often tending to replace the interatomic bonds in the qualitative
chemical thinking.

For some specific series of molecules (e.g. polyenes), the H\"{u}ckel type
Hamiltonian matrices are known to take definite common forms. Accordingly,
the so-called closed solutions of the relevant eigenvalue equations are
obtainable that represent the entire series of the compounds concerned (see
Ref.[9] for an overview). To a certain extent, the same refers also to the
well-known alternant conjugated hydrocarbons. Indeed, the eigenvectors (MOs)
and the eigenvalues of the relevant matrix $\mathbf{H}$\ are characterized
by common properties found in quantum chemistry textbooks (see e.g. [8, 10,
11]). In the above-enumerated special cases, a subsequent employment of a
perturbative approch really yields important general results, e.g. the
famous rule of the alternating polarity `[12-15]. The overall potential of
the usual PMO theory, however, is rather limited in respect of going beyond
individual molecules and their pairs. The reason certainly lies in the
nature of the secular (eigenvalue) equation itself that is designed to
represent a single quantum mechanical system.

As is well-known, it is quite usual for chemists to think in terms of broad
classes of compounds (e.g. alkanes, arenes, etc.), as well as of the
so-called effects manifesting themselves in all representatives of a certain
class (e.g. the inductive effect, conjugation of unsaturated fragments,
etc.) [16-19]. In this connection, it seemed reasonable to look for a more
general version of the PMO theory. To this end, alternative (the so-called
non-canonical) one-electron problems have been taken as a basis instead of
the secular (eigenvalue) equation, namely the commutation condition for the
one-electron density matrix (DM) [14] and the (local) Brillouin theorem
[20-24]. Perturbative solutions of these two problems in the framework of
the H\"{u}ckel type approximation resulted into a new version of the PMO
theory called perturbational non-canonical MO (PNCMO) theory.

The above-mentioned two principal non-canonical one-electron problems are
discussd below in a detail (Subsections 1.1 and 1.2). Let us only note here
that the solutions of these problems proved to be obtainable for certain
classes of Hamiltonian matrices in terms of entire submatrices (blocks) of
the latter playing the role of multi-dimensional (and thereby
non-commutative) parameters. The standard RSPT has been accordingly replaced
by a certain generalized version of the perturbation theory, wherein
non-commutative quantities stand instead of usual (commutative) ones [25,
26]. The relevant results have been consequently interpreted in terms of
entire subsets of basis orbitals and the intersubset interaction. Thus, the
PNCMO theory seems to offer a novel way of qualitative chemical thinking.

Despite the above-mentioned distinctive features of the generalized theory,
a certain analogy between the latter and the usual PMO theory still remains,
which is useful for understanding both the nature of the newly- introduced
generalizations and the overall scheme of presentation chosen in this
article. Before clarifying this analogy, however, let us recall some
well-known facts.

In its standard form [4, 5], the RSPT is applicable to systems characterized
by relatively large intervals between individual energy levels as compared
to the extent of perturbation, usually referred to as the standard
"perturbational" case (i). Meanwhile, the opposite extreme situation
(embracing systems described by small or even zero inter-level energy
intervals and known as a (quasi)degenerate case (ii)) needs a certain
special treatment (namely, passing to linear combinations of
(quasi)degenerate orbitals). If we take a simple two-level system at the H%
\"{u}ckel level as an example, the above-mentioned cases (i) and (ii) are
correspondingly characterized by small ratios $\beta /(\alpha _{1}-\alpha
_{2})$ and $(\alpha _{1}-\alpha _{2})/\beta ,$\ where $\alpha _{1},\alpha
_{2}$ and $\beta $\ stand for the Coulomb and resonance parameters,
respectively.

The PNCMO theory overviewed in this article is constructed using the
above-specified extreme two-level systems as prototypes. Thus, the standard
(also alternatively called implicit) version of the theory (Sections 1 and
2) addresses systems representable by two weakly-interacting well-separated
subsets of basis orbitals in analogy with the two-level system of the first
type (i). Thereupon, the quasi-degenerate version of the PNCMO theory is
overviewed in Section 3 that embraces systems described by two
strongly-interacting subsets of orbitals.\bigskip

\textbf{1. The standard (implicit) form of the PNCMO theory\bigskip }

Let us consider a certain molecule or molecular system containing an even
number of electrons ($2n$). Let this system be represented by an orthogonal
basis set $\{\varphi \}$\ (the case of a non-orthogonal basis set has been
studied in Ref.[27]). The total number of basis functions will be denoted by 
$N$. It deserves an immediate emphasizing that neither the numbers $n$ and $%
N $, nor the nature of basis orbitals $\{\varphi \}$\ are specified at the
present stage of our discussion.

Let us now introduce a certain zero order approximation for our system,
wherein $n$ basis orbitals of lowest one-electron energies are
double-occupied, whilst the remaining $s$ functions are vacant, where $s=N-n$%
. Let these two types of orbitals be correspondingly denoted by $\varphi
_{(+)i}$ and $\varphi _{(-)l}$, where $i=1,2...n$ and $l=n+1,n+2,...N$.
Since the occupation numbers of these basis orbitals are no longer equal to
2 and 0 after taking into account the perturbation (as discussed below),
orbitals $\varphi _{(+)i}$ and $\varphi _{(-)l}$ will be referred to as
initially-occupied and initially-vacant ones. Let the above-defined two
types of orbitals be collected into separate $n-$and $s-$dimensional subsets
denoted by $\{\varphi _{(+)}\}$ and $\{\varphi _{(-)}\},$ respectively.
Another important property of our zero order approximation consists in
absence of interactions between orbitals of different subsets. Meanwhile,
interactions of the zero order magnitude are allowed inside subsets both $%
\{\varphi _{(+)}\}$ and $\{\varphi _{(-)}\}.$ At the H\"{u}ckel level, the
interactions concerned coincide with respective resonance parameters.

Further, let us assume the above-specified zero order system to undergo a
certain perturbation, generally embracing alterations in both Coulomb and
resonance parameters including those of intra- and intersubset types. The
newly-emerging interorbital interactions are supposed to be weak as compared
to the energy gap between subsets $\{\varphi _{(+)}\}$ and $\{\varphi
_{(-)}\}.$ Finally, let our energy reference point be chosen in the middle
of the above-mentioned gap. One-electron energies of orbitals $\varphi
_{(+)i}$ and $\varphi _{(-)l}$ will be denoted by $\varepsilon _{(+)i}$ and $%
-\varepsilon _{(-)l},$ respectively, where the minus sign in front of $%
\varepsilon _{(-)l}$\ is introduced for further convenience. Given that a
negative energy unit is accepted as usual, energetic parameters both $%
\varepsilon _{(+)i}$\ and $\varepsilon _{(-)l}$\ take positive values, i.e. $%
\varepsilon _{(+)i}>0$\ and $\varepsilon _{(-)l}>0.$

As a result, the total Hamiltonian matrix of our system ($\mathbf{H}$)
consists of the zero order member ($\mathbf{H}_{(0)}$) and of the first
order one ($\mathbf{H}_{(1)}$), the former taking a block-diagonal form,
viz. 
\begin{equation}
\mathbf{H=H}_{(0)}+\mathbf{H}_{(1)}=\left\vert 
\begin{array}{cc}
\mathbf{E}_{(+)} & \mathbf{0} \\ 
\mathbf{0} & -\mathbf{E}_{(-)}%
\end{array}%
\right\vert +\left\vert 
\begin{array}{cc}
\mathbf{T} & \mathbf{R} \\ 
\mathbf{R}^{+} & \mathbf{Q}%
\end{array}%
\right\vert .  \tag{1.1}
\end{equation}
Submatrices (blocks) $\mathbf{E}_{(+)}+\mathbf{T},$ $-\mathbf{E}_{(-)}+%
\mathbf{Q},$\textbf{\ }and\textbf{\ }$\mathbf{R}$ correspond here to
individual subsets $\{\varphi _{(+)}\}$\ and \ $\{\varphi _{(-)}\}$\ \ and
to their interaction, respectively, and correspondingly are $n\times n-$ and 
$s\times s-$dimensional matrices. The supersript + represents here and below
the Hermitian- conjugate (transposed) counterpart of the relevant matrix.%
\textbf{\ }Diagonal elements\textbf{\ }$E_{(+)ii}$\textbf{\ }and $E_{(-)ll}$%
\ of the zero order (sub)matrices $\mathbf{E}_{(+)}$ and $\mathbf{E}_{(-)}$
coincide with the above-introduced energetic parameters $\varepsilon _{(+)i}$
and $\varepsilon _{(-)l},$ respectively, whereas the off-diagonal ones $%
E_{(+)ij}(i\neq j)$\textbf{\ }and $E_{(-)lm}(m\neq l)$ represent the zero
order intrasubset interactions. Accordingly, $n\times n-$\ and $s\times s-$
dimensional (sub)matrices $\mathbf{T}$ and $\mathbf{Q}$ contain certain
corrections to the latter due to perturbation. Finally, newly-emerging
intersubset interactions are included into the $n\times s-$ and $s\times n-$%
dimensional (sub)matrices $\mathbf{R}$\ and $\mathbf{R}^{+}.$\ As with
numbers $n$ and $N$, the "internal" constitutions of (sub)matrices $\mathbf{E%
}_{(+)},\mathbf{T},$ $\mathbf{E}_{(-)},\mathbf{Q},$\textbf{\ }and\textbf{\ }$%
\mathbf{R}$ are not specified. \ Consequently, Eq.(1.1) actually embraces a
large class of specific Hamiltonian matrices and thereby a wide variety of
of underlying systems (although both $\mathbf{H}$ and its blocks are
referred to below as matrices for simplicity). It is also evident that the
above-exhibited matrix $\mathbf{H}$ is a generalization of a simple
Hamiltonian matrix of a two-level two-electron system [28], viz. 
\begin{equation}
\mathbf{h=h}_{(0)}+\mathbf{h}_{(1)}=\left\vert 
\begin{array}{cc}
\epsilon _{(+)} & 0 \\ 
0 & -\epsilon _{(-)}%
\end{array}%
\right\vert +\left\vert 
\begin{array}{cc}
\tau & \rho \\ 
\rho & \kappa%
\end{array}%
\right\vert ,  \tag{1.2}
\end{equation}%
where $\epsilon _{(+)}+\tau ,$\ $-\epsilon _{(-)}+\kappa $ and $\rho $\
coincide with the usual Coulomb and resonance parameters, respectively.
Accordingly, our principal condition about a weak intersubset interaction
turns into that of a small ratio $\rho /(\epsilon _{(+)}+\epsilon _{(-)}).$
In this connection, the 2$\times $2$-$dimensional matrix $\mathbf{h}$\ may
be called the prototype of the matrix $\mathbf{H.}$ It is also evident that
the matrix $\mathbf{h}$\ refers to the usual "perturbational" case (i) of
the Introduction that is characterized by a small resonance parameter as
compared to the energy gap. Again, it is seen that passing from the simple 2$%
\times $2$-$dimensional matrix $\mathbf{h}$\ of Eq.(1.2) to the
multidimensional one ($\mathbf{H}$ of Eq.(1.1)) is accompanied by replacing
the usual (one-dimensional) parameters ($\epsilon _{(+)},$\ $\epsilon
_{(-)},\tau ,\kappa $ and $\rho $) by multidimensional ones ($\mathbf{E}%
_{(+)},\mathbf{E}_{(-)},\mathbf{T,Q},$\textbf{\ }and\textbf{\ }$\mathbf{R,}$
respectively). Thus, we actually have to do here with a non-trivial
generalization. It is then no surprise that this step has important
consequences.

First of all, solution of the usual secular (eigenvalue) equation is no
longer possible for the matrix $\mathbf{H}$ [28]\textbf{. }This implies that
neither the canonical MOs (CMOs) nor their one-electron energies may be
derived and used as the principal quantum-chemical characteristic when
desribing the class of compounds embraced by the matrix $\mathbf{H.}$ Thus,
we necessarily have to abandon the usual (canonical) MO method and turn to
alternative (non-canonical) ones.

As opposed to the unique eigenvalue equation, various forms of the
non-canonical one-electron problem are possible. The commutation equation
[14]\ for the one-electron density matrix (DM) (Subsect. 1.1) was chosen as
the principal problem in the PNCMO theory under discussion for the following
reasons: First, the DM (and thereby the charge-bond order (CBO) matrix) is
among the most fundamental unique quantum-chemical characteristics of
molecule [29] describing the relevant charge distribution and related to
numerous observed properties. Second, solution of the above-mentioned
equation yields the CBO matrix ($\mathbf{P}$) directly without any reference
to MOs and thereby allows us to study the relation between individual
elements of the matrix $\mathbf{P,}$\ on the one hand, and those of the
respective Hamiltonian matrix $\mathbf{H}$, on the other hand, i.e. between
charge distribution and the structure of the given system. The third and the
most important reason, however, is that the commutation equation for the DM
proves to be solvable for the whole class of matrices embraced by Eq.(1.1)
as discussed in the Subsection 1.1 in a detail.

An alternative non-canonical one-electron problem originates from the
Brillouin theorem [20-24] and determines the so-called localized
(non-canonical) MOs (LMOs) directly without invoking the usual canonical MOs
(CMOs). This problem is a generalization of the eigenvalue equation for the
prototypal matrix $\mathbf{h}$ of Eq.(1.2) [28] to the case of
multidimensional elements (Subsect. 1.2). Apart from solution of the
commutation equation, the DM (CBO matrix) is also known to follow from the
projector to the subspace of occupied one-electron states [29] including
either CMOs or non-canonical MOs (NCMOs). Employment of just this fact
allowed us to relate the overall charge (re)distribution to shapes
(reshapings) of particular NCMOs (Subsect. 1.3). Finally, expressions for
the total energy (Subsect. 1.4) also are obtainable using either the DM or
NCMOs.

\textbf{1.1. The direct way of derivation of the one-electron DM}\ \ \ \ 

The non-canonical one-electron problem for the DM (charge-bond order matrix) 
$\mathbf{P}$\ consists in solution of the following system of matrix
equations [14, 30-34] 
\begin{equation}
\lbrack \mathbf{H,P}]_{-}=\mathbf{0,\quad P}^{2}=2\mathbf{P};\quad Tr\mathbf{%
P}=2n,  \tag{1.1.1}
\end{equation}%
where $\mathbf{H}$\ is the initial Hamiltonian matrix coinciding with that
of Eq.(1.1) in our case. Notations $[\mathbf{..,..}]_{-}$ and $Tr$ stand
here and below for the commutator of matrices and for a \textit{Trace} of
the latter, respectively. The first relation of Eq.(1.1.1) (the commutation
condition) is the main physical requirement determining the matrix $\mathbf{P%
}$ [35] and resulting from the Dirac equation for the time-independent
Hamiltonian [14]. The remaining relations are additional system-structure-
independent restrictions following from the idempotence requirement ($%
\mathbf{\Pi }^{2}\mathbf{=\Pi }$) for the projector $\mathbf{\Pi }=\frac{1}{2%
}\mathbf{P}$ [4, 29, 35] and from the charge conservation condition,
respectively.

As opposed to the most well-known way of derivation of the CBO matrix $%
\mathbf{P}$ on the basis of projector to the subspace of occupied
one-electron (molecular) orbitals [29], the above-exhibited problem may be
solved directly, i.e. without any reference to molecular orbitals (MOs). The
solution concerned [36] was based on the following points: First, the CBO
matrix $\mathbf{P}$ has been sought in the form of power series%
\begin{equation}
\mathbf{P}=\sum\limits_{k=0}^{\infty }\mathbf{P}_{(k)}=\mathbf{P}_{(0)}+%
\mathbf{P}_{(1)}+\mathbf{P}_{(2)}+\mathbf{P}_{(3)}...  \tag{1.1.2}
\end{equation}%
i.e. as a sum of increments ($\mathbf{P}_{(k)}$) of increasing orders ($k$)
with respect to the first order Hamiltonian matrix $\mathbf{H}_{(1)}$
[Parameters contained within the latter and underlying the expansion should
be specified in each particular case separately]. Second, both the total CBO
marrix $\mathbf{P}$ and its individual increments $\mathbf{P}%
_{(k)},k=0,1,2,...$ have been initially represented in terms of four
submatrices (blocks) of appropriate dimensions (coinciding with those of
blocks of matrices $\mathbf{H}$, $\mathbf{H}_{(0)}$ and $\mathbf{H}_{(1)}$
of Eq.(1.1)), viz. 
\begin{equation}
\mathbf{P=}\left\vert 
\begin{array}{cc}
\mathbf{P}_{11} & \mathbf{P}_{12} \\ 
\mathbf{P}_{21}^{{}} & \mathbf{P}_{22}%
\end{array}%
\right\vert ,\qquad \mathbf{P}_{(k)}\mathbf{=}\left\vert 
\begin{array}{cc}
\mathbf{P}_{11}^{(k)} & \mathbf{P}_{12}^{(k)} \\ 
\mathbf{P}_{21}^{(k)} & \mathbf{P}_{22}^{(k)}%
\end{array}%
\right\vert ,  \tag{1.1.3}
\end{equation}%
where the subscripts 11, 22 and 12(21) here and below refer to separate
subsets of basis orbitals $\{\varphi _{(+)}\}$\ and $\{\varphi _{(-)}\}$ and
to their interaction, respectively. Accordingly, the blocks of the matrix $%
\mathbf{P}$ ( i.e. $\mathbf{P}_{11},\mathbf{P}_{12},\mathbf{P}_{21}^{{}}$
and $\mathbf{P}_{22}$) also take the form of power series like that of
Eq.(1.1.2). It deserves emphasising that no need arises here for specifying
either the internal constitutions of the blocks concerned nor their
dimensions ($n$ and $s$) as it was the case with submatrices of the initial
matrix $\mathbf{H}$\ of Eq.(1.1). Quite the reverse, these blocks are
considered as "indivisible" elements of respective matrices. It should be
kept in mind, however, that these are non-commutative quantities in contrast
to usual (one-dimensional) matrix elements.

The above-specified form of the CBO matrix $\mathbf{P}$ has been
subsequently substituted into the matrix problem of Eq.(1.1.1) and terms of
each order ($k$) separately have been collected (as is usual in
perturbational approaches). As a result, the solution of Eq.(1.1.1) (i.e.
the matrix $\mathbf{P)}$ has been obtained in terms of entire submatrices
(blocks) of our matrix $\mathbf{H}$, i.e. via entire blocks $\mathbf{E}%
_{(+)},$ $\mathbf{E}_{(-)},$ $\mathbf{T,R}$ and \ $\mathbf{Q}$. Before
passing to this solution, however, let us introduce some additional
definitions and notations.

As is well-known, diagonal elements of any CBO matrix represent populations
(occupation numbers) of the rel\textbf{e}vant basis orbitals. In our case,
diagonal elements $P_{11,ii}$\textbf{\ }and\textbf{\ }$P_{22,ll}$\textbf{\ }%
of submatrices\textbf{\ }(blocks)\textbf{\ }$\mathbf{P}_{11}$\textbf{\ }and%
\textbf{\ }$\mathbf{P}_{22}$ yield occupation numbers of orbitals $\varphi
_{(+)i}$ and $\varphi _{(-)l}$ belonging to subsets $\{\varphi _{(+)}\}$ and 
$\{\varphi _{(-)}\},$ respectively, where $i=1,2...n$ and $l=n+1,n+2,...N$.
\ Thus, the blocks $\mathbf{P}_{11}$\textbf{\ }and\textbf{\ }$\mathbf{P}%
_{22} $\ will be called the intrasubset population matrices and
alternatively denoted by $\mathbf{X}_{(+)}$ and $\mathbf{X}_{(-)}$. The
relevant terms of the $k$th order will be accordingly designated by $\mathbf{%
X}_{(+)}^{(k)}$ and $\mathbf{X}_{(-)}^{(k)},$ respectively, so that the
following relations are valid, viz.%
\begin{equation}
\mathbf{X}_{(+)}=\sum\limits_{k=0}^{\infty }\mathbf{X}_{(+)}^{(k)},\qquad 
\mathbf{X}_{(-)}=\sum\limits_{k=0}^{\infty }\mathbf{X}_{(-)}^{(k)}. 
\tag{1.1.4}
\end{equation}

Meanwhile, the off-diagonal (intersubset) blocks of our matrix $\mathbf{P}$
(i.e. $\mathbf{P}_{12}$\textbf{\ }and\textbf{\ }$\mathbf{P}_{21})$ represent
bond orders between basis orbitals of opposite initial occupation, whereas
the relevant blocks of the correction $\mathbf{P}_{(k)}$ (i.e. $\mathbf{P}%
_{12}^{(k)}$\textbf{\ }and\textbf{\ }$\mathbf{P}_{21}^{(k)})$ yield the
increments of the $k$th order \ to the same quantities. Let us introduce the
convenient designations $-2\mathbf{G}_{(k)}$ and $-2\mathbf{G}_{(k)}^{+}$
for submatrices (blocks) $\mathbf{P}_{12}^{(k)}$\textbf{\ }and\textbf{\ }$%
\mathbf{P}_{21}^{(k)}$, respectively, and note in advance that matrices $%
\mathbf{G}_{(k)}$\ play an important role in the PNCMO theory. These are
called below the intersubset coupling matrices (of the $k$th order). In the
above-defined terms, the standard form of the correction $\mathbf{P}_{(k)}$
is as follows 
\begin{equation}
\mathbf{P}_{(k)}\mathbf{=}\left\vert 
\begin{array}{cc}
\mathbf{X}_{(+)}^{(k)} & -2\mathbf{G}_{(k)} \\ 
-2\mathbf{G}_{(k)}^{+} & \mathbf{X}_{(-)}^{(k)}%
\end{array}%
\right\vert .  \tag{1.1.5}
\end{equation}%
Let us now turn to particular values of the order parameter $(k)$.

The zero order member ($\mathbf{P}_{(0)}$) of the power series of Eq.(1.1.2)
follows from solution of the matrix problem shown in Eq.(1.1.1), where the
zero order Hamiltonian matrix ($\mathbf{H}_{(0)})$ stands instead of $%
\mathbf{H}$. As it may be easily shown, the matrix $\mathbf{P}_{(0)}$\ of
the following simple form 
\begin{equation}
\mathbf{P}_{(0)}\mathbf{=}\left\vert 
\begin{array}{cc}
2\mathbf{I} & \mathbf{0} \\ 
\mathbf{0} & \mathbf{0}%
\end{array}%
\right\vert  \tag{1.1.6}
\end{equation}%
meets the relevant three equations, where $\mathbf{I}$ is an $n-$dimensional
unit matrix. At the same time, $\mathbf{P}_{(0)}$\ of Eq.(1.1.6) reflects
the initial occupation numbers of our basis orbitals $\varphi _{(+)i}$ ($%
i=1,2...n$) and $\varphi _{(-)l}$ ($l=n+1,n+2,...N$) equal to 2 and 0,
respectively, and thereby corresponds to the ground state of our system. In
terms of designations used in Eq. (1.1.5), we alternatively obtain that%
\begin{equation}
\mathbf{X}_{(+)}^{(0)}=2\mathbf{I},\qquad \mathbf{X}_{(-)}^{(0)}=\mathbf{%
0,\qquad G}_{(0)}=\mathbf{0}.  \tag{1.1.7}
\end{equation}%
Expressions for the subsequent members $\mathbf{P}_{(k)},k=1,2,3...$ of the
series concerned, however, take a much more involved form. In particular,
the intersubset coupling matrices $\mathbf{G}_{(k)}$ ($k=1,2,3...$) and
thereby the off-diagonal blocks of any correction $\mathbf{P}%
_{(k)}(k=1,2,3...)$ are determined by the following matrix equations\ \ 
\begin{equation}
\mathbf{E}_{(+)}\mathbf{G}_{(k)}+\mathbf{G}_{(k)}\mathbf{E}_{(-)}+\mathbf{W}%
_{(k)}\mathbf{=0,}  \tag{1.1.8}
\end{equation}%
the last $\mathbf{G}_{(k)}$-free terms of which are expressible via the
relevant matrices of lower orders ($\mathbf{G}_{(k-1)},$ $\mathbf{G}%
_{(k-2)}, $ $etc$), as well as via blocks $\mathbf{T,R}$ and \ $\mathbf{Q}$
of the first order Hamiltonian matrix $\mathbf{H}_{(1)}$ of Eq.(1.1)
[37-40], e.g. 
\begin{align}
\mathbf{W}_{(1)} =&\mathbf{R,\qquad W}_{(2)}=\mathbf{TG}_{(1)}-\mathbf{G}%
_{(1)}\mathbf{Q,}  \tag{1.1.9} \\
\mathbf{W}_{(3)} =&\mathbf{TG}_{(2)}-\mathbf{G}_{(2)}\mathbf{Q-(RG}%
_{(1)}^{+}\mathbf{G}_{(1)}+\mathbf{G}_{(1)}\mathbf{G}_{(1)}^{+}\mathbf{%
R),\quad }etc\mathbf{.}  \nonumber
\end{align}%
By contrast, the intrasubset population matrices of the $k$th order ($%
\mathbf{X}_{(+)}^{(k)}$ and $\mathbf{X}_{(-)}^{(k)}$) are representable
algebraically via sums of products of matrices $\mathbf{G}_{(k)}$ of lower
orders [37-40], e.g.%
\begin{align}
\mathbf{X}_{(+)}^{(1)} =&\mathbf{X}_{(-)}^{(1)}=\mathbf{0,\quad X}%
_{(+)}^{(2)}=-2\mathbf{G}_{(1)}\mathbf{G}_{(1)}^{+},\quad \mathbf{X}%
_{(-)}^{(2)}=2\mathbf{G}_{(1)}^{+}\mathbf{G}_{(1)},  \tag{1.1.10} \\
\mathbf{X}_{(+)}^{(3)} =&-2(\mathbf{G}_{(1)}\mathbf{G}_{(2)}^{+}+\mathbf{G}%
_{(2)}\mathbf{G}_{(1)}^{+}),\quad \mathbf{X}_{(-)}^{(3)}=2(\mathbf{G}%
_{(1)}^{+}\mathbf{G}_{(2)}+\mathbf{G}_{(2)}^{+}\mathbf{G}_{(1)}^{{}}),\quad
etc.  \nonumber
\end{align}%
Moreover, the relation%
\begin{equation}
Tr\mathbf{X}_{(+)}^{(k)}=-Tr\mathbf{X}_{(-)}^{(k)}  \tag{1.1.11}
\end{equation}%
may be easily proven for any $k$. This relation implies coincidence between
the population lost by all orbitals of the subset $\{\varphi _{(+)}\}$\ and
\ that acquired by all orbitals of its counterpart $\{\varphi _{(-)}\}$ (or
vice versa) within each order $k$ separately due to the intersubset
interaction. Thus, it may be interpreted as the charge conservation
condition for subsets $\{\varphi _{(+)}\}$\ and $\{\varphi _{(-)}\}.$ An
important implication of relations like those of Eq. (1.1.10) consists in
representability of population alterations of basis orbitals $\varphi
_{(+)i} $ and $\varphi _{(-)l}$ ($X_{(+)ii}^{(k)}$ and $X_{(-)ll}^{(k)}$) of
each order ($k$) in the form of sums of increments of orbitals of the
opposite subset [37, 39, 41, 42], viz. 
\begin{equation}
X_{(+)ii}^{(k)}=-\mathop{\displaystyle \sum }%
\limits_{(-)m}x_{(+)i,(-)m}^{(k)},\quad X_{(-)ll}^{(k)}=\mathop{%
\displaystyle \sum }\limits_{(+)j}x_{(-)l,(+)j}^{(k)},  \tag{1.1.12}
\end{equation}%
where $x_{(+)i,(-)m}^{(k)}$\ (and $x_{(-)l,(+)j}^{(k)}$) are expressible as
follows\ 
\begin{align}
x_{(+)i,(-)m}^{(2)} =&2(G_{(1)im})^{2},\quad
x_{(+)i,(-)m}^{(3)}=4G_{(1)im}G_{(2)im},  \tag{1.1.13} \\
x_{(+)i,(-)m}^{(4)} =&4G_{(1)im}G_{(3)im}+2G_{(1)im}(\mathbf{G}_{(1)}%
\mathbf{G}_{(1)}^{+}\mathbf{G}_{(1)})_{im}+2(G_{(2)im})^{2},\quad etc. 
\nonumber
\end{align}%
and describe partial populations of the $k$th order transferred between
individual orbitals of opposite subsets. Sums of Eq.(1.1.12) over $(-)m$ and
over $(+)j$ embrace all initially-vacant orbitals $\varphi _{(-)m}$ $%
m=n+1,n+2,...N$ \ and all initially-occupied ones $\varphi _{(+)j}$ $%
j=1,2...n$, respectively. Moreover, it may be easily shown that 
\begin{equation}
x_{(+)i,(-)m}^{(k)}=x_{(-)m,(+)i}^{(k)}  \tag{1.1.14}
\end{equation}%
in accordance with the expectation. For total populations of orbitals $%
\varphi _{(+)i}$ and $\varphi _{(-)l},$ we accordingly obtain 
\begin{align}
X_{(+)ii} =&2-\mathop{\displaystyle \sum }\limits_{k=2}^{\infty }%
\mathop{\displaystyle \sum }\limits_{(-)m}x_{(+)i,(-)m}^{(k)}=2-%
\mathop{\displaystyle \sum }\limits_{(-)m}x_{(+)i,(-)m}^{{}},  \nonumber \\
\quad X_{(-)ll} =&\mathop{\displaystyle \sum }\limits_{k=2}^{\infty }%
\mathop{\displaystyle \sum }\limits_{(+)j}x_{(-)l,(+)j}^{(k)}=%
\mathop{\displaystyle \sum }\limits_{(+)j}x_{(-)l,(+)j}^{{}}, 
\tag{1.1.15}
\end{align}%
where $x_{(+)i,(-)m}^{{}}$\ and $x_{(-)l,(+)j}^{{}}$\ are the overall
partial populations transferred between the relevant pairs of orbitals [Note
that Eq.(1.1.7) also is invoked here along with the first relation of
Eq.(1.1.10)].

As is seen from the first relation of Eq.(1.1.13), the second order
increment ($x_{(+)i,(-)m}^{(2)}$) to the partial population ($%
x_{(+)i,(-)m}^{{}}$) transferred between orbitals $\varphi _{(+)i}$ and $%
\varphi _{(-)m}$ is an \textit{a priori} positive quantity. This implies
that any initially-occupied orbital $\varphi _{(+)i}$ looses its population
due to perturbation in accordance with the expectation (the minus sign of
Eq.(1.1.12) should be taken into consideration). Accordingly, any
initially-vacant orbital ($\varphi _{(-)m}$) acquires an additional
population and this result also causes no surprise. Meanwhile, the signs of
the relevant increments of higher orders ($x_{(+)i,(-)m}^{(3)},$ $%
x_{(+)i,(-)m}^{(4)},$ $etc.$) cannot be established \textit{a priori}.
Nevertheless, the expressions of Eq.(1.1.13) allow us to formulate the
following rule: The increment $x_{(+)i,(-)m}^{(k)}$\ is a positive
(negative) quantity, if the elements $G_{(k-1)im},G_{(k-2)im},etc.$
contained within the relevant definition are of the same (opposite) signs
[41], For example, $x_{(+)i,(-)m}^{(3)}>0,$ if both $G_{(1)im}$\ and $%
G_{(2)im}$\ are either of positive signs or of negative ones [Given that an
element $G_{(k)im}$\ contains two additive components $G_{(k)im}^{a}$\ and $%
G_{(k)im}^{b},$\ coinciding (opposite) signs of the latter become of
importance in the formation of the absolute value of square $(G_{(k)im})^{2}$%
, e.g. of ($G_{(2)im})^{2}$]. This rule forms the basis for conditions
determining stabilization (destabilization) of a certain system due to
perturbation (Subsect. 2.3).

As is seen from the above overview, the series of intersubset coupling
matrices $\mathbf{G}_{(k)},k=1,2,3...$ prove to be the principal quantities
of the present theory, and these are determined by matrix equations shown in
Eq.(1.1.8). The latter are easily solvable and yield the usual
"perturbational" fractions containing resonance parameters and energy
diffences in their numerators and denominators, respectively, under an
assumption that $n=s=1$, i.e. in the particular case of one-dimensional
parameters $\varepsilon _{(+)},$ $\varepsilon _{(-)},$ $\tau ,\rho $ and \ $%
\kappa $ instead of (sub)matrices $\mathbf{E}_{(+)},$ $\mathbf{E}_{(-)},$ $%
\mathbf{T,R}$ and \ $\mathbf{Q.}$ Generally, however, we have to solve
matrix equations of the following form 
\begin{equation}
\mathbf{AX}+\mathbf{XB}=a\mathbf{C,}  \tag{1.1.16}
\end{equation}%
where $\mathbf{X}$ is the matrix being sought, $\mathbf{A}$ and $\mathbf{B}$
are known matrices and $a$ is a constant. According to the theory of these
equations [43], a unique solution exists under certain conditions imposed on
matrices $\mathbf{A}$ and $\mathbf{B.}$ Using the designations underlying
Eq.(1.1.8), the solution concerned is representable in the form of the
following integral [25, 36]%
\begin{equation}
\mathbf{G}_{(k)}=\int\limits_{0}^{\infty }\exp [\mathbf{E}_{(+)}t]\mathbf{W}%
_{(k)}\exp [\mathbf{E}_{(-)}t]dt,  \tag{1.1.17}
\end{equation}%
which evidently yields no explicit expressions for matrices $\mathbf{G}%
_{(k)} $. In this connection, the present version of the PNCMO theory is
called the implicit one. Nevertheless, explicit solutions are obtainable in
some particular cases as discussed in Section 2 in a detail. Meanwhile, the
present section is devoted to some relations between corrections $\mathbf{P}%
_{(k)},$ on the one hand, and other characteristics of the same systems, on
the other hand, that have been found in the framework of the implicit form
of the PNCMO theory.

As already mentioned, importance of the CBO matrix $\mathbf{P}$\ as a
quantum chemical characteristic is beyond any doubt. This matrix, however,
yields the overall charge (re)distribution only, but not those of individual
pairs of electrons. Meanwhile, just the shifts (displacements) of separate
electron pairs form the basis of the qualitative chemical thinking (cf. the
well-known 'curly arrow chemistry' (see e.g. [18])). Moreover, these pairs
are assumed to be predominantly localized on individual chemical bonds
and/or to build up the so-called lone electron pairs usually belonging to
heteroatoms.

It is evident that squares of individual one-electron orbitals (MOs) may be
expected to yield partial charge (re)distributions associated with separate
electron pairs. The standard (canonical) MOs (CMOs), however, usually are
delocalized over the whole system under study [4, 11] and thereby cannot be
ascribed to individual chemical bonds. As a result, CMOs hardly are suitable
for the above-specified purpose. Fortunately, one-electron orbitals as such
are not unique (as opposed to the DM). This implies existance of other sets
of MOs that offer a quantum chemical analogue of the above-described
classical perspective on molecular systems. These alternative MOs are
usually referred to as localized MOs(LMOs) and/or non-canonical MOs (NCMOs)
[20-24, 44-46]. The most popular way of derivation of the latter consists in
transforming the set of CMOs using various localization criteria [20,
47-50]. Again, NCMOs are also alternatively obtainable directly by invoking
the Brillouin theorem [20-26, 28, 36, 51-55]. Just the latter approach is
invoked below in this article for the following reasons: First, general
expressions for NCMOs prove to be then obtainable for any system described
by the Hamiltonian matrix $\mathbf{H}$ of Eq.(1.1). Second, terms of the
power series for the NCMO representation matrix are expressible in terms of
entire blocks of the matrix $\mathbf{H}$\ and thereby these are directly
comparable [36, 51, 52] to those for the matrix $\mathbf{P}$ of
Eqs.(1.1.6)-(1.1.10). Finally, zero order NCMOs may be chosen to coincide
with individual basis orbitals. Consequently, changing extents of their
subsequent delocalization due to perturbation are expected to reflect the
classical shifts (displacements) of separate electron pairs. Thus, let us
now turn to application of the Brillouin theorem to our Hamiltonian matrix
of Eq.(1.1).

\textbf{1.2. The block-diagonalization problem for the Hamiltonian matrix,
and its perturbative solution}

Among particular forms of the Brillouin theorem there is a zero value
requirement for an off-diagonal element of the Fockian operator referring to
an occupied and a vacant MO [11]. In its matrix form, this requirement
resolves itself into the zero matrix condition for the occupied-vacant
off-diagonal block (submatrix) of the total Fockian (or Hamiltonian) matrix
in the basis of non-canonical MOs (NCMOs) being sought [21-23, 25-28, 36,
51-55].

Let us now dwell on the case of the Hamiltonian matrix $\mathbf{H}$\ of
Eq.(1.1). Let us look for an unitary matrix $\mathbf{C}$\ that transforms
the above-mentioned matrix $\mathbf{H}$\ into a block-diagonal form
containing a direct sum of $n\times n-$ and $s\times s-$ dimensional
submatrices $\mathbf{E}_{1}^{{}}$\ and $\mathbf{E}_{2},$ viz.%
\begin{equation}
\mathbf{H}^{\prime }=\mathbf{C}^{+}\mathbf{HC=}\left\vert 
\begin{array}{cc}
\mathbf{E}_{1}^{{}} & \mathbf{0} \\ 
\mathbf{0}^{{}} & \mathbf{E}_{2}^{{}}%
\end{array}%
\right\vert .  \tag{1.2.1}
\end{equation}%
Moreover, the unitarity condition of the following form%
\begin{equation}
\mathbf{C}^{+}\mathbf{C=I}  \tag{1.2.2}
\end{equation}%
ensures the ortogonality of NCMOs. It also deserves mention that Eq.(1.2.1)
turns into the diagonality requirement under an assumption that $n=s=1$
[28], i.e. for the two-electron prototype described by the matrix $\mathbf{h}
$ of Eq.(1.2). Further, our problem may be reformulated as follows%
\begin{equation}
\mathbf{HC=CE,}  \tag{1.2.3}
\end{equation}%
where $\mathbf{E}$ stands for the total block-diagonal matrix of the
right-hand side of Eq.(1.2.1). This alternative form, in turn, proves to be
a generalization of the usual eigenvalue equation for the matrix $\mathbf{h}$
to the case of multidimensional (non-commutative) elements. In this
connection, the submatrices $\mathbf{E}_{1}^{{}}$\ and $\mathbf{E}_{2}$ of
Eq.(1.2.1) may be conveniently called the eigenblocks of the matrix $\mathbf{%
H}$ [28]$\mathbf{.}$ Accordingly, the problem of Eq.(1.2.3) will be referred
to as the eigenblock equation for the same matrix.

As with basis orbitals $\{\varphi \}$, the NCMOs also may be collected into
two subsets, viz. the $n-$dimensional subset of occupied NCMOs and the $s-$%
dmensional one of their vacant counterparts correspondingly denoted below by 
$\{\psi _{1}\}$ and $\{\psi _{2}\}$ [In some cases, the subscripts $(+)$ and 
$(-)$ are more convenient and these are used instead of 1 and 2].
Accordingly, $\ $notations $\mid \Psi _{1}>$ and $\mid \Psi _{2}>$ will
stand for respective ket-vectors (row-matrices). Passing from the initial
basis set $\{\varphi \}$ to that of NCMOs $\{\psi \}$ will be then
represented by the following relation 
\begin{equation}
\mid \Psi >=\mid \Phi >\mathbf{C}  \tag{1.2.4}
\end{equation}%
where $\mid \Psi >$\ and $\mid \Phi >$ are total row-matrices (ket-vectors)
of NCMOs and of basis functions, respectively, i.e.
\begin{align}
&\mid \Psi >=\mid \mid \Psi _{1}>\mid \Psi _{2}>>\equiv \mid \mid \Psi
_{(+)}>\mid \Psi _{(-)}>>\mathbf{,}  \nonumber \\
&\mid \Phi >=\mid \mid \Phi _{1}>\mid \Phi _{2}>>\equiv \mid \mid \Phi
_{(+)}>\mid \Phi _{(-)}>>.  \tag{1.2.5}
\end{align}%
As with the CBO matrix $\mathbf{P}$ (Subsect. 1.1), the transformation
matrix $\mathbf{C}$ also has been sought in the form of four submatrices
(blocks) $\mathbf{C}_{11}^{{}}$, $\mathbf{C}_{12}^{{}},$ $\mathbf{C}%
_{21}^{{}}$ and $\mathbf{C}_{22}^{{}}$ of appropriate dimensions, each of
them consisting of sums over $k$ of the relevant increments of the $k$th
order ($\mathbf{C}_{11}^{(k)}$ , $\mathbf{C}_{12}^{(k)},$ $\mathbf{C}%
_{21}^{(k)}$ and $\mathbf{C}_{22}^{(k)},$ respectively). Moreover, the
eigenblocks $\mathbf{E}_{1}^{{}}$\ and $\mathbf{E}_{2}$ also have been
represented similarly [55], i.e. 
\begin{equation}
\mathbf{E}_{1}=\sum\limits_{k=0}^{\infty }\mathbf{E}_{1}^{(k)},\qquad 
\mathbf{E}_{2}=\sum\limits_{k=0}^{\infty }\mathbf{E}_{2}^{(k)},  \tag{1.2.6}
\end{equation}%
where $\mathbf{E}_{1}^{(k)}$\ and $\mathbf{E}_{2}^{(k)}$\ are the increments
of the $k$th order. Since the zero order member $\mathbf{H}_{(0)}$\ of our
Hamiltonian matrix $\mathbf{H}$\ of Eq.(1.1) complies with the
block-diagonality condition from the outset, the following equalities%
\begin{equation}
\mathbf{C}_{11}^{(0)}=\mathbf{C}_{22}^{(0)}=\mathbf{I},\qquad \mathbf{C}%
_{12}^{(0)}=\mathbf{C}_{21}^{(0)}=\mathbf{0},\qquad \mathbf{E}_{1}^{(0)}=%
\mathbf{E}_{(+)},\qquad \mathbf{E}_{2}^{(0)}=-\mathbf{E}_{(-)}  \tag{1.2.7}
\end{equation}%
may be accepted \textit{a priori} and and these imply coincidence of the
total zero order matrix $\mathbf{C}_{(0)}^{{}}$ with the unit matrix $%
\mathbf{I}$. Individual ket-vectors $\mid \Psi _{1}>$ and $\mid \Psi _{2}>$
take then the following form 
\begin{align}
&\mid \Psi _{1}>=\mid \Phi _{1}>\mathbf{C}_{11}+\mid \Phi _{2}>\mathbf{C}%
_{21}\mathbf{,}  \nonumber \\
&\mid \Psi _{2}>=\mid \Phi _{1}>\mathbf{C}_{12}+\mid \Phi _{2}>\mathbf{C}%
_{22}\mathbf{,}  \tag{1.2.8}
\end{align}%
where 
\begin{equation}
\mathbf{C}_{11}=\mathbf{I+}\sum\limits_{k=1}^{\infty }\mathbf{C}%
_{11}^{(k)},\quad \mathbf{C}_{22}=\mathbf{I+}\sum\limits_{k=1}^{\infty }%
\mathbf{C}_{22}^{(k)},\quad \mathbf{C}_{12}=\sum\limits_{k=1}^{\infty }%
\mathbf{C}_{12}^{(k)},\quad \mathbf{C}_{21}=\sum\limits_{k=1}^{\infty }%
\mathbf{C}_{21}^{(k)}.  \tag{1.2.9}
\end{equation}

The expression of Eq.(1.2.8) is a generalization of the well-known LCAO form
for the two MOs of our prototype described by the simple matrix $\mathbf{h}$
of Eq.(1.2), where submatrices of the matrix $\mathbf{C}$ stand instead of
usual (one-dimensional) coefficients. It is also seen that (sub)matrices $%
\mathbf{C}_{12}$ and $\mathbf{C}_{21}$ represent the intersubset
delocalization (intersubset tails) of NCMOs, whereas $\mathbf{C}_{11}$ and $%
\mathbf{C}_{22}$ reflect the consequent renormalization of the latter. In
this connection, $\mathbf{C}_{11}$ and $\mathbf{C}_{22}$ are referred to
below as renormalization matrices. Analogously, $\mathbf{C}_{11}^{(k)}$ and $%
\mathbf{C}_{22}^{(k)}$ are called renormalization matrices of the $k$th
order. Finally, substituting Eq.(1.2.9) into Eq.(1.2.8) shows that our NCMOs
are of the basis-orbital-and-tail constitution. In other words, one-to-one
correspondence takes place between individual NCMOs $\psi _{1,i}$\ ($\psi
_{2,l}$) and basis orbitals $\varphi _{(+)i}$\ ($\varphi _{(-)l}$) so that
delocalization of the former is a small correction only. Thus, the
assumption of Eq.(1.2.7) is equivalent to confinement to NCMOs of the
above-described particular structure. As a result, we largely reduce the
extent of ambiguity in determining NCMOs.\ \ 

The very procedure of solution of the present matrix problem for NCMOs [36,
52] also closely resembles that for the CBO matrix $\mathbf{P}$ (Subsect.
1.1): The above-specified forms of matrices $\mathbf{C}$\ and $\mathbf{E}$\
have been substituted into Eqs.(1.2.1) and (1.2.2) and terms of the same
order ($k$) have been collected. Moreover, the similarity concerned embraces
also the resulting matrices $\mathbf{C}$\ and $\mathbf{P.}$ This important
and somewhat surprising point deserves a more detailed discussion.

Comparative analysis of the two non-canonical problems [36, 52] showed that
the off-diagonal blocks of the correction $\mathbf{C}_{(k)}$\ (i.e.
submatrices $\mathbf{C}_{12}^{(k)}$ and $\mathbf{C}_{21}^{(k)}$) actually
are determined by matrix equations of the form shown in Eq.(1.1.8), as it
was the case with the relevant blocks of the matrix $\mathbf{P}$ (Subsect.
1.1). Moreover, equations for blocks of the first and second orders (i.e.
for $\mathbf{C}_{12}^{(1)}$ and $\mathbf{C}_{21}^{(1)}$, as well as for $%
\mathbf{C}_{12}^{(2)}$ and $\mathbf{C}_{21}^{(2)}$) contain the same $%
\mathbf{G}_{(k)}-$free members $\mathbf{W}_{(1)}$\ and $\mathbf{W}_{(2)}$\
shown in Eq.(1.1.9). Consequently, the off-diagonal blocks of corrections $%
\mathbf{C}_{(1)}$\ and $\mathbf{C}_{(2)}$\ are representable via our former
principal matrices \ $\mathbf{G}_{(1)}$\ and $\mathbf{G}_{(2)},$\
respectively, viz. 
\begin{equation}
\mathbf{C}_{12}^{(1)}=-\mathbf{C}_{21}^{(1)+}=\mathbf{G}_{(1)},\qquad 
\mathbf{C}_{12}^{(2)}=-\mathbf{C}_{21}^{(2)+}=\mathbf{G}_{(2)}.  \tag{1.2.10}
\end{equation}%
Analogously, diagonal blocks of the same corrections are expressible
algebraically as follows%
\begin{align}
\mathbf{C}_{11}^{(1)} =&\mathbf{C}_{22}^{(1)}=\mathbf{0},\qquad \mathbf{C}%
_{11}^{(2)}=-\frac{1}{2}\mathbf{G}_{(1)}\mathbf{G}_{(1)}^{+},\qquad \mathbf{C%
}_{22}^{(2)}=-\frac{1}{2}\mathbf{G}_{(1)}^{+}\mathbf{G}_{(1)},  \nonumber \\
\mathbf{C}_{11}^{(3)} =&-\frac{1}{2}(\mathbf{G}_{(1)}\mathbf{G}_{(2)}^{+}+%
\mathbf{G}_{(2)}^{{}}\mathbf{G}_{(1)}^{+}),etc.  \tag{1.2.11}
\end{align}%
and these relations closely resemble their counterparts of Eq.(1.1.10) for $%
k\leqslant 3$. These results evidently imply proportionalities between
respective submatrices of corrections $\mathbf{C}_{(k)}$\ and $\mathbf{P}%
_{(k)}$\ and thereby a high extent of similarity between entire matrices $%
\mathbf{C}$ and $\mathbf{P}$. Consequently, both NCMOs (LMOs) and rows
(columns) of the CBO matrix are expected to be characterized by a similar
dependence upon the specific structure of the system under interest. On the
same basis, both NCMOs (LMOs) and the one-electron DM have been concluded to
belong to the localized way of describing electronic structures in general.

The above-described simple proportionalities between individual blocks of
corrections \ $\mathbf{C}_{(k)}$\ and $\mathbf{P}_{(k)},$ however, become
replaced by more involved interdependences with the growing value of the
order parameter ($k$). The immediate reason for this state of things is that
the $\mathbf{G}_{(k)}-$free terms of matrix equations of Eq. (1.1.8)
determining the off-diagonal submatrices of matrices $\mathbf{C}_{(k)}$\ and 
$\mathbf{P}_{(k)}$\ are no longer uniform for $k>2$. For example, the
equation for the submatrix $\mathbf{C}_{12}^{(3)}$\ contains the following
new matrix [52, 55] 
\begin{equation}
\widetilde{\mathbf{W}}_{(3)}=\mathbf{TG}_{(2)}-\mathbf{G}_{(2)}\mathbf{Q-}%
\frac{1}{2}\mathbf{(RG}_{(1)}^{+}\mathbf{G}_{(1)}+\mathbf{G}_{(1)}\mathbf{G}%
_{(1)}^{+}\mathbf{R+\mathbf{G}}_{(1)}\mathbf{R\mathbf{G}}_{(1)}^{+}\mathbf{)}
\tag{1.2.12}
\end{equation}%
instead of the former matrix $\mathbf{W}_{(3)}$\ of Eq.(1.1.9) and thereby
yields another solution denoted below by $\widetilde{\mathbf{G}}_{(3)}.$\
Nevertheless, matrices $\mathbf{G}_{(k)}$ and $\widetilde{\mathbf{G}}_{(k)}$
prove to be interrelated as discussed in the next Subsection. Before
finishing the comparison of matrices $\mathbf{C}$ and $\mathbf{P}$\ it
deserves mention that both the very fact of their similarity and the limited
extent of the latter may be easily understood by invoking the prototype of
our systems described by the $2\times 2-$dimensional Hamiltonian matrix $%
\mathbf{h}$ of Eq.(1.2). Indeed, the CBO matrix of this simple system ($%
\mathbf{p}$) is determined by the shape of its only occupied MO and,
consequently, matrices $\mathbf{c}$ and $\mathbf{p}$\ contain the same
perturbational fraction and thereby are interrelated. To a certain extent,
this interrelation is retained also when passing to the case of two
multidimensional interacting subsets $\{\varphi _{(+)}\}$ and $\{\varphi
_{(-)}\}.$

Finally, the eigenblocks of the matrix $\mathbf{H}$\ ($\mathbf{E}_{1}^{{}}$\
and $\mathbf{E}_{2}$) also are expressible algebraically via entire matrices 
$\mathbf{G}_{(1)},$ $\mathbf{G}_{(2)}$\ and $\widetilde{\mathbf{G}}_{(3)}$
as follows [55]%
\begin{align}
\mathbf{E}_{1}^{{}} =&\mathbf{E}_{(+)}^{{}}+\mathbf{T-}\frac{1}{2}(\mathbf{%
RG}_{(1)}^{+}+\mathbf{G}_{(1)}\mathbf{R}^{+})\mathbf{-}\frac{1}{2}(\mathbf{RG%
}_{(2)}^{+}+\mathbf{G}_{(2)}\mathbf{R}^{+})\mathbf{-}\frac{1}{2}(\mathbf{R}%
\widetilde{\mathbf{G}}_{(3)}^{+}+\widetilde{\mathbf{G}}_{(3)}\mathbf{R}^{+})
\nonumber \\
&-\frac{1}{8}(\mathbf{RG}_{(1)}^{+}\mathbf{G}_{(1)}\mathbf{G}_{(1)}^{+}+%
\mathbf{G}_{(1)}\mathbf{R}^{+}\mathbf{G}_{(1)}\mathbf{G}_{(1)}^{+}+\mathbf{G}%
_{(1)}\mathbf{G}_{(1)}^{+}\mathbf{RG}_{(1)}^{+}+\mathbf{G}_{(1)}\mathbf{G}%
_{(1)}^{+}\mathbf{G}_{(1)}\mathbf{R}^{+})...  \nonumber \\
\mathbf{E}_{2}^{{}} =&-\mathbf{E}_{(-)}^{{}}+\mathbf{Q+}\frac{1}{2}(\mathbf{%
G}_{(1)}^{+}\mathbf{R+R}^{+}\mathbf{G}_{(1)})\mathbf{+}\frac{1}{2}(\mathbf{G}%
_{(2)}^{+}\mathbf{R+R}^{+}\mathbf{G}_{(2)})\mathbf{+}\frac{1}{2}(\widetilde{%
\mathbf{G}}_{(3)}^{+}\mathbf{R+R}^{+}\widetilde{\mathbf{G}}_{(3)})  \nonumber
\\
&+\frac{1}{8}(\mathbf{R}^{+}\mathbf{G}_{(1)}\mathbf{G}_{(1)}^{+}\mathbf{G}%
_{(1)}+\mathbf{G}_{(1)}^{+}\mathbf{RG}_{(1)}^{+}\mathbf{G}_{(1)}+\mathbf{G}%
_{(1)}^{+}\mathbf{G}_{(1)}\mathbf{R}^{+}\mathbf{G}_{(1)}+\mathbf{G}_{(1)}^{+}%
\mathbf{G}_{(1)}\mathbf{G}_{(1)}^{+}\mathbf{R)...}  \nonumber \\
&  \tag{1.2.13}
\end{align}

\begin{center}
\textbf{1.3. Interrelations between occupation numbers of basis orbitals and
delocalization coefficients of NCMOs}
\end{center}

As already mentioned, matrices $\mathbf{P}$ and $\mathbf{C}$ contain
information about the overall charge (re)distribution and about that
associated with individual electron pairs, respectively. Again, the same
matrices were shown to resemble one another in the above subsection. This
allows us to expect a certain interdependence to exist between the
above-mentioned two patters of charge (re)distribution. Now, we will turn to
verification of this expectation.

It deserves an immediate emphasizing that the interrelation being sought is
of a non-trivial nature. Indeed, matrices $\mathbf{P}$ and $\mathbf{C}$
contain distinct submatrices $\mathbf{G}_{(k)}$ $\ $and $\widetilde{\mathbf{G%
}}_{(k)}$ for higher values of the order parameter ($k$) that are determined
by different matrix equations (Subsect. 1.2). Thus, we will invoke here an
alternative (indirect) way of derivation of the one-electron DM on the basis
of projector to the subspace of occupied one-electron orbitals (MOs).
Instead of usual (canonical) MOs [29], however, we will use the subset of
occupied NCMOs $\{\psi _{1}\}$ of subsection 1.2. Comparison of expressions
for the matrix $\mathbf{P}$ derived directly (as described in the subsection
1.1) and indirectly (using NCMOs) is then expected to yield interrelations
between submatrices $\mathbf{G}_{(k)}$ $\ $and $\widetilde{\mathbf{G}}_{(k)}$
for $k$=3,4,... and thereby between charge (re)distributions under our
interest.

Let us start with the general definition of the one-electron DM $\mathbf{P}(%
\mathbf{r}\mid \mathbf{r}^{\prime })$\ as a projector [29] to the subset of
occupied NCMOs $\{\psi _{1}\},$\ viz. 
\begin{equation}
\mathbf{P}(\mathbf{r}\mid \mathbf{r}^{\prime })=2\left\vert \Psi _{1}(%
\mathbf{r)}\right\rangle \left\langle \Psi _{1}(\mathbf{r}^{\prime
})\right\vert ,  \tag{1.3.1}
\end{equation}%
where $\left\vert \Psi _{1}(\mathbf{r)}\right\rangle $\ and $\left\langle
\Psi _{1}(\mathbf{r}^{\prime })\right\vert $\ correspondingly stand for the
\ ket-vector of occupied NCMOs and for the relevant bra-vector (i.e. row-
and column matrices of the latter, respectively), whilst $\mathbf{r}$ and $%
\mathbf{r}^{\prime }$ are vectors representing positions of an electron in
the real space. After substituting the expressions like that of Eq.(1.2.8)
into the right-hand side of Eq.(1.3.1), a new representation of the same DM $%
\mathbf{P}(\mathbf{r}\mid \mathbf{r}^{\prime })$\ follows in terms of entire
subsets (vectors) $\left\vert \Phi _{1}\right\rangle $ and $\left\vert \Phi
_{2}\right\rangle $ (or $\left\vert \Phi _{(+)}\right\rangle $ and $%
\left\vert \Phi _{(-)}\right\rangle $), viz. 
\begin{equation}
\mathbf{P}(\mathbf{r}\mid \mathbf{r}^{\prime })=\mathop{\displaystyle \sum }%
\limits_{L,M=1}^{2}\left\vert \Phi _{L}(\mathbf{r)}\right\rangle \mathbf{P}%
_{LM}\left\langle \Phi _{M}(\mathbf{r}^{\prime })\right\vert ,  \tag{1.3.2}
\end{equation}%
where $\mathbf{P}_{LM}$ ($L=1,2;M=1,2)$\ are multidimensional elements
(blocks) of the representation of the DM in terms of two subsets of basis
functions $\{\varphi _{1}\}$ and $\{\varphi _{2}\}$ introduced in the
subsection 1.1. The expression of Eq.(1.3.2) evidently is a matrix
generalization of the well-known bilinear form of the DM in terms of
individual basis functions [29]. At the same time, Eq.(1.3.2) yields the
following formulae for $\mathbf{P}_{LM}$\ in terms of submatrices of the
NCMO representation matrix $\mathbf{C}$ of subsection 1.2, viz.%
\begin{equation}
\mathbf{P}_{11}=2\mathbf{C}_{11}\mathbf{C}_{11}^{+};\quad \mathbf{P}_{22}=2%
\mathbf{C}_{21}\mathbf{C}_{21}^{+};\quad \mathbf{P}_{12}=2\mathbf{C}_{11}%
\mathbf{C}_{21}^{+},  \tag{1.3.3}
\end{equation}%
where $\mathbf{C}_{11}$\textbf{\ }and\textbf{\ }$\mathbf{C}_{21}$ are
defined by Eqs.(1.2.8) and (1.2.9).

If we recall now that the one-electron DM $\mathbf{P}(\mathbf{r}\mid \mathbf{%
r}^{\prime })$\ and thereby its particular representation in terms of
subsets $\{\varphi _{(+)}\}$ and $\{\varphi _{(-)}\}$ are unique quantum
chemical characteristics of molecules, the elements $\mathbf{P}_{LM}$\ of
Eq.(1.3.2) necessarily coincide with submatrices of the CBO matrix $\mathbf{P%
}$, i.e. $\mathbf{P}_{11},\mathbf{P}_{22}$ and $\mathbf{P}_{12}$ shown in
the left-hand side of Eq.(1.1.3). The relations of Eq.(1.3.3) then form the
basis of the indirect way of derivation of the matrix $\mathbf{P.}$
Moreover, the relevant results are expected to coincide with those obtained
directly and described in the subsection 1.1. For $k$=1 and $k$=2, such a
coincidence may be easily verified. Meanwhile, for higher values of the
order parameter ($k$), the last relation of Eq.(1.3.3) yields non-trivial
connections between pairs of matrices $\mathbf{G}_{(k)}$ $\ $and $\widetilde{%
\mathbf{G}}_{(k)}$ [52]. For example, after substituting $-2\mathbf{G}_{(3)}$
$\ $and 
\begin{equation}
2[\mathbf{C}_{21}^{(3)+}+\mathbf{C}_{11}^{(2)}\mathbf{C}_{21}^{(1)+}]\equiv
-2[\mathbf{C}_{12}^{(3)}+\mathbf{C}_{11}^{(2)}\mathbf{C}_{12}^{(1)}] 
\tag{1.3.4}
\end{equation}%
into the left- and right-hand sides of the above-mentioned relation,
respectively, we obtain that 
\begin{equation}
\mathbf{G}_{(3)}=\widetilde{\mathbf{G}}_{(3)}-\frac{1}{2}\mathbf{G}_{(1)}%
\mathbf{G}_{(1)}^{+}\mathbf{G}_{(1)}.  \tag{1.3.5}
\end{equation}%
For k=4, an analogous procedure yields the following result 
\begin{equation}
\mathbf{G}_{(4)}=\widetilde{\mathbf{G}}_{(4)}-\frac{1}{2}(\mathbf{G}_{(1)}%
\mathbf{G}_{(1)}^{+}\mathbf{G}_{(2)}+\mathbf{G}_{(1)}\mathbf{G}_{(2)}^{+}%
\mathbf{G}_{(1)}+\mathbf{G}_{(2)}\mathbf{G}_{(1)}^{+}\mathbf{G}_{(1)}). 
\tag{1.3.6}
\end{equation}

Apart from the CBO matrix $\mathbf{P}$\ itself, the overall charge
(re)distribution has been described in the Subsection 1.1 by a certain
series of additional characteristics starting with the intrasubset
population matrices $\mathbf{X}_{(+)}^{{}}$ and $\mathbf{X}_{(-)}^{{}}$ and
their diagonal elements and ending with partial populations \ $%
x_{(+)i,(-)m}^{{}}$\ transferred between separate pairs of basis orbitals $%
\varphi _{(+)i}$\ and $\varphi _{(-)m}$. Thus, before turning to
implications of the first two relations of Eq.(1.3.3), let us now introduce
an analogous series of quantities representing the shapes (reshapings) of
individual pairs of electrons. To this end, let us take explicit expressions
for an occupied NCMO $\psi _{1,i}(\psi _{(+)i})$\ and for a vacant one $\psi
_{2,l}(\psi _{(-)l})$\ resulting from Eq.(1.2.8), viz. 
\begin{align}
\psi _{(+)i} =&\mathop{\displaystyle \sum }\limits_{(+)j}\varphi
_{(+)j}C_{11,ji}+\mathop{\displaystyle \sum }\limits_{(-)m}\varphi
_{(-)m}C_{21,mi},  \nonumber \\
\psi _{(-)l} =&\mathop{\displaystyle \sum }\limits_{(+)j}\varphi
_{(+)j}C_{12,jl}+\mathop{\displaystyle \sum }\limits_{(-)m}\varphi
_{(-)m}C_{22,ml},  \tag{1.3.7}
\end{align}%
where sums over $(+)j$ and $(-)m$ correspondingly embrace all orbitals of
subsets $\{\varphi _{(+)}\}$ and $\{\varphi _{(-)}\}$ as previously. Let us
now introduce the so-called partial (intersubset) delocalization
coefficients of NCMOs $\psi _{(+)i}$\ and $\psi _{(-)l}$\ over particular
basis orbitals $\varphi _{(-)m}$ and $\varphi _{(+)j},$ respectively [28].
Let these coefficients be correspondingly denoted by $d_{(+)i,(-)m}$ and $%
d_{(-)l,(+)j}$\ and defined as follows%
\begin{align}
d_{(+)i,(-)m} =&\mid C_{21,mi}\mid ^{2}=\mid G_{(1)mi}^{+}+G_{(2)mi}^{+}+%
\widetilde{G}_{(3)mi}^{+}+\widetilde{G}_{(4)mi}^{+}+...\mid ^{2},  \nonumber
\\
d_{(-)l,(+)j} =&\mid C_{12,jl}\mid ^{2}=\mid G_{(1)jl}+G_{(2)jl}+\widetilde{%
G}_{(3)jl}+\widetilde{G}_{(4)jl}+...\mid ^{2}.  \tag{1.3.8}
\end{align}%
As is easily seen, the partial delocalization coefficients concerned also
are expressible in the form of power series, i.e. as sums over parameter $k$
of contributions $d_{(+)i,(-)m}^{(k)}$ and $d_{(-)l,(+)j}^{(k)}$,
respectively, that, in turn, coincide one with another for the same pair of
NCMOs (e.g. for $\psi _{(+)i}$ and $\psi _{(-)m}$). We then obtain that%
\begin{equation}
d_{(+)i,(-)m}^{(k)}=d_{(-)m,(+)i}^{(k)},\qquad
d_{(-)l,(+)j}^{{}}=d_{(+)j,(-)l}^{{}}.  \tag{1.3.9}
\end{equation}%
Further, let us define total delocalization coefficients of the same NCMOs ($%
D_{(+)i}$\ and $D_{(-)l}$) over basis orbitals of subsets $\{\varphi
_{(+)}\} $ and $\{\varphi _{(-)}\},$ respectively, viz.%
\begin{equation}
D_{(+)i}=\mathop{\displaystyle \sum }\limits_{(-)m}d_{(+)i,(-)m},\quad
D_{(-)l}=\mathop{\displaystyle \sum }\limits_{(+)j}d_{(-)l,(+)j}. 
\tag{1.3.10}
\end{equation}%
Finally, complete delocalization coefficients of occupied NCMOs $\{\psi
_{(+)}\}$\ and of vacant ones $\{\psi _{(-)}\}$ will be defined as follows%
\begin{equation}
D_{(+)}=\mathop{\displaystyle \sum }\limits_{(+)i}D_{(+)i},\qquad D_{(-)}=%
\mathop{\displaystyle \sum }\limits_{(-)l}D_{(-)l}.  \tag{1.3.11}
\end{equation}%
Besides, substituting Eq.(1.3.10) into Eq.(1.3.11) and employment of
Eq.(1.3.9) shows that $D_{(+)}$\ and $D_{(-)}$\ coincide one with another.
Thus, the overall extent of delocalization of occupied NCMOs over the subset 
$\{\varphi _{(-)}\}$ coincides with that of vacant NCMOs over $\{\varphi
_{(+)}\}.$\ [Note that such a "symmetry" of MOs is peculiar to our prototype
described by the Hamiltonian matrix $\mathbf{h}$ of Eq.(1.2)].

Let us now introduce the so-called intersubset delocalization matrices $%
\mathbf{D}_{(+)}$\ and $\mathbf{D}_{(-)}$ [52]$,$\ viz. 
\begin{equation}
\mathbf{D}_{(+)}=\mathbf{C}_{21}^{+}\mathbf{C}_{21},\quad \mathbf{D}_{(-)}=%
\mathbf{C}_{12}^{+}\mathbf{C}_{12},  \tag{1.3.12}
\end{equation}%
diagonal elements of the latter ($D_{(+)ii}$\ and $D_{(-)ll}$) coinciding
with total delocalization coefficients of NCMOs $\psi _{(+)i}$\ and $\psi
_{(-)l}$\ of Eq.(1.3.10). Meanwhile, traces of matrices $\mathbf{D}_{(+)}$\
and $\mathbf{D}_{(-)}$\ yield complete delocalization coefficients of
Eq.(1.3.11) and%
\begin{equation}
Tr\mathbf{D}_{(+)}\ =Tr\mathbf{D}_{(-)}=D_{(+)}\ =D_{(-)}.  \tag{1.3.13}
\end{equation}%
If we substitute the power series for (sub)matrices $\mathbf{C}_{21}$ and $%
\mathbf{C}_{12}$ (Subsection 1.2) into the relations of Eq.(1.3.12),
matrices $\mathbf{D}_{(+)}$ and $\mathbf{D}_{(-)}$ also become expressed in
the form of an analogous series. The starting members of the latter are
proportional to renormalization matrices of the same order, e.g.%
\begin{equation}
\mathbf{D}_{(+)}^{(2)}=-2\mathbf{C}_{11}^{(2)},\quad \mathbf{D}%
_{(+)}^{(3)}=-2\mathbf{C}_{11}^{(3)}.  \tag{1.3.14}
\end{equation}%
Meanwhile, expressions for the next terms of the same series are of a
somewhat more involved constitutions [52], e.g. 
\begin{equation}
\mathbf{D}_{(+)}^{(4)}=-2\mathbf{C}_{11}^{(4)}-(\mathbf{C}_{11}^{(2)})^{2}. 
\tag{1.3.15}
\end{equation}%
As already mentioned, our final step consists in employment of the first two
relations of Eq.(1.3.3). The left-hand sides of these relations (i.e. $%
\mathbf{P}_{11}$\ and $\mathbf{P}_{22}$) coincide with the intrasubset
population matrices $\mathbf{X}_{(+)}$\ and $\mathbf{X}_{(-)},$
respectively, as discussed above (subsection 1.1). Thus, the power series
for the latter of Eqs.(1.1.4), (1.1.7) and (1.1.10) may be substituted into
the relations concerned. The same refers also to the analogous series of
Eqs.(1.2.9), (1.2.10) and (1.2.11) for (sub)matrices $\mathbf{C}_{11}$ and $%
\mathbf{C}_{21}$\ contained within the right-hand sides of the same
relations. Thereupon, interdependences between matrices $\mathbf{D}%
_{(+)}^{(k)}(\mathbf{D}_{(-)}^{(k)})$ and $\mathbf{C}_{11}^{(k)}(\mathbf{C}%
_{22}^{(k)})$\ like those shown in Eqs.(1.3.14) and (1.3.15) should be
invoked. We then obtain that 
\begin{equation}
\mathbf{X}_{(+)}^{(k)}=-2\mathbf{D}_{(+)}^{(k)},\qquad \mathbf{X}%
_{(-)}^{(k)}=2\mathbf{D}_{(-)}^{(k)}  \tag{1.3.16}
\end{equation}%
for $k=2,3...$ Consequently, the total matrices $\mathbf{X}_{(+)}(\mathbf{X}%
_{(-)})$\ and $\mathbf{D}_{(+)}(\mathbf{D}_{(-)})$\ are interrelated as
follows 
\begin{equation}
\mathbf{X}_{(+)}=2(\mathbf{I}-\mathbf{D}_{(+)}),\qquad \mathbf{X}_{(-)}=2%
\mathbf{D}_{(-)}.  \tag{1.3.17}
\end{equation}%
Thus, intrasubset population matrices prove to be proportional to the
relevant intersubset delocalization matrices. For traces of matrices of the $%
k$th order we accordingly obtain 
\begin{equation}
Tr\mathbf{X}_{(+)}^{(k)}=-2Tr\mathbf{D}_{(+)}^{(k)},\qquad Tr\mathbf{X}%
_{(-)}^{(k)}=2Tr\mathbf{D}_{(-)}^{(k)}  \tag{1.3.18}
\end{equation}%
and 
\begin{equation}
Tr\mathbf{D}_{(+)}^{(k)}=Tr\mathbf{D}_{(-)}^{(k)}  \tag{1.3.19}
\end{equation}%
is the analogue of Eq.(1.1.11) for the intersubset delocalization matrices.
Meanwhile, traces of total matrices $\mathbf{X}_{(+)}(\mathbf{X}_{(-)})$\
and $\mathbf{D}_{(+)}(\mathbf{D}_{(-)})$\ meet the following relations\ 
\begin{equation}
Tr\mathbf{X}_{(+)}=2(n-Tr\mathbf{D}_{(+)}),\qquad Tr\mathbf{X}_{(-)}=2Tr%
\mathbf{D}_{(-)},  \tag{1.3.20}
\end{equation}%
where the term $2n$ originates from the matrix $\mathbf{X}_{(+)}^{(0)}=2%
\mathbf{I}$\ of Eq.(1.1.7). [Note that the series for $\mathbf{D}_{(+)}$\
starts with the second order member $\mathbf{D}_{(+)}^{(2)}$\ of
Eq.(1.3.14)]. Using Eq.(1.3.13) we then obtain that%
\begin{equation}
Tr(\mathbf{X}_{(+)}+\mathbf{X}_{(-)})=Tr\mathbf{P}=2n  \tag{1.3.21}
\end{equation}%
in accordance with the last relation of Eq.(1.1.1). Similarly, we may start
with Eqs.(1.3.16) and (1.3.17) and turn to diagonal elements of matrices
contained there. We then obtain\ 
\begin{equation}
X_{(+)ii}=2(1-D_{(+)ii})\equiv 2(1-D_{(+)i}),\quad
X_{(-)ll}=2D_{(-)ll}\equiv 2D_{(-)l}  \tag{1.3.22}
\end{equation}%
and \ 
\begin{equation}
X_{(+)ii}^{(k)}=-2D_{(+)ii}^{(k)}\equiv -2D_{(+)i}^{(k)},\qquad
X_{(-)ll}^{(k)}=2D_{(-)ll}^{(k)}\equiv 2D_{(-)l}^{(k)},  \tag{1.3.23}
\end{equation}%
where $D_{(+)i}$\ and $D_{(-)l}$ are total delocalization coefficients
defined by Eq.(1.3.10). These relations indicate that the actual population
of any basis orbital $\varphi _{(+)i}$\ (or $\varphi _{(-)l}$) is determined
only by the shape of a single NCMO, namely of the NCMO $\psi _{(+)i}$\ (or $%
\psi _{(-)l}$) associated with just the basis orbital concerned [One-to-one
correspondence between basis orbitals and NCMOs should be recalled here
(Subsect. 1.2)]. More precisely, the population of the given basis orbital
lost (acquired) due to a certain interaction is proportional to the
delocalization coefficient of the respective single NCMO (LMO) (and not to
the sum of contributions of all occupied MOs as usual). This principal
result forms the basis for interpretation of the overall charge
(re)distribution in terms of displacements (shifts) of individual pairs of
electrons. \ Another form of the same rule is as follows: The more
delocalized the orbital $\varphi _{(+)i}$\ becomes when building up the
respective NCMO (LMO) $\psi _{(+)i},$ the more charge it loses and vice
versa. This result may be also interpreted as a kind of simultaneous
separability of both charge redistribution and delocalization into
increments of separate pairs of electrons. Finally, invoking the expressions
like those of Eqs.(1.1.12) and (1.3.10) for $%
X_{(+)ii}^{(k)}(X_{(-)ll}^{(k)}) $\ and $D_{(+)i}^{(k)}(D_{(-)l}^{(k)}),$\
respectively, yields proportionalities between the relevant partial
increments, viz. 
\begin{equation}
x_{(+)i,(-)m}^{(k)}=2d_{(+)i,(-)m}^{(k)},\quad
x_{(-)l,(+)j}^{(k)}=2d_{(-)l,(+)j}^{(k)}.  \tag{1.3.24}
\end{equation}%
[The minus sign of Eq. (1.1.12) is taken into consideration here]. Thus, the
more population is transferred between orbitals $\varphi _{(+)i}$\ and $%
\varphi _{(-)m},$\ the larger is the partial delocalization coefficient of
the NCMO (LMO)\ $\psi _{(+)i}$ over the same vacant basis function and vice
versa. If we recall the definition of this coefficient as square of the
relevant tail of the NCMO $\psi _{(+)i}$ (see Eq.(1.3.8)), we may also
conclude a more significant tail of this NCMO over the vacant orbital $%
\varphi _{(-)m}$\ to correspond to a more efficient charge transfer between
basis functions $\varphi _{(+)i}$\ and $\varphi _{(-)m}$\ and vice versa.
Irrelevance of the intrasubset delocalization in the formation of charge
redistributions also is among the conclusions. This result evidently causes
no surprise. Immediate reasons why the respective terms vanish in the
expressions for occupation numbers are clarified in Ref.[52]. It deserves
mention finally that the relations of Eqs.(1.3.22) and (1.3.23) indicate
both populations of basis orbitals and delocalization coefficients of NCMOs
to be characterized by the same dependence upon the structure of the given
system.

On the whole, the results of this subsection may be summarized as
parallelism between charge (re)distribution and delocalization. Moreover,\
we obtain a certain quantum-chemical analogue of the Lewis perspective on
charge (re)distribution [56] and thereby of the 'curly arrow chemistry'[18].
Two differences between this analogue and its classical version deserve
mention: First, NCMOs (LMOs) are not localized completely in contrast to the
relevant classical model. Second, the approach suggested allows comparisons
of relative extents of shifts (reshapings) of separate pairs of electrons
for related compounds and/or for alternative routes of the same process.
Just the second point may be regarded as an important advantage of the new
approach over the classical one.

Finally, the principal result of this subsection may be alternatively
formulated as a kind of one-orbital representation for populations lost
(acquired) by individual basis orbitals. This state of things closely
resembles the one-orbital representation of ionization potentials in the
canonical MO method known as the Koopmans' theorem [57]. If we recall that
representations of this type are not achievable for ionization potentials
and for charge (re)distributions in the NCMO and the CMO methods,
respectively, the present results support the complementary nature of the
above-mentioned principal approaches of quantum chemistry [58].

\begin{center}
\textbf{1.4. The principal relations concerning total energies }
\end{center}

Total energies of molecules also rank among the most popular quantum-
chemical characteristics. Moreover, these are comparable to experimental
data even more directly, viz. to heats of formation and/or atomization.

As is well-known, total energies of molecules ($\mathcal{E}$) usually are
defined as sums of one-electron energies referring to occupied MOs (and/or
of the relevant part of eigenvalues of the respective Hamiltonian matrix)\
multiplied by their occupation number 2 [1, 4, 13, 29]. As is seen from the
above subsections 1.1 and 1.2, however, neither matrix equations for the DM
of Eq.(1.1.1) nor the block-diagonalization problem of Eqs.(1.2.1) and
(1.2.2) yield eigenvalues of particular Hamiltonian matrices. Thus, we need
a more general definition of the total energy when looking for general
expressions for the same quantity referring to the Hamiltonian matrix $%
\mathbf{H}$ of Eq.(1.1). Fortunately, alternative definitions of this
characteristic exist both in terms of the CBO matrix $\mathbf{P}$ and via
eigenblocks of the matrix $\mathbf{H}$. These definitions correspondingly
take the following forms [11, 14]%
\begin{equation}
\mathcal{E}=Tr(\mathbf{PH})  \tag{1.4.1}
\end{equation}%
and 
\begin{equation}
\mathcal{E}=2Tr\mathbf{E}_{1}=2Tr\{\mathbf{C}^{+}\mathbf{HC}\}_{11}, 
\tag{1.4.2}
\end{equation}%
where $\mathbf{E}_{1}^{{}}$\ is the $n\times n-$dimensional eigenblock of
the Hamiltonian matrix referring to the subset of occupied NCMOs and the
symbol $\{...\}_{11}$\ \ stands here for the submatrix of the matrix product 
$\mathbf{C}^{+}\mathbf{HC}$\ taking the first diagonal position. It is
evident that both Eq.(1.4.1) and (1.4.2) embrace the above-mentioned
standard definition as a particular case [Note that matrices $\mathbf{P}$
and $\mathbf{H}$\ take diagonal forms in the CMO basis. Moreover, diagonal
elements of the former are 2 and 0 for occupied and vacant MOs]. It is also
obvious that expressions for matrices $\mathbf{P,}$ $\mathbf{H}$\ \ and $%
\mathbf{E}_{1}$ as sums over parameter $k$ of increments $\mathbf{P}_{(k)}%
\mathbf{,}$ $\mathbf{H}_{(k)}$\ \ and $\mathbf{E}_{1}^{(k)}$ of Eqs.(1.1.2),
(1.1) and (1.2.6), respectively, may be substituted into Eqs.(1.4.1) and
(1.4.2). Collecting of terms of the same order $k$ within the right-hand
sides of the latter relations then consequently yields the relevant power
series for the energy $\mathcal{E}$.

Instead, we will start with some general properties of the energy expansion.
Let us note first that presence of the zero order member ($\mathbf{H}_{(0)}$%
) and of the first order one ($\mathbf{H}_{(1)}$) in our Hamiltonian matrix
of Eq.(1.1) gives birth to two additive components [37] within any energy
correction \ $\mathcal{E}_{(k)}^{{}},$ namely the $\mathbf{H}_{(0)}-$%
containing component$\ (\mathcal{E}_{(k)}^{(\alpha )})$ and the $\mathbf{H}%
_{(1)}-$containing one ($\mathcal{E}_{(k)}^{(\beta )}),$ viz. 
\begin{equation}
\mathcal{E}_{(k)}=\mathcal{E}_{(k)}^{(\alpha )}+\mathcal{E}_{(k)}^{(\beta )}.
\tag{1.4.3}
\end{equation}%
Given that the definition of Eq.(1.4.1) is used, the components concerned
are expressible as follows 
\begin{equation}
\mathcal{E}_{(k)}^{(\alpha )}=Tr\mathbf{(P}_{(k)}\mathbf{H}_{(0)}\mathbf{)}%
,\quad \mathcal{E}_{(k)}^{(\beta )}=Tr\mathbf{(P}_{(k-1)}\mathbf{H}_{(1)}%
\mathbf{)}.  \tag{1.4.4}
\end{equation}%
[An analogous partition of the right-hand side of Eq.(1.4.2) may be found in
Ref.[52]]. Moreover, the above-exhibited components were shown to be
interrelated as follows%
\begin{equation}
(k-1)\mathcal{E}_{(k)}^{(\beta )}=-k\mathcal{E}_{(k)}^{(\alpha )} 
\tag{1.4.5}
\end{equation}%
for any $k.$ This principal relation has been originally derived for $k\leq
4 $\ [37] by substituting into Eq.(1.4.4) the expressions for $\mathbf{P}%
_{(k)} $ of Eqs.(1.1.5)-(1.1.10) followed by a definite algebraic procedure
based on employment of Eq.(1.1.8). Later, the same result was rederived
using Eq.(1.4.2) and verified for higher values of the order parameter $k$
[52]. Implications of Eq.(1.4.5) are as follows: First, this relation points
to opposite signs of the components $\mathcal{E}_{(k)}^{(\alpha )}$ and $%
\mathcal{E}_{(k)}^{(\beta )}.$ Thus, the total $k$th order energy actually
depends upon the difference between absolute values of these components.
Second, the absolute value of $\mathcal{E}_{(k)}^{(\beta )}$ is foreseen to
exceed that of $\mathcal{E}_{(k)}^{(\alpha )},$ i.e. $\mid \mathcal{E}%
_{(k)}^{(\beta )}\mid >\mid \mathcal{E}_{(k)}^{(\alpha )}\mid .$
Consequently, it is the sign\ of $\mathcal{E}_{(k)}^{(\beta )}$\ that
determines the actual sign of the total $k$th order energy $\mathcal{E}%
_{(k)} $ and thereby its stabilizing or destabilizing nature. Thus, the $%
\mathbf{H}_{(1)}-$containing component is expected to be the principal
(decisive) one, whereas $\mathcal{E}_{(k)}^{(\alpha )}$\ is likely to play a
secondary role in the formation of the $k$th order energy $\mathcal{E}_{(k)}$%
. The actual state of things, however, is somewhat more complicated as
clarified below.

Let us note first that the relations of Eqs.(1.4.3) and (1.4.5) allow the
total correction $\mathcal{E}_{(k)}(k>1)$\ to be alternatively represented
as follows%
\begin{equation}
\mathcal{E}_{(k)}=-\frac{1}{k-1}\mathcal{E}_{(k)}^{(\alpha )},\qquad 
\mathcal{E}_{(k)}=\frac{1}{k}\mathcal{E}_{(k)}^{(\beta )}  \tag{1.4.6}
\end{equation}%
and these expressions evidently are formally equivalent. Actually, however,
the first relation of Eq.(1.4.6) yields chemically-meaningful formulae for
the whole correction $\mathcal{E}_{(k)}$\ but not the second one. This
circumstance, in turn, may be traced back to relative simplicity of the
expression for $\mathcal{E}_{(k)}^{(\alpha )}$\ itself vs. that for $%
\mathcal{E}_{(k)}^{(\beta )}$, viz. 
\begin{equation}
\mathcal{E}_{(k)}^{(\alpha )}=Tr[\mathbf{X}_{(+)}^{(k)}\mathbf{E}_{(+)}^{{}}-%
\mathbf{X}_{(-)}^{(k)}\mathbf{E}_{(-)}^{{}}],  \tag{1.4.7}
\end{equation}%
\begin{equation}
\mathcal{E}_{(k)}^{(\beta )}=Tr[\mathbf{X}_{(+)}^{(k-1)}\mathbf{T}+\mathbf{X}%
_{(-)}^{(k-1)}\mathbf{Q}]-2Tr[\mathbf{G}_{(k-1)}\mathbf{R}^{+}+\mathbf{G}%
_{(k-1)}^{+}\mathbf{R}],  \tag{1.4.8}
\end{equation}%
where Eqs.(1.1) and (1.1.5) are correspondingly substituted for $\mathbf{H}%
_{(0)},\mathbf{H}_{(1)}$ and $\mathbf{P}_{(k-1)}.$\ As is easily seen,
employment just of Eq.(1.4.7) along with the first relation of Eq.(1.4.6)
yields a compact expression for the $k$th order energy ($\mathcal{E}_{(k)}$)
in terms of intrasubset population matrices of the same order only, viz. 
\begin{equation}
\mathcal{E}_{(k)}=-\frac{1}{k-1}Tr[\mathbf{X}_{(+)}^{(k)}\mathbf{E}%
_{(+)}^{{}}-\mathbf{X}_{(-)}^{(k)}\mathbf{E}_{(-)}^{{}}].  \tag{1.4.9}
\end{equation}%
After invoking proportionalities between $\mathbf{X}_{(+)}^{(k)}$ and\textbf{%
\ \ }$\mathbf{D}_{(+)}^{(k)},$ as well as between $\mathbf{X}_{(-)}^{(k)}$
and\textbf{\ }$\mathbf{D}_{(-)}^{(k)}$\ shown in Eq.(1.3.16), the above
formula takes the following form 
\begin{equation}
\mathcal{E}_{(k)}=\frac{2}{k-1}Tr[\mathbf{D}_{(+)}^{(k)}\mathbf{E}%
_{(+)}^{{}}+\mathbf{D}_{(-)}^{(k)}\mathbf{E}_{(-)}^{{}}],  \tag{1.4.10}
\end{equation}%
which points to a dependence of $\mathcal{E}_{(k)}$\ upon the intersubset
delocalization matrices $\mathbf{D}_{(+)}^{(k)}$\textbf{\ }and $\mathbf{D}%
_{(-)}^{(k)}.$\ Hence, we obtain an analogue of the well-known
intuition-based assumption about a relation between stabilization (or
destabilization) of a certain system and the relevant extent of
delocalization. By contrast, employment of Eq.(1.4.8) along with the second
relation of Eq.(1.4.6) yields no simple interpretation of $\mathcal{E}%
_{(k)}. $\ More details concerning this intriguing situation may be found in
the subsection 2.3.

The last and practically even more important implication of the relation of
Eq.(1.4.5) consists in the possibility of summing up the components $%
\mathcal{E}_{(k)}^{(\alpha )}$ and $\mathcal{E}_{(k)}^{(\beta )}$\ within
Eq.(1.4.3) so that the $k$th order energy $\mathcal{E}_{(k)}$\ becomes
expressible directly via the principal matrices $\mathbf{G}_{(k)}$\ and
(sub)matrices (blocks) of our Hamiltonian matrix [37].%
\begin{align}
\mathcal{E}_{(0)} =&2Tr\mathbf{E}_{(+)},\qquad \mathcal{E}_{(1)}=2Tr\mathbf{%
T,}  \nonumber \\
\mathcal{E}_{(2)} =&-2Tr(\mathbf{G}_{(1)}\mathbf{R}^{+}),\qquad \mathcal{E}%
_{(3)}=-2Tr(\mathbf{G}_{(2)}\mathbf{R}^{+}),  \nonumber \\
\mathcal{E}_{(4)} =&-2Tr[(\mathbf{G}_{(3)}+\mathbf{G}_{(1)}\mathbf{G}%
_{(1)}^{+}\mathbf{G}_{(1)})\mathbf{R}^{+}],  \tag{1.4.11} \\
\mathcal{E}_{(5)} =&-2Tr[(\mathbf{G}_{(4)}+\mathbf{G}_{(2)}\mathbf{G}%
_{(1)}^{+}\mathbf{G}_{(1)}+\mathbf{G}_{(1)}\mathbf{G}_{(2)}^{+}\mathbf{G}%
_{(1)}+\mathbf{G}_{(1)}\mathbf{G}_{(1)}^{+}\mathbf{G}_{(2)})\mathbf{R}%
^{+}],etc.  \nonumber
\end{align}%
[Derivations of $\mathcal{E}_{(5)}$\ and $\mathcal{E}_{(6)}$\ may be found
in Refs.[38] and [40], respectively. The same expressions are obtainable
also using Eq.(1.4.2) [52]]. It is seen that the sum of zero and first order
members of Eq.(1.4.11) coincides with the total energy of the subset of
isolated occupied basis functions $\{\varphi _{(+)}\}$\ in accordance with
the expectation, whereas that of the remaining corrections describes
stabilization (or destabilization) of the given system vs. the
above-specified set. Comparison of the above-exhibited expressions to the
usual perturbational formulae for energy corrections may be found in the
subsection 2.3.

\begin{center}
\textbf{1.5. Indirect applications of the PNCMO theory}
\end{center}

In this subsection, we are about to demonstrate that the actual scope of
applicability of the above-described PNCMO theory is not limited to
Hamiltonian matrices of Eq.(1.1) containing a zero order term ($\mathbf{H}%
_{(0)}$) of the block-diagonal constitution.

Suppose that we start with a certain $N\times N-$dimensional Hamiltonian
matrix $\widetilde{\mathbf{H}}$ that does not meet the above requirement.
Let the orthogonal basis set underlying the matrix \ $\widetilde{\mathbf{H}}$
be denoted by $\{\chi \}$. Meanwhile, the total number of electrons
coincides with $2n$ as previously.

Let us now assume that an unitary matrix $\mathbf{U}$ is found that allows
us to transform the matrix $\widetilde{\mathbf{H}}$ into the nearly
block-diagonal form, viz.%
\begin{equation}
\mathbf{U}^{+}\widetilde{\mathbf{H}}\mathbf{U}=\mathbf{H,}  \tag{1.5.1}
\end{equation}%
where $\mathbf{H}$\ coincides with the Hamiltonian matrix of Eq.(1.1). It is
evident that the above relation represents passing from our initial basis
set $\{\chi \}$\ to the former one $\{\varphi \}.$\ Employment of the
standard PNCMO theory of subsections 1.1-1.4 to the transformed matrix $%
\mathbf{H}$\ allows us to find the relevant CBO matrix $\mathbf{P}$\ and the
respective NCMO representation matrix $\mathbf{C.}$\ The final step of the
overall procedure then consists in retransforming the above-mentioned
matrices $\mathbf{P}$\ and $\mathbf{C}$\ into the initial basis $\{\chi \}$\
again by means of the following relations%
\begin{equation}
\widetilde{\mathbf{P}}=\mathbf{UPU}^{+},\qquad \widetilde{\mathbf{C}}=%
\mathbf{UC.}  \tag{1.5.2}
\end{equation}%
Let us recall now that the requirements underlying the standard PNCMO theory
actually refer to constitution of the zero order Hamiltonian matrix $\mathbf{%
H}_{(0)}$\ only, viz. the latter should be of the block-diagonal
constitution. Meanwhile, the first order member $\mathbf{H}_{(1)}$\ is free
from additional conditions. Given that the initial matrix $\widetilde{%
\mathbf{H}}$\ also consists of a zero order term ($\widetilde{\mathbf{H}}%
_{(0)}$) and of the first order one ($\widetilde{\mathbf{H}}_{(1)}$), i.e. 
\begin{equation}
\widetilde{\mathbf{H}}=\widetilde{\mathbf{H}}_{(0)}+\widetilde{\mathbf{H}}%
_{(1)},  \tag{1.5.3}
\end{equation}%
we may confine ouselves to finding an unitary matrix $\mathbf{U}$\ that
transforms the zero order member $\widetilde{\mathbf{H}}_{(0)}$\ into a
block-diagonal form shown in Eq.(1.1). As a result, we then arrive at the
block-diagonalization problem for the matrix $\widetilde{\mathbf{H}}_{(0)}$,
viz.%
\begin{equation}
\mathbf{U}^{+}\widetilde{\mathbf{H}}_{(0)}\mathbf{U}=\mathbf{H}%
_{(0)}=\left\vert 
\begin{array}{cc}
\mathbf{E}_{(+)} & \mathbf{0} \\ 
\mathbf{0} & -\mathbf{E}_{(-)}%
\end{array}%
\right\vert \mathbf{,}  \tag{1.5.4}
\end{equation}%
where $\mathbf{E}_{(+)}$\ and $\mathbf{E}_{(-)}$\ now play the role of the
relevant eigenblocks. At the first sight, the problem of Eq.(1.5.4)
resembles that of Eq.(1.2.1). Nevertheless, an important difference between
these seemingly similar problems arises due to the zero order of the matrix
concerned (i.e. of $\widetilde{\mathbf{H}}_{(0)}$). Consequently, the
problem of Eq.(1.5.4) cannot be solved perturbatively in contrast to that of
Eq.(1.2.1). Fortunately, non-perturbative solutions of block-diagonalization
problems also are possible in some cases as exemplified below (subsection
3.1). Given that such a solution is found and thereby the matrix $\mathbf{U}$%
\ is known, the remaining (first order) member of our initial Hamiltonian
matrix ($\widetilde{\mathbf{H}}_{(1)}$) may be transformed without any
restriction using the following relation%
\begin{equation}
\mathbf{U}^{+}\widetilde{\mathbf{H}}_{(1)}\mathbf{U}=\mathbf{H}_{(1)}. 
\tag{1.5.5}
\end{equation}%
Application of the above-formulated standard PNCMO theory to the sum $%
\mathbf{H}_{(0)}\mathbf{+H}_{(1)}$ then yields matrices $\mathbf{P}$\ and $%
\mathbf{C}$\ in the form of power series, individual members of which (i.e. $%
\mathbf{P}_{(k)}$\ and $\mathbf{C}_{(k)}$) may be retransformed into the
initial basis $\{\chi \}$\ separately, i.e. on the basis of the following
expressions 
\begin{equation}
\widetilde{\mathbf{P}}_{(k)}=\mathbf{UP}_{(k)}\mathbf{U}^{+},\qquad 
\widetilde{\mathbf{C}}_{(k)}=\mathbf{UC}_{(k)}.  \tag{1.5.6}
\end{equation}

It deserves adding here that all relations concerning total energies and
overviewed in the subsection 1.4 are invariant against unitary
transformations of basis sets. This implies that energy components $\mathcal{%
E}_{(k)}^{(\alpha )}$ and $\mathcal{E}_{(k)}^{(\beta )}$\ may be derived
either using the original relations of Eqs.(1.4.3)-(1.4.10) or their
analogues containing members of the power series for matrices $\widetilde{%
\mathbf{H}}$\ and $\widetilde{\mathbf{P}},$\ i.e. 
\begin{equation}
\mathcal{E}_{(k)}^{(\alpha )}=Tr\mathbf{(}\widetilde{\mathbf{P}}_{(k)}%
\widetilde{\mathbf{H}}_{(0)}\mathbf{)},\quad \mathcal{E}_{(k)}^{(\beta )}=Tr%
\mathbf{(}\widetilde{\mathbf{P}}_{(k-1)}\widetilde{\mathbf{H}}_{(1)}\mathbf{)%
}.  \tag{1.5.7}
\end{equation}%
Analogously, the total correction $\mathcal{E}_{(k)}$ is alternatively
representable as follows 
\begin{equation}
\mathcal{E}_{(k)}=-\frac{1}{k-1}Tr\mathbf{(}\widetilde{\mathbf{P}}_{(k)}%
\widetilde{\mathbf{H}}_{(0)}\mathbf{)},\quad \mathcal{E}_{(k)}=\frac{1}{k}Tr%
\mathbf{(}\widetilde{\mathbf{P}}_{(k-1)}\widetilde{\mathbf{H}}_{(1)}\mathbf{)%
}.  \tag{1.5.8}
\end{equation}

\begin{center}
\bigskip

\textbf{2. Particular cases of the PNCMO theory\bigskip }
\end{center}

In this Section, we will discuss two particular cases of the standard PNCMO
theory of section 1 when explicit solutions may be obtained for principal
matrix equations shown in Eq.(1.1.8). We start with the case of diagonal
zero order matrices $\mathbf{E}_{(+)}$\ and $\mathbf{E}_{(-)}$\ of Eq.(1.1),
when individual elements $G_{(k)im}$\ are expressible in terms of usual
perturbational fractions (subsection 2.1). Properties of charge
(re)distribution and delocalization referring just to this case are then
discussed along with those of the relevant total energy (subsections 2.2 and
2.3, respectively). Thereupon, we turn to the case of matrices $\mathbf{E}%
_{(+)}$\ and $\mathbf{E}_{(-)}$\ proportional to the unit matrix ($\mathbf{I}
$) and discuss the consequent explicit expressions for entire principal
matrices $\mathbf{G}_{(k)}$\ along with the relevant implications
(subsection 2.4).

\begin{center}
\textbf{2.1. The case of a diagonal zero order Hamiltonian matrix.
Expressions for particular elements }$G_{(k)im}$\textbf{\ and their
interpretation}
\end{center}

The present subsection addresses the case of absence of the zero order
intrasubset interactions in our system so that both inter- and intrasubset
resonance parameters are of the first order magnitude. In other words,
submatrices $\mathbf{E}_{(+)}$\ and $\mathbf{E}_{(-)}$\ of the zero order
Hamiltonian matrix of Eq.(1.1) (and thereby the whole member $\mathbf{H}%
_{(0)}$)\ are assumed to take diagonal forms, viz.%
\begin{equation}
E_{(+)ij}=\varepsilon _{(+)i}\delta _{ij},\qquad E_{(-)lm}=\varepsilon
_{(-)m}\delta _{lm},  \tag{2.1.1}
\end{equation}%
where $\varepsilon _{(+)i}$ and $-\varepsilon _{(-)m}$\ coincide with
one-electron energies of basis orbitals $\varphi _{(+)i}$ and $\varphi
_{(-)m},$\ respectively (Section 1). Under the above condition, elements $%
G_{(k)im}$\ of matrices $\mathbf{G}_{(k)}$ are expressible in terms of the
following fractions [27, 36] \ 
\begin{equation}
G_{(k)im}^{{}}=-\frac{W_{(k)im}}{\varepsilon _{(+)i}+\varepsilon _{(-)m}} 
\tag{2.1.2}
\end{equation}%
instead of the integral of Eq.(1.1.17). Using Eq.(1.1.9), we then obtain
that \ 
\begin{equation}
G_{(1)im}^{{}}=-\frac{R_{im}}{\varepsilon _{(+)i}+\varepsilon _{(-)m}}, 
\tag{2.1.3}
\end{equation}%
\begin{equation}
G_{(2)im}^{{}}=\frac{1}{\varepsilon _{(+)i}+\varepsilon _{(-)m}}\left\{ %
\mathop{\displaystyle \sum }\limits_{(+)j}\frac{T_{ij}R_{jm}}{\varepsilon
_{(+)j}+\varepsilon _{(-)m}}-\mathop{\displaystyle \sum }\limits_{(-)r}\frac{%
R_{ir}Q_{rm}}{\varepsilon _{(+)i}+\varepsilon _{(-)r}}\right\} ,  \tag{2.1.4}
\end{equation}%
whilst elements of higher orders take somewhat more cumbersome forms (see
e.g.[59, 60]). Sums over $(+)j$\ and over $(-)r$ correspondingly embrace
here all orbitals of subsets $\{\varphi _{(+)}\}$\ and $\{\varphi _{(+)}\}$\
as previously. Let us turn now to interpretation of the above expressions.

\ Let us start with the first order element $G_{(1)im}.$ As is seen from
Eq.(2.1.3), this element is proportional to the resonance parameter ($R_{im}$%
) between basis orbitals $\varphi _{(+)i}$ and $\varphi _{(-)m}$\ and
inversely proportional to the energy gap between the latter ($\varepsilon
_{(+)i}+\varepsilon _{(-)m}$) [The minus sign in front of $\varepsilon
_{(-)m}$ and/or $\mathbf{E}_{(-)}$ of Eq.(1.1) should be taken into
consideration]. Thus, the element $G_{(1)im}$\ represents the direct
interaction between the orbitals concerned that is often alternatively
called the through-space one. It is also evident that the element $%
G_{(1)im}\ $does not vanish if the orbitals concerned (i.e. $\varphi _{(+)i}$
and $\varphi _{(-)m}$) are characterized by a non-zero resonance parameter.
These orbitals are referred to below as neighboring ones. Meanwhile, the
expression of Eq.(2.1.4) for the second order element $G_{(2)im}$\ is
somewhat more involved and contains sums of products of pairs of mutually
"connected" resonance parameters. Moreover, the remaining orbitals of the
same system (i.e. $\varphi _{(+)j}$ and $\varphi _{(-)r}$) also participate
in the formation of the element $G_{(2)im}$\ and play the role of mediators
in the second order interaction between orbitals $\varphi _{(+)i}$ and $%
\varphi _{(-)m}.$It is also seen that the orbitals $\varphi _{(+)j}$ and $%
\varphi _{(-)r}$ should overlap directly both with $\varphi _{(+)i}$ and
with $\varphi _{(-)m}$ to be efficient mediators. In this connection $%
G_{(2)im}$\ has been interpreted as the indirect interaction between
orbitals $\varphi _{(+)i}$ and $\varphi _{(-)m\text{ }}$through individual
mediators. Similarly, the formula for the element $G_{(3)im}$\ [59, 60]
contains sums of products of triplets of "connected" resonance parameters.
Consequently, this element represents an analogous indirect interaction via
pairs of mediators. Generally, the number of mediators coincides with $k-1$
for an element $G_{(k)im}.$ Given that a certain diagonal element (say $%
T_{ii}$) takes a non-zero value, we accordingly have to do with the
self-mediating effect of the orbital $\varphi _{(+)i}$. Furthermore, an
analogous interpretation may be easily extended also to elements of matrix
products contained within definitions of intrasubset population matrices $%
\mathbf{X}_{(+)}^{(k)}$\textbf{\ }and\textbf{\ }$\mathbf{X}_{(-)}^{(k)}$\ of
Eq.(1.1.10), e.g. $\mathbf{G}_{(1)}\mathbf{G}_{(1)}^{+},$\textbf{\ }$\mathbf{%
G}_{(1)}\mathbf{G}_{(2)}^{+},etc.$ For example, the element ($\mathbf{G}%
_{(1)}\mathbf{G}_{(1)}^{+})_{ij}$\ is interpretable as the indirect
interaction between basis orbitals $\varphi _{(+)i}$ and $\varphi _{(+)j},$
where orbitals of the opposite subset $(\varphi _{(-)r})$ play the role of
mediators.

Let us now formulate a general non-zero-value condition for indirect
interorbital interactions [38, 61]. To this end, let us introduce the
following explicit designations for resonance parameters contained within
Eqs.(2.1.3) and\ (2.1.4), viz. \ 
\begin{align}
T_{ij} =&<\varphi _{(+)i}\mid \widehat{H}\mid \varphi _{(+)j}>,\qquad
R_{ir}=<\varphi _{(+)i}\mid \widehat{H}\mid \varphi _{(-)r}>,  \tag{2.1.5}
\\
Q_{rm} =&<\varphi _{(-)r}\mid \widehat{H}\mid \varphi _{(-)m}>,  \nonumber
\end{align}%
where the orbitals concerned are shown inside the bra- and ket-vectors. The
necessary condition for an element $G_{(k)im}$\ to take a non-zero value
then consists in the presence of at least a single non-zero product of $k$
"connected" resonance parameters in the system concerned, i.e. 
\begin{equation}
<\varphi _{(+)i}\mid \widehat{H}\mid \varphi _{1}><\varphi _{1}\mid \widehat{%
H}\mid \varphi _{2}>...<\varphi _{k-2}\mid \widehat{H}\mid \varphi
_{k-1}><\varphi _{k-1}\mid \widehat{H}\mid \varphi _{(-)m}>\neq 0, 
\tag{2.1.6}
\end{equation}%
where the notations $\varphi _{1},\varphi _{2},...\varphi _{k-1}$\ stand for
mediating orbitals. In other words, presence of a chain-like set of $k-1$
mediators in between the interacting orbitals $\varphi _{(+)i}$ and $\varphi
_{(-)m\text{ }}$is required here. Given that the condition of Eq.(2.1.6) is
met, we will say that in the given system there is a pathway of the $k$th
order between orbitals $\varphi _{(+)i}$ and $\varphi _{(-)m}.$\ Moreover,
additivity of the element $G_{(k)im}$\ with respect to contributions of
individual pathways easily follows from expressions like that of Eq.(2.1.4).
For an element of a matrix product like ($\mathbf{G}_{(1)}\mathbf{G}%
_{(1)}^{+})_{ij},$\ we accordingly have to do with a set of pathways between
pairs of orbitals of the same subset (i.e. between $\varphi _{(+)i}$ and $%
\varphi _{(+)j}$). It is also evident that increments of some pathways may
cancel one another so that the indirect interaction concerned vanishes even
if the relevant requirement of the form shown in Eq.(2.1.6) is met. That is
why the conditions under discussion are of a necessary (but not sufficient)
nature.

Finally, the well-known extinction of resonance parameters when the distance
between the relevant orbitals grows [62] allows us to expect an analogous
behaviour of interorbital interactions too. Nevertheless, indirect
interactions of higher orders are likely to take significant values even for
remote orbitals if the system concerned offers appropriate sets of
mediators. Implications of Eq.(2.1.1) upon properties of charge
(re)distribution and total energy are discussed in the next subsections (2.2
and 2.3).

\begin{center}
\textbf{2.2. Semilocal nature of charge (re)distribution and delocalization}
\end{center}

This subsection addresses specific properties of charge (re)distribution
and/or of delocalization for systems described by diagonal matrices $\mathbf{%
E}_{(+)}$\ and $\mathbf{E}_{(-)}$\ of Eq.(1.1).

As discussed already in the subsection 1.1, populations of basis orbitals $%
\varphi _{(+)i}$ and $\varphi _{(-)l\text{ }}$($X_{(+)ii}$ and $X_{(-)ll}$)
are expressible in terms of the relevant partial contributions ($%
x_{(+)i,(-)m}$ and $x_{(-)l,(+)j}$) referring to separate pairs of orbitals
of opposite initial occupation (see Eq.(1.1.15)). These contributions, in
turn, take the form of sums over the order parameter ($k$) of increments $%
x_{(+)i,(-)m}^{(k)}$ and $x_{(-)l,(+)j}^{(k)}.$ If we recall now the
above-established proportionality between $x_{(+)i,(-)m}^{(k)}$\ and the
respective partial delocalization coefficient $d_{(+)i,(-)m}^{(k)}$\ (see
Eq.(1.3.24) along with the equalities shown in Eqs.(1.1.14) and (1.3.9)), we
may actually confine ourselves to analysis of properties of increments $%
x_{(+)i,(-)m}^{(k)}$\ of different orders $k$, i.e. of partial populations
of the $k$th order transferred between orbitals $\varphi _{(+)i}$ and $%
\varphi _{(-)m}.$\ To this end, the expressions for $G_{(k)im}$\ like those
of Eq.(2.1.3) and (2.1.4) should be substituted into definitions of $%
x_{(+)i,(-)m}^{(k)}$\ shown in Eq.(1.1.13).

Let us start with the increment of the second order $x_{(+)i,(-)m}^{(2)}$.
As is seen from the first relation of Eq.(1.1.13), this increment is
determined by square of the first order element $G_{(1)im}$\ and thereby of
the direct interaction between orbitals concerned (i.e. $\varphi _{(+)i}$
and $\varphi _{(-)m}$). As with the above-specified interaction, the
increment $x_{(+)i,(-)m}^{(2)}$\ is then expected to take a non-zero value
for neighboring pairs of basis orbitals only. Moreover, dependence of the
increment concerned upon the mutual arrangement of only two orbitals in the
real space is foreseen along with its rapid extinction when the distance
between these orbitals increases. The \textit{a priori} positive sign of $%
x_{(+)i,(-)m}^{(k)}$ (Subsection 1.1) also deserves recalling here.
Consequently, the overall contribution of the second order to the population
of an initially-occupied orbital $\varphi _{(+)i}$\ lost due to perturbation
($X_{(+)ii}^{(2)}$ of Eq.(1.1.12)) originates mainly from increments ($%
x_{(+)i,(-)m}^{(2)}$) of the nearest-neighboring initially-vacant orbitals ($%
\varphi _{(-)m\text{ }}$). The same evidently refers to the second order
term ($D_{(+)i}^{(2)}$) of the analogous power series for the delocalization
coefficient ($D_{(+)i}$) of the NCMO $\psi _{(+)i}\ $(Subsection 1.3), as
well as to the relevant characteristics of initially-vacant orbitals.
Moreover, transferability of contributions like $X_{(+)ii}^{(2)}$\ and/or $%
X_{(-)ll}^{(2)}$\ is predicted if the nearest neighborhoods of orbitals $%
\varphi _{(+)i}$ and/or $\varphi _{(-)l\text{ }}$are similar.

The third order increment $x_{(+)i,(-)m}^{(3)}$\ to the same partial
transferred population ($x_{(+)i,(-)m}$) is determined by the product of
elements $G_{(1)im}$\ and $G_{(2)im}$\ as the second expression of
Eq.(1.1.13) shows. Consequently, the increment concerned does not vanish if
the relevant orbitals \ $\varphi _{(+)i}$ and $\varphi _{(-)m\text{ }}$%
interact both directly and indirectly by means of a single mediator and
thereby these orbitals are both first- and second-neighboring ones. In\
other words, at least a single triplet of neighboring orbitals is required
to ensure non-vanishing third order increments to charge (re)distribution
and/or to delocalization. Moreover, additivity of these characteristics with
respect to contributions of individual triplets of the above-specified type
also is among anticipations. So far as the overall population of the orbital 
$\varphi _{(+)i}$\ is concerned, the third order member of the relevant
expansion ($X_{(+)ii}^{(3)}$) depends upon a somewhat more extended
neighborhood of the orbital under consideration as compared to the
above-discussed second order term ($X_{(+)ii}^{(2)}$). Moreover, the
increment $X_{(+)ii}^{(3)}$\ contributes to an increase (decrease) of the
total population of the orbital $\varphi _{(+)i}$ if the direct and indirect
interactions between the latter and the neighboring initially-vacant
orbitals ($\varphi _{(-)m}$) are of the same (opposite) signs (subsection
1.1).

Increments of higher orders to the partial transferred population $%
x_{(+)i,(-)m}$\ also may be analyzed similarly. To be able to draw general
conclusions regarding the increment $x_{(+)i,(-)m}^{(k)}$\ of any order $k$,
let us recall the non-zero-value condition for elements $G_{(k)im}$\ in
terms of pathways over basis orbitals (subsection 2.1) and note that cyclic
(self-returning) pathways [38] correspond to separate members of definitions
of $x_{(+)i,(-)m}^{(k)}$\ like those of Eq.(1.1.13). For example, a cyclic
pathway of the $k$th order embracing orbitals $\varphi _{(+)i}$ and $\varphi
_{(-)m\text{ }}$refers to the product $G_{(k-1)im}G_{(1)im}.$\ The necessary
condition for the increment $x_{(+)i,(-)m}^{(k)}$\ to take a non-zero value
then consists in the presence of at least a single pathway of the
above-specified type in the system concerned. It is also evident that
contributions of separate pathways to $x_{(+)i,(-)m}^{(k)}$\ can cancel out
one another. [Rules governing the signs of increments of different pathways
have been formulated in Ref.[42]]. Finally, a cyclic pathway of the $k$th
order embraces no more than $k$ orbitals [Note that a certain orbital may be
"visited" more than once in this pathway]. This implies that no more than $k$
basis orbitals participate in the formation of the increment $%
x_{(+)i,(-)m}^{(k)}$.

In summary, the higher is the order parameter ($k$) of the given increment
to the overall charge (re)distribution and/or delocalization, the more
extended fragments of the system concerned are embraced by this increment.
Again, convergence of the relevant power series [58] ensures decrease of
increments with increasing values of $k$. Thus, a semilocal nature of both
charge (re)distribution and delocalization may be finally concluded for all
systems described by a diagonal zero order Hamiltonian matrix.

The results of this subsection have been successfully applied to systems
consisting of weakly-interacting two-center chemical bonds and lone electron
pairs. Bonding and antibonding orbitals of individual bonds along with those
of lone pairs played the role of basis functions $\{\varphi \}$ in this
case. Accordingly, the weak interbond interaction allowed us to assume the
diagonal constitutions of the relevant submatrices $\mathbf{E}_{(+)}$\ and $%
\mathbf{E}_{(-)}$\ of Eq.(1.1).

Heteroatom influence in substituted alkanes (the so-called inductive effect
[19]) has been studied [63] in the first place. The above-discussed
proportionalities between $x_{(+)i,(-)m}^{(k)}$\ and $d_{(+)i,(-)m}^{(k)}$\
along with other relations of subsection 1.3 formed the basis for two
mutually equivalent interpretations of the effect concerned, namely in terms
of a perturbed electron density distribution (as usual) and via changes in
the degrees of delocalization of separate NCMOs and thereby of individual
pairs of electrons. Moreover, the latter changes proved to be proportional
to delocalization coefficients of NCMOs in the respective parent alkane.
Finally, increments of the second order (i.e. $x_{(+)i,(-)m}^{(2)}$ and $%
d_{(+)i,(-)m}^{(2)}$) were shown to play the most important role in the
formation of the electron density redistribution and/or reshaping of NCMOs
due to emergence of a heteroatom. As a result, the short-range nature and
other peculiarities of the inductive effect have been concluded to originate
from a weak interbond delocalization in respective parent alkanes.
Analogously, additivity of the total effect of two heteroatoms upon the
hydrocarbon fragment has been revealed and traced back to the relevant
additivity of delocalization. The general nature of the above-overviewed
conclusions also deserves emphasizing, viz. these refer to any alkane and
its derivative.

More subtle aspects of the heteroatom influence have been studied in
Ref.[64], namely the so-called trans-effect revealing itself as
non-equivalence of the cis- and trans-arranged C$_{\beta }-$C$_{\gamma }$
(or C$_{\beta }-$H) bonds with respect to the heteroatom-containing (X$-$C$%
_{\alpha }$) bond. The third order corrections to occupation numbers of
basis orbitals proved to play the decisive role in this case. Moreover,
contributions [63] and [64] provide us with a unified description of the
inductive and trans-effects. Indeed, these two aspects of the heteroatom
influence prove to be representable by members of different orders of the
same power series.

In this context, investigation of non-additive effects of two substituents
(intersubstituent interaction) in D,A-disubstituted benzenes [65] also
deserves attention (D and A correspondingly stand here for an
electron-donating substituent and for the electron-accepting one). The usual
(canonical) MOs of benzene played the role of basis functions $\{\varphi \}$%
\ here along with orbitals localized on individual substituents. The
intersubstituent interaction was shown to manifest itself as two effects,
viz. (i) as an alteration in the extent of the intramolecular charge
transfer between an individual substituent and the phenyl ring owing to the
indirect participation of orbitals of the another substituent and (ii) as an
additional indirect charge transfer between substituents D and A by means of
orbitals of the phenyl ring playing the role of mediators. Moreover, these
two components of the intersubstituent interaction have been represented by
terms of the fourth order of the expansion for the relevant CBO matrix $%
\mathbf{P}$.

Therefore, classification of separate aspects of heteroatom influence proved
to be possible on the basis of the order parameter ($k$) of the decisive
terms of the power series applied. An analogous classification of chemical
reactions also was among the achievements that are overviewed elsewhere in a
detail [41, 42, 52, 66].

\begin{center}
\textbf{2.3. Interpretation of total energies as charge transfer energies. A
new insight into the Dewar formula}
\end{center}

Let us now discuss the implications of Eq.(2.1.1) upon the relevant total
energy ($\mathcal{E}$). To this end, let us return to general results of
subsection 1.4 and employ our principal assumption of Eq.(2.1.1) in the
expressions contained there.

The zero order member ($\mathcal{E}_{(0)}$) of the power series for the
total energy ($\mathcal{E)}$\ follows directly from the first relation of
Eq.(1.4.11) and coincides with the double sum of one-electron energies ($%
\varepsilon _{(+)i}$) of all initially-occupied orbitals $\varphi _{(+)i}$\
and thereby with their total energy before the interaction in accordance
with the expectation. Given that the actual one-electron energies of basis
orbitals are entirely included into the above-mentioned parameters $%
\varepsilon _{(+)i}$\ so that diagonal elements ($T_{ii}$) of the matrix $%
\mathbf{T}$ vanish (this always may be done without restrictions), a zero
value follows from the second relation of Eq.(1.4.11) for the first order
member of the same series ($\mathcal{E}_{(1)}$) in accordance with original
result of M. J. S.Dewar [13, 67, 68]. Thus, we may turn immediately to
increments $\mathcal{E}_{(k)}$\ of higher orders ($k=2,3,4...$).

As discussed already (Subsect. 1.4), each correction $\mathcal{E}_{(k)}$\ is
representable as an algebraic sum of two components $\mathcal{E}%
_{(k)}^{(\alpha )}$\ and $\mathcal{E}_{(k)}^{(\beta )}$\ defined by
Eqs.(1.4.3) and (1.4.4) and expressed as shown in Eqs.(1.4.7) and (1.4.8),
respectively. Due to the diagonal constitution of our matrices $\mathbf{E}%
_{(+)}$\ and $\mathbf{E}_{(-)}$\ (see Eq.(2.1.1)), only diagonal elements of
the intrasubset population matrices $\mathbf{X}_{(+)}^{(k)}$\ and $\mathbf{X}%
_{(-)}^{(k)}$\ actually contribute to the component $\mathcal{E}%
_{(k)}^{(\alpha )}$\ as Eq.(1.4.7) shows. As a result, this component
depends only upon increments of the $k$th order to the populations of basis
orbitals ($X_{(+)ii}^{(k)}$ and $X_{(-)ll}^{(k)}$). After representing the
latter as sums of partial transferred populations of the respective order $k$
(see Eq.(1.1.12)) and invoking the relations of Eq.(1.1.14), we obtain the
following formula for $\mathcal{E}_{(k)}^{(\alpha )}$\ [37], viz.\ 

\begin{equation}
\mathcal{E}_{(k)}^{(\alpha )}=-\mathop{\displaystyle \sum }\limits_{(+)i}%
\mathop{\displaystyle \sum }\limits_{(-)m}x_{(+)i,(-)m}^{(k)}(\varepsilon
_{(+)i}+\varepsilon _{(-)m}).  \tag{2.3.1}
\end{equation}%
Thus, the component $\mathcal{E}_{(k)}^{(\alpha )}$\ is now expressible in
terms of products of partial populations of the $k$th order transferred
between orbitals of opposite initial occupation and energy intervals between
the latter. Consequently, this component is interpretable as the charge
transfer energy. The minus sign in front of the sum of Eq.(2.3.1) also has a
clear meaning, namely it points to the destabilizing nature of $\mathcal{E}%
_{(k)}^{(\alpha )},$\ if $x_{(+)i,(-)m}^{(k)}$\ is a positive quantity and
vice versa. [Note that $\varepsilon _{(+)i}>0$ and $\varepsilon _{(-)m}>0$
and thereby $\varepsilon _{(+)i}+\varepsilon _{(-)m}>0$\ in our negative
energy units]. This result causes no surprise if we bear in mind that the
positive sign of $x_{(+)i,(-)m}^{(k)}$\ implies a certain charge transfer
from an orbital of the lower energy ($\varphi _{(+)i}$) to that of a higher
energy ($\varphi _{(-)m}).$

Let us now turn to the remaining component of the $k$th order energy ($%
\mathcal{E}_{(k)}^{(\beta )}$) defined by Eq.(1.4.8). Since diagonal
elements of the matrix $\mathbf{Q}$ also may be assumed to take zero values
for analogous reasons, the expression of Eq.(1.4.8) consequently contains no
increments of population alterations $X_{(+)ii}^{(k-1)}$ and $%
X_{(-)ll}^{(k-1)}$ in contrast to Eq.(1.4.7). Hence, the component $\mathcal{%
E}_{(k)}^{(\beta )}$\ describes the effect of formation of bond orders
between basis orbitals upon the $k$th order energy ($\mathcal{E}_{(k)}$). It
is also noteworthy that bond orders determining the component $\mathcal{E}%
_{(k)}^{(\beta )}$\ originate from members of the ($k-1$)th order of the
power series for the DM $\mathbf{P}$. If we recall now that the intrasubset
population matrices of the $k$th order (i.e. $\mathbf{X}_{(+)}^{(k)}$\ and $%
\mathbf{X}_{(-)}^{(k)}$) are expressed in terms of products of the principal
matrices of lower orders ($\mathbf{G}_{(k-1)},\mathbf{G}_{(k-2)},etc.$)
that, in turn, represent intersubset bond orders being formed within
previous terms of the same series (see Eq.(1.1.5)), the above-established
interdependence between components $\mathcal{E}_{(k)}^{(\alpha )}$ and $%
\mathcal{E}_{(k)}^{(\beta )}$ shown in Eq.(1.4.5) causes no surprise.
Moreover, charge redistribution of the $k$th order may be then considered as
a consequence (or counter-effect) of formation of intersubset bond orders
within terms of lower orders of the same series for the DM $\mathbf{P}$.
Thus, we have to do here with a certain gradual reorganization of bonding,
the energetic increments of which are interrelated and governed by
Eq.(1.4.5). Furthermore, the above-drawn conclusions concerning the signs
and absolute values of $\mathcal{E}_{(k)}^{(\alpha )}$ and $\mathcal{E}%
_{(k)}^{(\beta )}$\ (subsection 1.4) allow us to expect that stabilization
of our system vs. the set of isolated initially-occupied basis orbitals (if
any) is entirely due to formation of new bond orders, and the subsequent
charge redistribution actually reduces this stabilizing effect.
Nevertheless, the absolute value of the total $k$th order energy ($\mathcal{E%
}_{(k)})$ is proportional to that of the respective charge transfer energy $%
\mathcal{E}_{(k)}^{(\alpha )}$ as the first relation of Eq.(1.4.6)
indicates. Indeed, the minus signs of Eqs.(1.4.6) and (2.3.1) cancel out one
another and we obtain [37]%
\begin{equation}
\mathcal{E}_{(k)}=\frac{1}{k-1}\mathop{\displaystyle \sum }\limits_{(+)i}%
\mathop{\displaystyle \sum }\limits_{(-)m}x_{(+)i,(-)m}^{(k)}(\varepsilon
_{(+)i}+\varepsilon _{(-)m}).  \tag{2.3.2}
\end{equation}%
On the basis of the above relation, an interpretation of the total $k$th
order energy $\mathcal{E}_{(k)}$\ as the charge transfer energy also becomes
acceptable (in spite of its somewhat oversimplified nature). This
interpretation provides us with a quantum- chemical analogue of the
intuition-based relation between stabilization and charge redistribution. If
we recall now the interrelations between $x_{(+)i,(-)m}^{(k)}$\ and $%
d_{(+)i,(-)m}^{(k)}$\ shown in Eq.(1.3.24), we obtain that [52] 
\begin{equation}
\mathcal{E}_{(k)}=\frac{2}{k-1}\mathop{\displaystyle \sum }\limits_{(+)i}%
\mathop{\displaystyle \sum }\limits_{(-)m}d_{(+)i,(-)m}^{(k)}(\varepsilon
_{(+)i}+\varepsilon _{(-)m}).  \tag{2.3.3}
\end{equation}%
Thus, the $k$th order energy $\mathcal{E}_{(k)}$\ is alternatively
interpretable as that of the delocalization of the initially-occupied basis
orbitals over the initially-vacant ones, i.e. as the intersubset
delocalization energy of all initially-localized pairs of electrons.
Additivity of $\mathcal{E}_{(k)}$\ with respect to contributions of
individual pairs of electrons also is easily seen from Eq.(2.3.3).\ Hence,
the well-known intuition- based relation between delocalization and
stabilization [16, 69] acquires a quantum- chemical support. It is also
evident that Eq.(2.3.3) is a particular case of the relation of Eq.(1.4.10)
referring to diagonal matrices $\mathbf{E}_{(+)}$\ and $\mathbf{E}_{(-)}$.

As discussed in the subsection 1.1, the signs of partial transferred
populations $x_{(+)i,(-)m}^{(k)}$\ depend upon those of elements $G_{(k)im}$
contained within the relevant definition of Eq.(1.1.13), the latter
describing particular interorbital interactions in the present case
(Subsection 2.1). The expression of Eq.(2.3.2) then allows us to formulate
the rule governing the signs of energy corrections $\mathcal{E}_{(k)}$\ and
thereby the nature of the respective effect. Thus, the \textit{a priori}
positive sign of any second order term $x_{(+)i,(-)m}^{(2)}$\ (Subsect. 1.1)
indicates that the correction $\mathcal{E}_{(2)}$\ always is of stabilizing
nature in accordance with the expectation. Meanwhile, the signs of the
subsequent energy increments $\mathcal{E}_{(k)},k=3,4,..$\ depend upon those
of the relevant partial transferred populations and thereby these are
governed by the following rule: The correction $\mathcal{E}_{(k)},k=3,4..$\
contributes to an additional stabilization (destabilization) of the given
system (vs. the sum $\mathcal{E}_{(0)}+\mathcal{E}_{(1)}$), if the
interorbital interactions concerned are of coinciding (opposite) signs for
all (or almost all) pairs of orbitals $\varphi _{(+)i}$ and $\varphi
_{(-)m}. $ This rule will be referred to below as the sign consistence
condition for interorbital interactions. Applications of the latter to
chemical reactions are overviewed in Refs.[41, 52, 66].

As already mentioned, the vanishing correction $\mathcal{E}_{(1)}$\ is in
line with the relevant result obtained by M.S.J.Dewar [13, 67, 68]. An
analogous parallelism reveals itself also for higher values of the order
parameter $k$. To show this, let us consider the second order term $\mathcal{%
E}_{(2)}$\ as an example. After substituting Eqs.(1.1.13) and (2.1.3) into
Eq.(2.3.2) for $k=2$, we obtain the following expression%
\begin{equation}
\mathcal{E}_{(2)}^{(D)}=2\mathop{\displaystyle \sum }\limits_{(+)i}%
\mathop{\displaystyle \sum }\limits_{(-)m}\frac{(R_{im})^{2}}{\varepsilon
_{(+)i}+\varepsilon _{(-)m}}>0,  \tag{2.3.4}
\end{equation}%
which coincides with the usual perturbational formula for the second order
energy $\mathcal{E}_{(2)}$\ originally applied to chemical problems by
M.J.S. Dewar. [Just this fact is reflected by the superscript $(D)$]. Again,
the above-discussed representation of any correction $\mathcal{E}_{(k)}$\ in
terms of two components $\mathcal{E}_{(k)}^{(\alpha )}$\ and $\mathcal{E}%
_{(k)}^{(\beta )}$ gives us a new insight into the content of the well-known
formula of Eq.(2.3.4). Indeed, substituting Eqs.(1.1.13) and (2.1.3) into
Eq.(2.3.1) shows that the absolute value of the negative second order charge
transfer energy $\mathcal{E}_{(2)}^{(\alpha )}$\ coincides with that of $%
\mathcal{E}_{(2)}^{(D)}$. At the same time, the remaining (positive)
component $\mathcal{E}_{(2)}^{(\beta )}$\ is twice as large as $\mathcal{E}%
_{(2)}^{(D)}$\ as the second relation of Eq.(1.4.6) for $k=2$ indicates.
Thus, the stabilizing (positive) Dewar energy $\mathcal{E}_{(2)}^{(D)}$\ of
Eq.(2.3.4) proves to be actually made up of a difference beween the
stabilization energy due to formation of intersubset bond orders ($\mathcal{E%
}_{(2)}^{(\beta )}$) and the destabilization energy related to the
consequent charge transfer and/or delocalization ($\mathcal{E}%
_{(2)}^{(\alpha )}$) [Note that the above-mentioned newly-formed bond orders
are contained within non-zero off-diagonal blocks of the matrix $\mathbf{P}%
_{(1)}$\ defined by Eq.(1.1.5)]. Accordingly, our previous expression for
the second order energy $\mathcal{E}_{(2)}$\ shown in Eq.(1.4.11) proves to
be a generalization of the Dewar formula to the case of intrasubset
interactions (resonance parameters) of the zero order magnitude [70].
Evidently, the same refers even more to energy increments of Eq.(1.4.11) of
higher orders ($k=3,4...$).

\begin{center}
\textbf{2.4. The case of uniform one-electron energies inside subsets of
basis orbitals. Explicit formulae for entire matrices }$\mathbf{G}_{(k)}$
\end{center}

In addition to our previous assumption of Eq.(2.1.1), let the one-electron
energies of all initially-occupied orbitals ($\varepsilon _{(+)i},i=1,2,3..$%
) and those of all initially-vacant ones ($\varepsilon _{(-)m},$ $%
m=n+1,n+2,..N$) to take uniform values denoted by $\varepsilon _{(+)}$ and $%
\varepsilon _{(-)},$\ respectively. After an appropriate choice of the
energy unit, we may then accept the equality $\varepsilon _{(+)}=\varepsilon
_{(-)}=1,$\ the matrix form of which is as follows 
\begin{equation}
\mathbf{E}_{(+)}\ =\mathbf{E}_{(-)}=\mathbf{I.}  \tag{2.4.1}
\end{equation}%
It is evident that Eq.(2.4.1) is a particular case of Eq.(2.1.1). This
implies that all the above-overviewed implications of the latter
(Subsections 2.1-2.3) automatically refer to the present case as well. In
this connection, we concentrate ourselves here mostly on new (additional)
implications of Eq.(2.4.1) vs. those of Eq.(2.1.1).

Let us start with the notation that our principal matrix equations shown in
Eq.(1.1.8) may be solved algebraically under the condition of Eq.(2.4.1)
[36] and, consequently, the relevant matrices $\mathbf{G}_{(k)}$\ are
expressible via entire blocks of our Hamiltonian matrix of Eq.(1.1), viz.%
\begin{equation}
\mathbf{G}_{(1)}=-\frac{1}{2}\mathbf{R,}\qquad \mathbf{G}_{(2)}=\frac{1}{4}(%
\mathbf{TR}-\mathbf{RQ}),etc.  \tag{2.4.2}
\end{equation}%
These matrices now correspondingly represent the direct intersubset
interaction and the indirect one, wherein matrices $\mathbf{T}$ and $\mathbf{%
Q}$\ describe the "self-mediating effects" of subsets $\{\varphi _{(+)}\}$
and $\{\varphi _{(-)}\},$ respectively. Moreover, analogous expressions are
easily obtainable for related characteristcs, e.g. for the relevant CBO
matrix $\mathbf{P}$\ and the NCMO representation matrix $\mathbf{C}$. For
example, the starting members of the series for the matrix $\mathbf{P}$\
follow from Eqs.(1.1.5)-(1.1.7) and (1.1.10) after employment of Eq.(2.4.2),
e.g.%
\begin{equation}
\mathbf{P}_{(1)}=\left\vert 
\begin{array}{cc}
\mathbf{0} & \mathbf{R} \\ 
\mathbf{R}^{+} & \mathbf{0}%
\end{array}%
\right\vert ,\quad \mathbf{P}_{(2)}=\frac{1}{2}\left\vert 
\begin{array}{cc}
-\mathbf{RR}^{+} & \mathbf{RQ-TR} \\ 
\mathbf{QR}^{+}-\mathbf{R}^{+}\mathbf{T} & \mathbf{R}^{+}\mathbf{R}%
\end{array}%
\right\vert .  \tag{2.4.3}
\end{equation}%
[Note that the zero order member $\mathbf{P}_{(0)}$\ is shown in
Eq.(1.1.6)]. Similarly, the ket-vector of occupied NCMOs $\mid \Psi _{1}>$\
(Subsection 1.2) takes the following form 
\begin{equation}
\mid \Psi _{1}>=\mid \Phi _{1}>\mathbf{(I-}\frac{1}{8}\mathbf{\mathbf{R}%
R^{+})}+\frac{1}{2}\mid \Phi _{2}>[\mathbf{R}^{+}+\frac{1}{2}(\mathbf{R}^{+}%
\mathbf{T-QR}^{+})]\mathbf{.}  \tag{2.4.4}
\end{equation}%
It is easily seen that Eqs.(2.4.3) and (2.4.4) contain the same combinations
of matrices $\mathbf{T,R}$ and $\mathbf{Q,}$\ namely $\mathbf{RR}^{+},$ $%
\mathbf{R}^{+}\mathbf{R,}$ \ $\mathbf{RQ-TR,}$ $etc\mathbf{.}$ This result
serves as an additional illustration of the interdependence between matrices 
$\mathbf{P}$ and $\mathbf{C}$. Besides, these matrices are comparatively
analyzed in Ref.[51] in a detail in connection with their relevance to
alkanes.

Let us now dwell on implications of Eq.(2.4.1) upon the relevant total
energies ($\mathcal{E}$). After substituting Eq.(2.4.1) into Eqs.(1.4.9) and
(1.4.10) and employment Eqs.(1.1.11), (1.3.11) and (1.3.19), respectively,
we obtain \ 
\begin{equation}
\mathcal{E}_{(k)}=-\frac{2}{k-1}Tr\mathbf{X}_{(+)}^{(k)}=\frac{4}{k-1}Tr%
\mathbf{D}_{(+)}^{(k)}\equiv \frac{4}{k-1}D_{(+)},  \tag{2.4.5}
\end{equation}%
i.e. the correction $\mathcal{E}_{(k)}$\ is now proportional to the complete
delocalization coefficient ($D_{(+)}$) of occupied NCMOs $\{\psi _{(+)}\}$
defined by Eq.(1.3.11). An alternative form of Eq.(2.4.5) is as follows%
\begin{equation}
\mathcal{E}_{(k)}=-\frac{2}{k-1}\mathop{\displaystyle \sum }\limits%
_{(+)i}X_{(+)ii}^{(k)}=\frac{4}{k-1}\mathop{\displaystyle \sum }\limits%
_{(+)i}D_{(+)ii}^{(k)}.  \tag{2.4.6}
\end{equation}%
Thus, the energy correction $\mathcal{E}_{(k)}$\ is expressible only via
traces of matrices $\mathbf{X}_{(+)}^{(k)}$\ and/or $\mathbf{D}_{(+)}^{(k)}$%
. After substituting the expressions for diagonal elements of these matrices
($X_{(+)ii}^{(k)}$ and $D_{(+)ii}^{(k)}$) in terms of the relevant partial
increments as shown in Eqs.(1.1.12) and (1.3.10), we accordingly obtain that
\ 
\begin{equation}
\mathcal{E}_{(k)}=\frac{2}{k-1}\mathop{\displaystyle \sum }\limits_{(+)i}%
\mathop{\displaystyle \sum }\limits_{(-)m}x_{(+)i,(-)m}^{(k)}=\frac{4}{k-1}%
\mathop{\displaystyle \sum }\limits_{(+)i}\mathop{\displaystyle \sum }\limits%
_{(-)m}d_{(+)i,(-)m}^{(k)}.  \tag{2.4.7}
\end{equation}%
It is also evident that the same results are obtainable from Eqs.(2.3.2) and
(2.3.3) after replacing all energy intervals $\varepsilon
_{(+)i}+\varepsilon _{(-)m}$ by 2. Finally, a simple proportionality between
matrices $\mathbf{G}_{(1)}$ and $\mathbf{R}$\ seen from the first relation
of Eq.(2.4.2) allows the corrections $\mathcal{E}_{(k)}$\ of Eq.(1.4.11) to
be represented in terms of matrices $\mathbf{G}_{(k)}$\ only [38, 40], e.g. 
\begin{align}
\mathcal{E}_{(2)} =&4Tr(\mathbf{G}_{(1)}\mathbf{G}_{(1)}^{+}),\quad 
\mathcal{E}_{(3)}=4Tr(\mathbf{G}_{(2)}\mathbf{G}_{(1)}^{+}),  \nonumber \\
\mathcal{E}_{(4)} =&4Tr[(\mathbf{G}_{(3)}+\mathbf{G}_{(1)}\mathbf{G}%
_{(1)}^{+}\mathbf{G}_{(1)})\mathbf{G}_{(1)}^{+}],  \tag{2.4.8} \\
\mathcal{E}_{(5)} =&4Tr[(\mathbf{G}_{(4)}+3\mathbf{G}_{(2)}\mathbf{G}%
_{(1)}^{+}\mathbf{G}_{(1)})\mathbf{G}_{(1)}^{+}],etc.  \nonumber
\end{align}%
Formulae of the above-exhibited type, however, proved to be somewhat
cumbersome for practical applications. In this connection, alternative
expressions for the same corrections $\mathcal{E}_{(k)}$\ have been derived,
wherein the principal matrices of the right-hand sides of Eq.(2.4.8) of the
highest order ($k$) have been eliminated. The derivations concerned [38, 40,
71] were based on recurrence relations for matrices $\mathbf{G}_{(k)}$\
following from Eqs.(1.1.8) and (1.1.9) under condition of Eq.(1.4.1), viz.%
\begin{equation}
\mathbf{G}_{(2)}=-\frac{1}{2}(\mathbf{TG}_{(1)}-\mathbf{G}_{(1)}\mathbf{Q}%
),\quad \mathbf{G}_{(3)}=-\frac{1}{2}(\mathbf{TG}_{(2)}-\mathbf{G}_{(2)}%
\mathbf{Q})-2\mathbf{G}_{(1)}\mathbf{G}_{(1)}^{+}\mathbf{G}_{(1)},etc. 
\tag{2.4.9}
\end{equation}%
and on their subsequent employment to reformulate $Tr(\mathbf{G}_{(3)}%
\mathbf{G}_{(1)}^{+}\mathbf{),}$ $Tr(\mathbf{G}_{(4)}\mathbf{G}_{(1)}^{+}%
\mathbf{)}$, etc. using cyclic transpositions of matrices inside the trace
signs. We then obtain the following relations 
\begin{align}
Tr(\mathbf{G}_{(3)}\mathbf{G}_{(1)}^{+}\mathbf{)} \mathbf{=}&Tr(\mathbf{G}%
_{(2)}\mathbf{G}_{(2)}^{+}\mathbf{)-}2Tr(\mathbf{G}_{(1)}\mathbf{G}_{(1)}^{+}%
\mathbf{G}_{(1)}\mathbf{G}_{(1)}^{+}\mathbf{),}  \nonumber \\
Tr(\mathbf{G}_{(4)}\mathbf{G}_{(1)}^{+}\mathbf{)} \mathbf{=}&Tr(\mathbf{G}%
_{(3)}\mathbf{G}_{(2)}^{+}\mathbf{)-}4Tr(\mathbf{G}_{(2)}\mathbf{G}_{(1)}^{+}%
\mathbf{G}_{(1)}\mathbf{G}_{(1)}^{+}\mathbf{)}  \tag{2.4.10}
\end{align}%
and alternative expressions for $\mathcal{E}_{(4)}$ [71] and $\mathcal{E}%
_{(5)}$ [38]$,$\ viz.%
\begin{align}
\mathcal{E}_{(4)} =&4Tr(\mathbf{G}_{(2)}\mathbf{G}_{(2)}^{+})-4Tr(\mathbf{G}%
_{(1)}\mathbf{G}_{(1)}^{+}\mathbf{G}_{(1)}\mathbf{G}_{(1)}^{+}), 
\tag{2.4.11} \\
\mathcal{E}_{(5)} =&4Tr[(\mathbf{G}_{(3)}-\mathbf{G}_{(1)}\mathbf{G}%
_{(1)}^{+}\mathbf{G}_{(1)})\mathbf{G}_{(2)}^{+}].  \nonumber
\end{align}%
Besides, individual terms of the right-hand side of the above expression for 
$\mathcal{E}_{(4)}$ contain positive-definite matrices $\mathbf{G}_{(2)}%
\mathbf{G}_{(2)}^{+}$\ and $\mathbf{G}_{(1)}\mathbf{G}_{(1)}^{+}\mathbf{G}%
_{(1)}\mathbf{G}_{(1)}^{+}$\ [Note that any matrix product of the form $%
\mathbf{AA}^{+}$ is a positive-definite matrix [72]]. Consequently, the
correction $\mathcal{E}_{(4)}$\ consists of a sum of a positive
(stabilizing) and a negative (destabilizing) components in contrast to the 
\textit{a priori} positive second order energy $\mathcal{E}_{(2)}$.

Corrections of even higher orders $(k=6,7..)$ also may be derived similarly.
For example, the sixth order energy $\mathcal{E}_{(6)}$\ takes the following
form [40]%
\begin{equation}
\mathcal{E}_{(6)}=4Tr(\mathbf{G}_{(3)}\mathbf{G}_{(3)}^{+})-8Tr(\mathbf{G}%
_{(1)}\mathbf{G}_{(2)}^{+}\mathbf{G}_{(1)}\mathbf{G}_{(2)}^{+})-8Tr(\mathbf{G%
}_{(1)}\mathbf{G}_{(1)}^{+}\mathbf{G}_{(1)}\mathbf{G}_{(1)}\mathbf{G}%
_{(1)}^{+}\mathbf{G}_{(1)}).  \tag{2.4.12}
\end{equation}%
The expressions of Eqs.(2.4.8), (2.4.11) and (2.4.12) have been successfully
applied in studies of relative stabilities of linear and branched isomers of
polyenes [73], as well as of individual Kekul\'{e} valence structures of
benzenoids [38] and phenylenes [61].\bigskip

\begin{center}
\textbf{3. An extension of the PNCMO theory to the case of two
quasi-degenerate subsets of basis orbitals\bigskip }
\end{center}

The standard PNCMO theory of Section 1 was formulated for a nearly
block-diagonal Hamiltonian matrix shown in Eq.(1.1) and referring to two
weakly-interacting (well-separated) subsets of basis orbitals $\{\varphi
_{(+)}\}$ and $\{\varphi _{(-)}\}$. Now, we will turn to the opposite case,
namely to systems characterized by a strong intersubset interaction as
compared to the energy gap between orbitals of different subsets. Due to a
prospective application of the results of the subsection 1.5, we will invoke
the designations used there. Thus, let $\{\chi \},\{\chi _{1}\}$ and $\{\chi
_{2}\}$\ stand for the $N\times N-$dimensional initial basis set and its
subsets, respectively, the latter generally embracing different numbers of
orbitals. The relevant total Hamiltonian matrix ($\widetilde{\mathbf{H}}$)
will contain a zero order member $(\widetilde{\mathbf{H}}_{(0)})$ and a
first order one ($\widetilde{\mathbf{H}}_{(1)}$) in accordance with
Eq.(1.5.3). As opposed to the zero order term ($\mathbf{H}_{(0)}$) of our
former Hamiltonian matrix of Eq.(1.1), however, the present one ($\widetilde{%
\mathbf{H}}_{(0)}$) is assumed to take an anti-block-diagonal form, i.e. 
\begin{equation}
\widetilde{\mathbf{H}}=\widetilde{\mathbf{H}}_{(0)}+\widetilde{\mathbf{H}}%
_{(1)}=\left\vert 
\begin{array}{cc}
\mathbf{0} & \mathbf{B} \\ 
\mathbf{B}^{+} & \mathbf{0}%
\end{array}%
\right\vert +\left\vert 
\begin{array}{cc}
\mathbf{A} & \mathbf{M} \\ 
\mathbf{M}^{+} & \mathbf{F}%
\end{array}%
\right\vert ,  \tag{3.1}
\end{equation}%
where the (sub)matrix $\mathbf{B}$\ embraces the most part of the
intersubset interaction (specified below), whilst $\mathbf{A,F}$ and $%
\mathbf{M}$\ stand for (sub)matrices of the first order magnitude vs. the
former. Accordingly, the two-dimensional prototype of the matrix $\widetilde{%
\mathbf{H}}$\ is as follows 
\begin{equation}
\widetilde{\mathbf{h}}=\widetilde{\mathbf{h}}_{(0)}+\widetilde{\mathbf{h}}%
_{(1)}=\left\vert 
\begin{array}{cc}
0 & \beta \\ 
\beta & 0%
\end{array}%
\right\vert +\left\vert 
\begin{array}{cc}
\alpha & \mu \\ 
\mu & \phi%
\end{array}%
\right\vert ,  \tag{3.2}
\end{equation}%
and represents a two-level system characterized by a large resonance
parameter ($\beta +\mu $) vs. the difference in the Coulomb ones ($\alpha
-\phi $), i.e. a system of the second type (ii) of the Introduction. Again,
two quasi-degenerate energy levels underly the simple matrix $\widetilde{%
\mathbf{h}}$. In this connection, systems embraced by the matrix $\widetilde{%
\mathbf{H}}$\ of Eq.(3.1) are referred to below as those of two
quasi-degenerate subsets of basis orbitals. Now, let us turn to derivation
of the CBO matrix $\widetilde{\mathbf{P}}$\ and of the NCMO representation
matrix $\widetilde{\mathbf{C}}$ corresponding to the above-specified systems.

\begin{center}
\textbf{3.1. General results for any pair of quasi-degenerate subsets}
\end{center}

As with the application of the usual RSPT to the simple matrix $\widetilde{%
\mathbf{h}}$\ of Eq.(3.2), we expect to get rid of the quasi-degeneracy
between subsets $\{\chi _{1}\}$ and $\{\chi _{2}\}$\ by finding an
appropriate unitary transformation of the basis set. As discussed in the
Subsection 1.5, we may confine ourselves to block-diagonalization of the
zero order member $\widetilde{\mathbf{H}}_{(0)}$\ in accordance with
Eq.(1.5.4). Non-perturbative solution of the block-diagonalization problem
for the matrix $\widetilde{\mathbf{H}}_{(0)}$\ of Eq.(3.1) is described in
Ref.[53] in a detail along with conditions imposed on the matrix $\mathbf{B.}
$\ Two alternative forms of the transformation matrix $\mathbf{U}$\ have
been derived in this study, viz. \ 
\begin{equation}
\mathbf{U=}\frac{1}{\sqrt{2}}\left\vert 
\begin{array}{cc}
\mathbf{I} & \mathbf{KB} \\ 
\mathbf{B}^{+}\mathbf{K} & -\mathbf{I}%
\end{array}%
\right\vert ,\quad \mathbf{U=}\frac{1}{\sqrt{2}}\left\vert 
\begin{array}{cc}
\mathbf{I} & \mathbf{BL} \\ 
\mathbf{LB}^{+} & -\mathbf{I}%
\end{array}%
\right\vert ,  \tag{3.1.1}
\end{equation}%
where $\mathbf{KB=}$\ $\mathbf{BL}$ and$\ \mathbf{K}$\ and $\mathbf{L}$\ are
Hermitian matrices defined as follows 
\begin{equation}
\mathbf{K}=\mathbf{(BB}^{+}\mathbf{)}^{-1/2},\qquad \mathbf{L=(B}^{+}\mathbf{%
B)}^{-1/2}.  \tag{3.1.2}
\end{equation}%
The eigenblocks of the matrix $\widetilde{\mathbf{H}}_{(0)},$\ in turn,
coincide with the inverse counterparts of matrices $\mathbf{K}$\ and $%
\mathbf{L}$\ , viz. \ 
\begin{equation}
\mathbf{E}_{(+)}\ =\mathbf{K}^{-1}=\mathbf{(BB}^{+}\mathbf{)}^{1/2},\qquad 
\mathbf{E}_{(-)}=\mathbf{L}^{-1}\mathbf{=(B}^{+}\mathbf{B)}^{1/2}. 
\tag{3.1.3}
\end{equation}%
Our next step consists in application of the matrix $\mathbf{U}$\ of
Eq.(3.1.1) to transform the first order member $\widetilde{\mathbf{H}}_{(1)}$%
\ of Eq.(3.1) in accordance with Eq.(1.5.5). The resulting total Hamiltonian
matrix then resembles that of Eq.(1.1), where submatrices of the new first
order member $\mathbf{H}_{(1)}$\ are expressible via those of $\widetilde{%
\mathbf{H}}_{(1)}$\ of Eq.(3.1) as follows%
\begin{align}
\mathbf{T} =&\frac{1}{2}[(\mathbf{A}+\mathbf{BLFLB}^{+})+(\mathbf{BLM}^{+}+%
\mathbf{MLB}^{+})],  \nonumber \\
\mathbf{Q} =&\frac{1}{2}[(\mathbf{F}+\mathbf{LB}^{+}\mathbf{ABL})-(\mathbf{M%
}^{+}\mathbf{BL}+\mathbf{LB}^{+}\mathbf{M})],  \tag{3.1.4} \\
\mathbf{R} =&\frac{1}{2}[(\mathbf{ABL}-\mathbf{BLF})+(\mathbf{BLM}^{+}%
\mathbf{BL-M})].  \nonumber
\end{align}%
[The second expression of Eq.(3.1.1) is used here]. Consequently, the
standard PNCMO theory of Section 1 may be applied to find formulae for
separate members ($\mathbf{P}_{(k)}$ and $\mathbf{C}_{(k)},$ $k=0,1,2,..$\ )
of power series for matrices \ $\mathbf{P}$ and $\mathbf{C.}$\ Due to the
generally non-diagonal constitution of (sub)matrices $\mathbf{K}$\ and $%
\mathbf{L}$\ and thereby of eigenblocks $\mathbf{E}_{(+)}$\ and $\mathbf{E}%
_{(-)}$\ (see Eq.(3.1.3)), the implicit form of the PNCMO theory should be
invoked here. In other words, matrix equations of Eq.(1.1.8) should be
solved for any order parameter ($k$), where $\mathbf{K}^{-1}$\ and $\mathbf{L%
}^{-1}$\ stand for $\mathbf{E}_{(+)}$\ and $\mathbf{E}_{(-)},$\
respectively. For example, the first order matrix $\mathbf{G}_{(1)}$\ is
obtainable from solution of the following equation%
\begin{equation}
\mathbf{G}_{(1)}\mathbf{K+LG}_{(1)}+\mathbf{KRL}=\mathbf{0,}  \tag{3.1.5}
\end{equation}%
where the (sub)matrix $\mathbf{R}$\ is shown in the last relation of
Eq.(3.1.4). The final step of the overall procedure consists in
retransforming the CBO matrix $\mathbf{P}$ and the NCMO representation
matrix\ $\mathbf{C}$\ obtained into the basis $\{\chi \}$ \ again.

As discussed already, each member of the expansions concerned (i.e. $\mathbf{%
P}_{(k)}$ and $\mathbf{C}_{(k)}$) may be retransformed separately as shown
in Eq.(1.5.6). For example, retransformation of the zero order member $%
\mathbf{P}_{(0)}$ of Eq.(1.1.6) of the expansion for the CBO matrix $\mathbf{%
P}$\ by means of Eqs.(3.1.1) and (1.5.6) yields the following result

\begin{equation}
\widetilde{\mathbf{P}}_{(0)}\mathbf{=}\left\vert 
\begin{array}{cc}
\mathbf{I} & \mathbf{KB} \\ 
\mathbf{B}^{+}\mathbf{K} & \mathbf{I}%
\end{array}%
\right\vert ,\quad \widetilde{\mathbf{P}}_{(0)}\mathbf{=}\left\vert 
\begin{array}{cc}
\mathbf{I} & \mathbf{BL} \\ 
\mathbf{LB}^{+} & \mathbf{I}%
\end{array}%
\right\vert .  \tag{3.1.6}
\end{equation}%
The above-exhibited matrix coincides with the CBO matrix of alternant
hydrocarbons derived in Ref.[74] using the known constitution of their
canonical MOs. Such a coincidence causes no surprise if we bear in mind that
these hydrocarbons are among systems embraced by the zero order matrix $%
\widetilde{\mathbf{H}}_{(0)}$\ of Eq.(3.1) as discussed in the next
subsection. Accordingly, the zero order member ($\widetilde{\mathbf{C}}%
_{(0)} $) of the relevant expansion for the NCMO representation matrix $%
\widetilde{\mathbf{C}}$\ \ follows after retransforming the matrix $\mathbf{C%
}_{(0)}=\mathbf{I}$ (subsection 1.2) and takes the form [53] \ \ 
\begin{equation}
\widetilde{\mathbf{C}}_{(0)}\mathbf{=}\frac{1}{\sqrt{2}}\left\vert 
\begin{array}{cc}
\mathbf{I} & \mathbf{KB} \\ 
\mathbf{B}^{+}\mathbf{K} & -\mathbf{I}%
\end{array}%
\right\vert ,\quad \widetilde{\mathbf{C}}_{(0)}\mathbf{=}\frac{1}{\sqrt{2}}%
\left\vert 
\begin{array}{cc}
\mathbf{I} & \mathbf{BL} \\ 
\mathbf{LB}^{+} & -\mathbf{I}%
\end{array}%
\right\vert .  \tag{3.1.7}
\end{equation}%
Comparison of Eqs.(3.1.6) and (3.1.7) indicates a high extent of similarity
between zero order members of the two expansions (as it was the case with
the CBO matrix and the NCMO representation matrix of Section 1). In
particular, columns of the matrix $\widetilde{\mathbf{P}}_{(0)}$\ coincide
with coefficients occupied NCMOs to within the factor $1/\sqrt{2}.$ A
similar result has been obtained also in Ref.[75]. Analogously, the relevant
members of higher orders ($\widetilde{\mathbf{P}}_{(k)}$ and $\widetilde{%
\mathbf{C}}_{(k)},$ $k=1,2,..$\ ) result after applying the matrix $\mathbf{U%
}$ of Eq.(3.1.1) to respective corrections $\mathbf{P}_{(k)}$ and $\mathbf{C}%
_{(k)},$ of subsections 1.1 and 1.2.

Let us concentrate ourselves here mostly on the CBO matrix $\widetilde{%
\mathbf{P}}.$ As is seen from Eq.(1.1.5), the $k$th order member ($\mathbf{P}%
_{(k)}$) of the series to be retransformed is representable as a sum of two
components denoted below by $\mathbf{P}_{(k)}^{(g)}$\ and $\mathbf{P}%
_{(k)}^{(x)}$\ and correspondingly containing the intersubset coupling
matrices ($\mathbf{G}_{(k)}$) and the intrasubset population matrices ($%
\mathbf{X}_{(+)}^{(k)}$ and $\mathbf{X}_{(-)}^{(k)}$), viz.%
\begin{equation}
\mathbf{P}_{(k)}=\mathbf{P}_{(k)}^{(g)}\ +\mathbf{P}_{(k)}^{(x)}, 
\tag{3.1.8}
\end{equation}%
where 
\begin{equation}
\mathbf{P}_{(k)}^{(g)}\mathbf{=}\left\vert 
\begin{array}{cc}
\mathbf{0} & -2\mathbf{G}_{(k)} \\ 
-2\mathbf{G}_{(k)}^{+} & \mathbf{0}%
\end{array}%
\right\vert ,\qquad \mathbf{P}_{(k)}^{(x)}\mathbf{=}\left\vert 
\begin{array}{cc}
\mathbf{X}_{(+)}^{(k)} & \mathbf{0} \\ 
0 & \mathbf{X}_{(-)}^{(k)}%
\end{array}%
\right\vert .  \tag{3.1.9}
\end{equation}%
It is evident that each of the components $\mathbf{P}_{(k)}^{(g)}$\ and $%
\mathbf{P}_{(k)}^{(x)}$\ also may be retransformed separately. Moreover, the
resulting total correction $\widetilde{\mathbf{P}}_{(k)}$\ accordingly
consists of a sum of two retransformed components as follows%
\begin{equation}
\widetilde{\mathbf{P}}_{(k)}=\widetilde{\mathbf{P}}_{(k)}^{(g)}\ +\widetilde{%
\mathbf{P}}_{(k)}^{(x)},  \tag{3.1.10}
\end{equation}%
where 
\begin{equation}
\widetilde{\mathbf{P}}_{(k)}^{(g)}=\mathbf{U\mathbf{P}}_{(k)}^{(g)}\mathbf{U}%
^{+},\qquad \widetilde{\mathbf{P}}_{(k)}^{(x)}=\mathbf{U\mathbf{P}}%
_{(k)}^{(x)}\mathbf{U}^{+}.  \tag{3.1.11}
\end{equation}%
Let us now introduce the following designations for individual submatrices
(blocks) of the retransformed matrices $\widetilde{\mathbf{P}}_{(k)}^{(g)}\ $%
\ and $\widetilde{\mathbf{P}}_{(k)}^{(x)},$\ viz. 
\begin{equation}
\widetilde{\mathbf{P}}_{(k)}^{(g)}\mathbf{=}\left\vert 
\begin{array}{cc}
\mathbf{Y}_{(k)} & \mathbf{V}_{(k)} \\ 
\mathbf{V}_{(k)}^{+} & \mathbf{Z}_{(k)}%
\end{array}%
\right\vert ,\qquad \widetilde{\mathbf{P}}_{(k)}^{(x)}\mathbf{=}\left\vert 
\begin{array}{cc}
\mathbf{S}_{(k)} & \mathbf{J}_{(k)} \\ 
\mathbf{J}_{(k)}^{+} & \mathbf{N}_{(k)}%
\end{array}%
\right\vert .  \tag{3.1.12}
\end{equation}%
As shown in Refs.[54, 76], submatrices $\mathbf{Y}_{(k)}$\ and $\mathbf{Z}%
_{(k)}$\ of the component $\widetilde{\mathbf{P}}_{(k)}^{(g)}$\ are
expressible as follows%
\begin{equation}
\mathbf{Y}_{(k)}=-\mathbf{BLG}_{(k)}^{+}-\mathbf{G}_{(k)}\mathbf{LB}%
_{{}}^{+},\qquad \mathbf{Z}_{(k)}=\mathbf{G}_{(k)}^{+}\mathbf{BL+\mathbf{LB}%
_{{}}^{+}G}_{(k)}  \tag{3.1.13}
\end{equation}%
and interrelated as exhibited below%
\begin{equation}
\mathbf{Y}_{(k)}=-\mathbf{BLZ}_{(k)}\mathbf{\mathbf{LB}_{{}}^{+}.} 
\tag{3.1.14}
\end{equation}%
Analogously, the remaining submatrix $\mathbf{V}_{(k)}$\ of the same
component $\widetilde{\mathbf{P}}_{(k)}^{(g)}$\ is representable via a
skew-symmetric matrix $\mathbf{\Delta }_{(k)}$, viz. 
\begin{equation}
\mathbf{V}_{(k)}=-\mathbf{BL\Delta }_{(k)},  \tag{3.1.15}
\end{equation}%
where 
\begin{equation}
\mathbf{\Delta }_{(k)}=-\mathbf{\Delta }_{(k)}^{+}=\mathbf{G}_{(k)}^{+}%
\mathbf{BL-\mathbf{LB}_{{}}^{+}G}_{(k)}.  \tag{3.1.16}
\end{equation}%
Meanwhile, submatrices of the intra-subset- population- matrices-containing
component $\widetilde{\mathbf{P}}_{(k)}^{(x)}$\ take the following form%
\begin{equation}
\mathbf{S}_{(k)}=\frac{1}{2}(\mathbf{X}_{(+)}^{(k)}+\mathbf{BLX}_{(-)}^{(k)}%
\mathbf{\mathbf{LB}_{{}}^{+}),}  \tag{3.1.17}
\end{equation}%
\begin{equation}
\mathbf{N}_{(k)}=\frac{1}{2}(\mathbf{X}_{(-)}^{(k)}+\mathbf{\mathbf{LB}%
_{{}}^{+}X}_{(+)}^{(k)}\mathbf{BL),}  \tag{3.1.18}
\end{equation}%
\begin{equation}
\mathbf{J}_{(k)}=-\frac{1}{2}\mathbf{BL}(\mathbf{X}_{(-)}^{(k)}-\mathbf{%
\mathbf{LB}_{{}}^{+}X}_{(+)}^{(k)}\mathbf{BL)=-}\frac{1}{2}(\mathbf{BLX}%
_{(-)}^{(k)}-\mathbf{X}_{(+)}^{(k)}\mathbf{BL).}  \tag{3.1.19}
\end{equation}%
Finally, the component $\widetilde{\mathbf{P}}_{(k)}^{(x)}$\ vanishes for $%
k=1$ due to the first relation of Eq.(1.1.10) and we obtain that%
\begin{equation}
\widetilde{\mathbf{P}}_{(1)}\mathbf{=}\left\vert 
\begin{array}{cc}
\mathbf{Y}_{(1)} & \mathbf{V}_{(1)} \\ 
\mathbf{V}_{(1)}^{+} & \mathbf{Z}_{(1)}%
\end{array}%
\right\vert ,  \tag{3.1.20}
\end{equation}%
where $\mathbf{Y}_{(1)},$\ $\mathbf{Z}_{(1)}$\ and $\mathbf{V}_{(1)}$\ are
defined by Eqs.(3.1.13)-(3.1.16). For comparison, the first order member ($%
\widetilde{\mathbf{C}}_{(1)})$\ of the expansion for the retransformed NCMO
representation matrix ($\widetilde{\mathbf{C}}$) is as follows%
\begin{equation}
\widetilde{\mathbf{C}}_{(1)}\mathbf{=}\frac{1}{\sqrt{2}}\left\vert 
\begin{array}{cc}
-\mathbf{BLG}_{(1)}^{+} & \mathbf{G}_{(1)} \\ 
\mathbf{G}_{(1)}^{+} & \mathbf{LB}^{+}\mathbf{G}_{(1)}%
\end{array}%
\right\vert ,  \tag{3.1.21}
\end{equation}%
and the extent of its similarity to $\widetilde{\mathbf{P}}_{(1)}$\ is much
lower (as compared to that between matrices $\widetilde{\mathbf{P}}_{(0)}$\
and $\widetilde{\mathbf{C}}_{(0)}$\ of Eqs.(3.1.6) and (3.1.7)). The same
refers even more to corrections $\widetilde{\mathbf{P}}_{(k)}$\ and $%
\widetilde{\mathbf{C}}_{(k)}$\ of higher orders ($k=2,3,..).$\ Nevertheless,
certain relations have been established between these corrections [54].

Let us now turn to power series for total energies of systems described by
the Hamiltonian matrix of Eq.(3.1). Thus, the zero order member of this
series ($\mathcal{E}_{(0)}$) takes the following form [53] 
\begin{equation}
\mathcal{E}_{(0)}=2Tr\mathbf{K}^{-1}=2Tr[\mathbf{(BB}^{+}\mathbf{)}^{1/2}]. 
\tag{3.1.22}
\end{equation}%
Further, the relevant first order term $\mathcal{E}_{(1)}$\ follows from the
second relation of Eq.(1.4.11) after substituting the first formula of
Eq.(3.1.4) and is expressible as follows%
\begin{equation}
\mathcal{E}_{(1)}=2Tr\mathbf{T=}Tr[\mathbf{A+F+BLM}^{+}\mathbf{+LB}^{+}%
\mathbf{M}],  \tag{3.1.23}
\end{equation}%
where the cyclic transposition of matrices inside the trace sign is used.
This correction generally does not vanish due to non-zero values of diagonal
elements of the matrix $\mathbf{T}$ defined by Eq.(3.1.4). Moreover, the
first order energy $\mathcal{E}_{(1)}$\ of Eq.(3.1.23) actually coincides
with $\mathcal{E}_{(1)}^{(\beta )}$\ (see Eqs.(1.4.4) and (1.5.7)), whereas $%
\mathcal{E}_{(1)}^{(\alpha )}$\ vanishes due to the skew-symmetric nature of
the matrix $\mathbf{\Delta }_{(1)}$\ defined by Eq.(3.1.16). Thus, the
expression of Eq.(3.1.23) is alternatively obtainable by substituting
matrices $\widetilde{\mathbf{H}}_{(1)}$\ of Eq.(3.1) and $\widetilde{\mathbf{%
P}}_{(0)}$\ of Eq.(3.1.6) into the second relation of Eq.(1.5.7).

Members of higher orders of the same expansion ($\mathcal{E}_{(k)},k=2,3,..)$
also may be derived analogously using any of the relations shown in
Eq.(1.5.8). Since the matrix $\widetilde{\mathbf{H}}_{(0)}$\ of Eq.(3.1) is
of a more simple constitution as compared to $\widetilde{\mathbf{H}}_{(1)},$%
\ employment of the first relation of Eq.(1.5.8) proves to be more
convenient for this purpose and, consequently, a more compact formula for $%
\mathcal{E}_{(k)}$\ follows. Indeed, only the off-diagonal blocks of the
correction $\widetilde{\mathbf{P}}_{(k)}$\ (i.e. (sub)matrices $\mathbf{V}%
_{(k)}$\ and $\mathbf{J}_{(k)}$) are able to contribute to the component $%
\mathcal{E}_{(k)}^{(\alpha )}$\ (the first relation of Eq.(1.5.8) is based
on). Moreover, the actual contributions of (sub)matrices $\mathbf{V}_{(k)}$\
vanish because of the skew-symmetric nature of (sub)matrices $\mathbf{\Delta 
}_{(k)}$\ of Eq.(3.1.16) as it was the case with the $\mathbf{V}_{(1)}-$%
related increment of the first order energy. The final formula for the $k$th
order energy is then as follows%
\begin{equation}
\mathcal{E}_{(k)}=-\frac{1}{k-1}Tr(\mathbf{J}_{(k)}\mathbf{B}^{+}+\mathbf{J}%
_{(k)}^{+}\mathbf{B}),\quad k=2,3..  \tag{3.1.24}
\end{equation}%
and contains the (sub)matrices $\mathbf{J}_{(k)}$\ only. Thus, the
correction $\mathcal{E}_{(k)}$\ proves to be expressible via alterations in
the intersubset bond orders caused by the (sub)matrices $\mathbf{J}_{(k)}$.
If we bear in mind that (sub)matrices $\mathbf{V}_{(k)}$\ also generally
yield non-zero increments to individual bond orders as demonstrated in
Ref.[76], the above outcome seems to be somewhat unexpected. The overall
situation here, however, closely resembles that described in the subsections
1.4 and 2.3. In particular, other submatrices of corrections $\widetilde{%
\mathbf{P}}_{(k)}$\ of Eq.(3.1.12) also participate in the formation of the $%
k$th order energy, namely of the relevant \ $\beta -$component ($\mathcal{E}%
_{(k)}^{(\beta )}$). To illustrate this point, let us dwell on components $%
\mathcal{E}_{(2)}^{(\alpha )}$\ and $\mathcal{E}_{(2)}^{(\beta )}$\ of the
second order energy $\mathcal{E}_{(2)}.$\ Thus, substituting matrices $%
\widetilde{\mathbf{H}}_{(1)}$\ and $\widetilde{\mathbf{P}}_{(1)}$\ of
Eqs.(3.1) and (3.1.20) into the definition of $\mathcal{E}_{(2)}^{(\beta )}$%
\ shown in Eq.(1.5.7) yields the following result%
\begin{equation}
\mathcal{E}_{(2)}^{(\beta )}=Tr[\mathbf{Y}_{(1)}\mathbf{A+\mathbf{Z}}_{(1)}%
\mathbf{F+V}_{(1)}\mathbf{M}^{+}\mathbf{+V}_{(1)}^{+}\mathbf{M].} 
\tag{3.1.25}
\end{equation}%
It is seen that the $\beta -$component \ of the second order energy
generally contains all submatrices of the correction $\widetilde{\mathbf{P}}%
_{(1)}$\ of Eq.(3.1.20) and thereby it is related to\ the overall primary
effect of our perturbation upon charge and bond order distribution.
Meanwhile, the relevant $\alpha -$component takes the form%
\begin{equation}
\mathcal{E}_{(2)}^{(\alpha )}=Tr(\mathbf{J}_{(2)}\mathbf{B}^{+}+\mathbf{J}%
_{(2)}^{+}\mathbf{B})  \tag{3.1.26}
\end{equation}%
and contains the (sub)matrix $\mathbf{J}_{(2)}$\ only in accordance with
Eq.(3.1.24). If we recall now opposite signs of components $\mathcal{E}%
_{(2)}^{(\alpha )}$\ and $\mathcal{E}_{(2)}^{(\beta )}$\ along with the
relation $\mid \mathcal{E}_{(2)}^{(\beta )}\mid >\mid \mathcal{E}%
_{(2)}^{(\alpha )}\mid $ of subsection 1.4, we arrive at a conclusion that
only a part of the overall secondary energetic effect of the perturbation $%
\widetilde{\mathbf{H}}_{(1)}$\ (namely, the increment of $\mathbf{J}_{(2)}-$%
dependent alterations in the intersubset bond orders) is actually subtracted
from the total primary effect when building up the correction $\mathcal{E}%
_{(2)}$. A more detailed illustration of these results is given below
(Subsection 3.2).

Now, we are about to consider some particular cases of the matrix $%
\widetilde{\mathbf{H}}$\ of Eq.(3.1) and draw more specific conclusions
concerning charge and bond order (re)distributions and energy alterations
due to perturbation.

\begin{center}
\textbf{3.2. Specific results for alternant conjugated hydrocarbons and
their derivatives }
\end{center}

As already mentioned, alternant conjugated hydrocarbons are among systems
underlying the anti-block-diagonal zero order matrix $\widetilde{\mathbf{H}}%
_{(0)}$\ of Eq.(3.1). The present subsection addresses just these compounds
and the relevant specific perturbation matrices.

As is well-known, the concept of alternant hydrocarbons (AHs) arises in the
framework of the simple H\"{u}ckel model of $\pi -$electron systems [6, 7,
10, 11, 62]. Thus, let us start with this model, wherein the 2p$_{z}$ AOs of
carbon atoms play the role of initial orbitals $\{\chi \}$. Moreover,
resonance parameters between AOs of chemically-bound pairs of atoms only are
taken into consideration here. An AH is then defined as a system, the
atoms(AOs) of which are divisible into two types (subsets) of the so-called
starred and unstarred atoms (AOs) so that no chemical bonds are present
between atoms (AOs) of the same type (subset). Under an appropriate
numbering of AOs, the above-mentioned subsets coincide with the former
subsets $\{\chi _{1}\}$\ and $\{\chi _{2}\}$\ (Subsect. 3.1). It is then
evident that the common H\"{u}ckel type Hamiltonian matrix of AHs takes an
anti-block-diagonal form coinciding with that of the zero order term of
Eq.(3.1). Suppose that resonance parameters of all chemical bonds take the
same value ($\beta $) in addition and the latter plays the role of our
energy unit, i.e. the equality $\beta =1$\ is accepted. Similarly, Coulomb
parameters of all AOs also are assumed to take the same value ($\alpha $)
that serves as an energy reference point, i.e. $\alpha =0$. Consequently,
non-zero elements of the (sub)matrix $\mathbf{B}$\ refer to chemical bonds
only and equal to 1. Moreover, formulae of Eqs.(3.1.6) and (3.1.7) reflect
the common CBO matrix of AHs and the relevant NCMO representation matrix
that have been comparatively analyzed in Ref.[53] in a detail. Finally, the
zero order energy $\mathcal{E}_{(0)}$\ of Eq.(3.1.22) coincides with the
total $\pi -$electron energy of AHs.

Accordingly, the total Hamiltonian matrix of Eq.(3.1) embraces perturbed
analogues (derivatives) of AHs. Corrections $\widetilde{\mathbf{P}}_{(k)}$\
defined by Eqs.(3.1.10)-(3.1.19) then describe charge and bond order
redistribution due to perturbation. In particular, elements of the
correction $\widetilde{\mathbf{P}}_{(1)}$\ of Eq.(3.1.20) yield the
classical polarizabilities of AHs. Indeed, the latter are known to be
definable in terms of derivatives of elements of the CBO matrix of the
perturbed system with respect to separate elements of the perturbation
matrix [10, 12, 13, 62]. Hence, polarizabilities of the atom-atom and
atom-bond type result from the following derivatives%
\begin{equation}
\pi _{ii}=\frac{\partial Y_{(1)ii}}{\partial A_{ii}},\quad \pi _{ij}=\frac{%
\partial Y_{(1)ii}}{\partial A_{jj}},\quad \pi _{pj}=\frac{\partial Z_{(1)pp}%
}{\partial A_{jj}},\quad \pi _{ip,j}=\frac{\partial V_{(1)ip}}{\partial
A_{jj}},  \tag{3.2.1}
\end{equation}%
where the subscripts $i$ an $j$ belong to the first subset $\{\chi _{1}\},$\
whereas $p$ refers to the second one $\{\chi _{2}\}.$\ Besides, $\pi _{ii}$\
is usually called the self-polarizability of the $i$th atom (AO). As is
well-known, perturbations of AHs are traditionally classified into those
associated with Coulomb and resonance parameters of 2p$_{z}$ AOs. In this
connection, two particular first order matrices ($\widetilde{\mathbf{H}}%
_{(1)}$) have been distinguished and studied separately in Ref.[54], viz. a
block-diagonal matrix ($\widetilde{\mathbf{H}}_{(1)}^{(bd)}$) corresponding
to $\mathbf{M=M}^{+}=\mathbf{0}$ and an anti-block-diagonal one ($\widetilde{%
\mathbf{H}}_{(1)}^{(ab)}$) containing zero matrices instead of both $\mathbf{%
A}$ and $\mathbf{F}$, i.e.%
\begin{equation}
\widetilde{\mathbf{H}}_{(1)}^{(bd)}=\left\vert 
\begin{array}{cc}
\mathbf{A} & \mathbf{0} \\ 
\mathbf{0} & \mathbf{F}%
\end{array}%
\right\vert ,\qquad \widetilde{\mathbf{H}}_{(1)}^{(ab)}=\left\vert 
\begin{array}{cc}
\mathbf{0} & \mathbf{M} \\ 
\mathbf{M}^{+} & \mathbf{0}%
\end{array}%
\right\vert .  \tag{3.2.2}
\end{equation}%
It is evident that the first perturbation matrix ($\widetilde{\mathbf{H}}%
_{(1)}^{(bd)})$\ embraces alterations both in Coulomb parameters and in the
resonance ones inside subsets of either starred or unstarred AOs. Meanwhile,
the second matrix ($\widetilde{\mathbf{H}}_{(1)}^{(ab)}$) refers to changes
in resonance parameters of the intersubset type so that the relevant total
matrix $\widetilde{\mathbf{H}}$\ remains to be of anti-block-diagonal
constitution.

To overview the relevant results, let us start with the perturbation matrix $%
\widetilde{\mathbf{H}}_{(1)}^{(bd)}.$\ The equality $\mathbf{M=M}^{+}=%
\mathbf{0}$ allowed us to derive the following relations%
\begin{equation}
\mathbf{G}_{(1)}^{(bd)+}\mathbf{BL=\mathbf{LB}_{{}}^{+}G}_{(1)}^{(bd)},%
\qquad \mathbf{BLG}_{(1)}^{(bd)+}=\mathbf{G}_{(1)}^{(bd)}\mathbf{LB}%
_{{}}^{+},  \tag{3.2.3}
\end{equation}%
where the relevant matrices are denoted here and below by an additional
superscript $(bd)$. The above relations, in turn, ensure vanishing matrices $%
\mathbf{\Delta }_{(1)}^{(bd)}$\ and $\mathbf{V}_{(1)}^{(bd)}$ of
Eqs.(3.1.15) and (3.1.16) so that the resulting first order correction $%
\widetilde{\mathbf{P}}_{(1)}^{(bd)}$\ of Eq.(3.1.20) also is a
block-diagonal matrix, viz.%
\begin{equation}
\widetilde{\mathbf{P}}_{(1)}^{(bd)}\mathbf{=}\left\vert 
\begin{array}{cc}
\mathbf{Y}_{(1)}^{(bd)} & \mathbf{0} \\ 
\mathbf{0} & \mathbf{Z}_{(1)}^{(bd)}%
\end{array}%
\right\vert .  \tag{3.2.4}
\end{equation}%
It is seen that the perturbation\ $\widetilde{\mathbf{H}}_{(1)}^{(bd)}$\
exerts influence upon populations of basis orbitals and upon intrasubset
bond orders within the first order approximation. Meanwhile, bond orders of
the intersubset type are not influenced by this perturbation. This result
may be regarded as the matrix generalization of the known zero values for
polarizabilities of AHs of the atom-bond type ($\pi _{ip,j}$) [12]. Further,
the submatrices $\mathbf{Y}_{(1)}^{(bd)}$\ and $\mathbf{Z}_{(1)}^{(bd)}$\
were shown to meet the following matrix equations%
\begin{align}
\mathbf{KY}_{(1)}^{(bd)}+\mathbf{Y}_{(1)}^{(bd)}\mathbf{K} \mathbf{=}&%
\mathbf{KAK-K}^{2}\mathbf{BFB}^{+}\mathbf{K}^{2},  \nonumber \\
\mathbf{LZ}_{(1)}^{(bd)}+\mathbf{Z}_{(1)}^{(bd)}\mathbf{L} \mathbf{=}&%
\mathbf{LFL-L}^{2}\mathbf{B}^{+}\mathbf{ABL}^{2}  \tag{3.2.5}
\end{align}%
that belong to the type shown in Eq.(1.1.16). Consequently, (sub)matrices $%
\mathbf{Y}_{(1)}^{(bd)}$\ and $\mathbf{Z}_{(1)}^{(bd)}$\ are expressible in
the form of an integral like that of Eq.(1.1.17). The latter allows us to
establish the signs of polarizabilities of the atom-atom type ($\pi _{ij}$
and $\pi _{pj}$ of Eq.(3.2.1)). As a result, we obtain a new integral
representation for the classical polarizabilities of AHs, as well as a
simple (re)derivation [54]\ of the well-known rule of the alternating
polarity [12-15].

Similarly, the first order correction $\widetilde{\mathbf{C}}_{(1)}^{(bd)}$\
to the NCMO representation matrix $\widetilde{\mathbf{C}}_{(bd)}$\ results
from Eq.(3.1.21) after employment of Eq.(3.2.3). This correction also is
representable in terms of matrices $\mathbf{Y}_{(1)}^{(bd)}$\ and $\mathbf{Z}%
_{(1)}^{(bd)},$\ viz. 
\begin{equation}
\widetilde{\mathbf{C}}_{(1)}^{(bd)}=\frac{1}{2\sqrt{2}}\left\vert 
\begin{array}{cc}
\mathbf{Y}_{(1)}^{(bd)} & -\mathbf{Y}_{(1)}^{(bd)}\mathbf{BL} \\ 
\mathbf{Z}_{(1)}^{(bd)}\mathbf{LB}^{+} & \mathbf{Z}_{(1)}^{(bd)}%
\end{array}%
\right\vert .  \tag{3.2.6}
\end{equation}%
This result forms the basis for interdependence [54] between charge and bond
order redistribution in AHs due to the perturbation concerned, on the one
hand, and the respective reshaping of NCMOs (LMOs), on the other hand.

For the anti-block-diagonal matrix $\widetilde{\mathbf{H}}_{(1)}^{(ab)},$\
the analogues of Eqs.(3.2.3) and (3.2.5) are as follows%
\begin{equation}
\mathbf{BLG}_{(1)}^{(ab)+}=-\mathbf{G}_{(1)}^{(ab)}\mathbf{LB}%
_{{}}^{+},\qquad \mathbf{G}_{(1)}^{(ab)+}\mathbf{BL=-\mathbf{LB}_{{}}^{+}G}%
_{(1)}^{(ab)},  \tag{3.2.7}
\end{equation}%
\begin{equation}
\widetilde{\mathbf{P}}_{(1)}^{(ab)}=\left\vert 
\begin{array}{cc}
\mathbf{0} & \mathbf{V}_{(1)}^{(ab)} \\ 
\mathbf{V}_{(1)}^{(ab)+} & \mathbf{0}%
\end{array}%
\right\vert  \tag{3.2.8}
\end{equation}%
and 
\begin{equation}
\mathbf{L\Delta }_{(1)}^{(ab)}+\mathbf{\Delta }_{(1)}^{(ab)}\mathbf{L}%
\mathbf{=L(\mathbf{M}^{+}B\mathbf{L-LB}}^{+}\mathbf{\mathbf{M})L,}^{{}} 
\tag{3.2.9}
\end{equation}%
where matrices $\mathbf{\Delta }_{(1)}^{(ab)}$\ and $\mathbf{V}_{(1)}^{(ab)}$%
\ are interrelated as shown in Eq. (3.1.15). As is seen from comparison of
Eqs.(3.2.4) and (3.2.8), the corrections $\widetilde{\mathbf{P}}%
_{(1)}^{(bd)} $\ and $\widetilde{\mathbf{P}}_{(1)}^{(ab)}$\ are of opposite
constitutions as it is the case with the respective perturbation matrices
shown in Eq.(3.2.2).

The second order corrections $\widetilde{\mathbf{P}}_{(2)}^{(bd)}$\ and $%
\widetilde{\mathbf{P}}_{(2)}^{(ab)}$\ corresponding to the perturbation
matrices $\widetilde{\mathbf{H}}_{(1)}^{(bd)}$\ and $\widetilde{\mathbf{H}}%
_{(1)}^{(ab)},$\ respectively, accordingly follow from Eqs.(3.1.10) and
(3.1.12). The relevant comparative analysis [76] showed that both
corrections are of the same constitution, namely of the anti-block-diagonal
one, viz.\ 
\begin{equation}
\widetilde{\mathbf{P}}_{(2)}^{(bd)}=\left\vert 
\begin{array}{cc}
\mathbf{0} & \mathbf{V}_{(2)}^{(bd)}+\mathbf{J}_{(2)}^{(bd)} \\ 
\mathbf{V}_{(2)}^{(bd)+}+\mathbf{J}_{(2)}^{(bd)+} & \mathbf{0}%
\end{array}%
\right\vert ,\quad \widetilde{\mathbf{P}}_{(2)}^{(ab)}=\left\vert 
\begin{array}{cc}
\mathbf{0} & \mathbf{V}_{(2)}^{(ab)}+\mathbf{J}_{(2)}^{(ab)} \\ 
\mathbf{V}_{(2)}^{(ab)+}+\mathbf{J}_{(2)}^{(ab)+} & \mathbf{0}%
\end{array}%
\right\vert ,  \tag{3.2.10}
\end{equation}%
where submatrices $\mathbf{V}_{(2)}^{(bd)}$\ and $\mathbf{V}_{(2)}^{(ab)}$\
are proportional to respective skew-symmetric matrices $\mathbf{\Delta }%
_{(2)}^{(bd)}$\ and $\mathbf{\Delta }_{(2)}^{(ab)},$\ the latter being
determined by matrix equations like that of Eq.(3.2.9). Again, submatrices $%
\mathbf{J}_{(2)}^{(bd)}$\ and $\mathbf{J}_{(2)}^{(ab)}$\ of Eq.(3.2.10) are
representable algebraically as follows [76] 
\begin{align}
\mathbf{J}_{(2)}^{(bd)} =&-2\mathbf{BLG}_{(1)}^{(bd)+}\mathbf{G}%
_{(1)}^{(bd)}=-\frac{1}{2}(\mathbf{Y}_{(1)}^{(bd)})^{2}\mathbf{BL=-}\frac{1}{%
2}\mathbf{BL}(\mathbf{Z}_{(1)}^{(bd)})^{2},  \tag{3.2.11} \\
\mathbf{J}_{(2)}^{(ab)} =&-2\mathbf{BLG}_{(1)}^{(ab)+}\mathbf{G}%
_{(1)}^{(ab)}=-\frac{1}{2}\mathbf{BL\mathbf{\Delta }}_{(1)}^{(ab)}\mathbf{%
\mathbf{\Delta }}_{(1)}^{(ab)+}\mathbf{=-}\frac{1}{2}\mathbf{\mathbf{V}}%
_{(1)}^{(ab)}\mathbf{\mathbf{V}}_{(1)}^{(ab)+}\mathbf{BL.}  \nonumber \\
&  \nonumber
\end{align}

Let us now turn to total energies of AHs and their derivatives. As already
mentioned, the parent AHs are characterized by the zero order energy $%
\mathcal{E}_{(0)}$ of Eq.(3.1.22). The first order corrections to the latter
due to perturbations $\widetilde{\mathbf{H}}_{(1)}^{(bd)}$\ and $\widetilde{%
\mathbf{H}}_{(1)}^{(ab)}$\ follow from Eq.(3.1.23) under assumptions that $%
\mathbf{M=M}^{+}=\mathbf{0}$ and $\mathbf{A=F}=\mathbf{0,}$ respectively.
The resulting simple formulae for $\mathcal{E}_{(1)}^{(bd)}$\ and $\mathcal{E%
}_{(1)}^{(ab)}$\ coincide with respective classical analogues in accordance
with the expectation. For example, the equality $\mathcal{E}%
_{(1)loc}^{(bd)}=\alpha $\ follows for a local perturbation ($\alpha )$\ of
a certain Coulomb parameter, e.g. of the AO $\chi _{1,1}$\ ($A_{11}=\alpha $%
). Similarly, $\mathcal{E}_{(1)loc}^{(ab)}=2p_{mn}\mu $ results for a local
perturbation of an intersubset resonance parameter, say, of that between AOs 
$\chi _{1,m}$ and $\chi _{2,n}$\ ($M_{mn}=\mu $), where $p_{mn}$ stands for
the bond order between the AOs concerned in the parent AH. Further, the
relevant corrections of the second order ($\mathcal{E}_{(2)}^{(bd)}$\ and $%
\mathcal{E}_{(2)}^{(ab)}$) consist of $\alpha -$ and $\beta -$components as
discussed in the above subsection, i.e. 
\begin{equation}
\mathcal{E}_{(2)}^{(bd)}=\mathcal{E}_{(2)}^{(bd,\alpha )}+\mathcal{E}%
_{(2)}^{(bd,\beta )},\qquad \mathcal{E}_{(2)}^{(ab)}=\mathcal{E}%
_{(2)}^{(ab,\alpha )}+\mathcal{E}_{(2)}^{(ab,\beta )},  \tag{3.2.12}
\end{equation}%
where terms of the right-hand sides of the above relations result from
Eq.(3.1.26) and (3.1.25), viz.%
\begin{equation}
\mathcal{E}_{(2)}^{(bd,\alpha )}=Tr[\mathbf{J}_{(2)}^{(bd)}\mathbf{B}^{+}+%
\mathbf{J}_{(2)}^{(bd)+}\mathbf{B}],\quad \mathcal{E}_{(2)}^{(bd,\beta )}=Tr[%
\mathbf{Y}_{(1)}^{(bd)}\mathbf{A+\mathbf{Z}}_{(1)}^{(bd)}\mathbf{F]} 
\tag{3.2.13}
\end{equation}%
and \ 
\begin{equation}
\mathcal{E}_{(2)}^{(ab,\alpha )}=Tr[\mathbf{J}_{(2)}^{(ab)}\mathbf{B}^{+}+%
\mathbf{J}_{(2)}^{(ab)+}\mathbf{B}],\quad \mathcal{E}_{(2)}^{(ab,\beta )}=Tr[%
\mathbf{V}_{(1)}^{(ab)}\mathbf{M}^{+}\mathbf{+\mathbf{V}}_{(1)}^{(ab)+}%
\mathbf{M].}  \tag{3.2.14}
\end{equation}%
To discuss these components in a more detail, let us start with the
block-diagonal perturbation matrix $\widetilde{\mathbf{H}}_{(1)}^{(bd)}$.\
It is seen that the component $\mathcal{E}_{(2)}^{(bd,\beta )}$\ generally
embraces the first order corrections both to populations of AOs and to
intrasubset bond orders. For the above-specified local alteration in the
Coulomb parameter of the AO $\chi _{1,1}$\ ($A_{11}=\alpha $), however, only
a single element of the submatrix $\mathbf{Y}_{(1)}^{(bd)}$\ remains, i.e.%
\begin{equation}
\mathcal{E}_{(2)loc}^{(bd,\beta )}=Y_{(1)11}^{(bd)}\alpha =\pi _{11}\alpha
^{2},  \tag{3.2.15}
\end{equation}%
where $\pi _{11}$\ is the self-polarizability of the AO under perturbation
defined as shown in Eq.(3.2.1). Thus, the component $\mathcal{E}%
_{(2)loc}^{(bd,\beta )}$\ is proportional to the population ($%
Y_{(1)11}^{(bd)}$) acquired by the AO $\chi _{1,1}$\ due to the given
perturbation and, consequently, it may be traced back to lowering of the
one-electron energy of electrons acquired by the AO $\chi _{1,1}$. It is no
surprise that this primary energetic effect of our local perturbation ($%
A_{11}=\alpha $) is stabilizing for $\alpha >0$\ [The \textit{a priori}
positive sign of any self-polarizability (in our negative energy units)
should be kept in mind here]. Finally, the effect concerned is twice as
large as the relevant total correction $\mathcal{E}_{(2)loc}^{(bd)}$\ equal
to $\pi _{11}\alpha ^{2}/2$ [62].

The $\alpha -$component of the same second order correction consists of a
sum of increments of individual carbon-carbon bonds as the first relation of
Eq.(3.2.13) indicates [One-to-one correspondence between non-zero elements
of the (sub)matrix $\mathbf{B}$ and chemical bonds should be recalled here].
Moreover, each of these increments is proportional to the respective element
of the (sub)matrix $\mathbf{J}_{(2)}^{(bd)}$. For the local perturbation ($%
A_{11}=\alpha $), the (sub)matrix $\mathbf{J}_{(2)}^{(bd)}$\ was shown to
give rise mostly to lowering of orders of bonds attached to the site (AO) of
perturbation ($\chi _{1,1}$) [76]. Thus, the above-concluded primary
stabilization of the system due to concentration of population on the
perturbed AO ($\chi _{1,1}$) becomes reduced twice because of the secondary
weakening of the neighboring bonds originating from the (sub)matrix $\mathbf{%
J}_{(2)}^{(bd)}.$\ [The latter, however, is not the only contributor to
orders of individual bonds: The (sub)matrix $\mathbf{V}_{(2)}^{(bd)}$\ also
gives rise to some alterations [76], but their total energetic effect
vanishes].

Before passing to components of the correction $\mathcal{E}_{(2)}^{(ab)}$\
referring to the anti-block-diagonal perturbation $\widetilde{\mathbf{H}}%
_{(1)}^{(ab)},$\ let us distinguish two possible types of elements of the
matrix $\mathbf{M}$, namely (i) elements representing (positive or negative)
alterations in resonance parameters of the former chemical bonds and (ii)
elements referring to formation of new weak bonds between AOs of different
subsets. Let us assume for simplicity that our perturbation matrix $%
\widetilde{\mathbf{H}}_{(1)}^{(ab)}$\ contains only some positive elements
of the second type, i.e. some new bonds are formed inside our AH. The
component $\mathcal{E}_{(2)}^{(ab,\beta )}$\ of Eq.(3.2.14) then represents
the overall primary stabilization of the system due to emergence of these
new bonds that is twice as large as the relevant total correction $\mathcal{E%
}_{(2)}^{(ab)}$. For example, formation of a single bond C$_{m}-$C$_{n}$ is
accompanied by the following energy increment \ 
\begin{equation}
\mathcal{E}_{(2)loc}^{(ab,\beta )}=2V_{(1)mn}^{(ab)}\mu =2\pi _{mn,mn}\mu
^{2},  \tag{3.2.16}
\end{equation}%
where $\pi _{mn,mn}$\ is the self-polarizability of the bond concerned.
Accordingly, the component $\mathcal{E}_{(2)}^{(ab,\alpha )}$\ of
Eq.(3.2.14) consists of increments of "initial" carbon-carbon bonds as it
was the case with $\mathcal{E}_{(2)}^{(bd,\alpha )}$ of Eq.(3.2.13). Thus, $%
\mathcal{E}_{(2)}^{(ab,\alpha )}$\ describes the secondary destabilizing
effect of the same perturbation upon orders of the already-existing bonds.
Analysis of some specific examples [76] showed that the secondary effect
consists in a predominant weakening of the former bonds in the nearest
neighborhood of the newly-emerging bond C$_{m}-$C$_{n}$. In other words,
formation of new bonds necessarily is accompanied by weakening of
already-existing bonds in accordance with the classical concept of the
limited valence. The above-described redistribution of bond orders has been
called the rebonding effect in Ref.[77]. Studies of more involved
perturbations in AHs may be found in Refs.[78,79].

\begin{center}
\textbf{3.3. The case of separate pairs of strongly-interacting basis
orbitals. Application to alkanes and their derivatives}
\end{center}

Perturbation matrices of two particular types (see Eq.(3.2.2)) have been
discussed in the above subsection 3.2. Let us now return again to the most
general initial Hamiltonian matrix of quasi-degenerated systems shown in Eq.
(3.1) and introduce some simplifications into the zero order member of the
latter ($\widetilde{\mathbf{H}}_{(0)}$). Meanwhile, the first order term
will now contain four non-zero submatrices $\mathbf{A}$, $\mathbf{F}$ and $%
\mathbf{M,}$ as shown in Eq.(3.1).

Let us start with an assumption that the total number of basis orbitals of
our system ($N$) coincides with that of electrons (2$n$) and thereby takes
an even value. Further, each subset ($\{\chi _{1}\}$ and $\{\chi _{2}\}$)
will contain the same number of orbitals equal to $n$. Most importantly,
orbitals of different subsets are now assumed to interact predominantly in
pairs. In other words, any orbital of the first subset (say, $\chi _{1,i}$)
is supposed to interact significantly only with a single orbital of the
second subset. Let the latter to acquire the number $n+i$\ and to be denoted
by $\chi _{2,n+i}$. Resonance parameters inside the above-specified pairs of
strongly-interacting orbitals $\chi _{1,i}$\ and $\chi _{2,n+i}$\ then
consequently take relatively large values as compared to the remaining
resonance parameters. Under the above-accepted numbering of basis orbitals,
these significant parameters occupy diagonal positions in the relevant
(sub)matrix $\mathbf{B}$. Let us assume finally that differences between
these parameters (if any) are of the first order magnitude so that these may
be incorporated into diagonal elements of the (sub)matrix $\mathbf{M}$. \
Given that the averaged value of the above-specified large parameters is
chosen as an energy unit, we ultimately arrive at the following simple
relation%
\begin{equation}
\mathbf{B=B}^{+}=\mathbf{I}.  \tag{3.3.1}
\end{equation}%
Substituting Eq.(3.3.1) into Eq.(3.1) shows that the resulting total
Hamiltonian matrix describes a set of $n$ weakly-interacting two-level
systems, each of them being represented by a matrix $\widetilde{\mathbf{h}}$
of Eq.(3.2). It is also evident that a slightly heteropolar (or almost
homopolar) two-center chemical bond may be ascribed to each individual
two-level system of the above-mentioned type. Thus, we actually have to do
here with a set of $n$ weakly-interacting almost homopolar bonds. Let the
latter to acquire the numbers $I=1,2,3..n.$ Accordingly, the bond between
orbitals $\chi _{1,i}$\ and $\chi _{2,n+i}$\ will be referred to as the $I$%
th bond. Systems under present discussion may be primarily exemplified by
saturated hydrocarbons (alkanes) represented in the basis of $sp^{3}-$hybrid
AOs (HAOs) of carbon atoms and $1s$ AOs of hydrogen atoms. Pairs of orbitals
pertinent to individual C$-$C and C$-$H bonds coincide with the
strongly-interacting ones in this case [80]. Indeed, resonance parameters
inside pairs of AOs(HAOs) pertinent to any C$-$C and C$-$H bond are of much
higher absolute values in alkanes as compared to the remaining resonance
parameters as the relevant estimations show. Moreover, Coulomb parameters of
all basis orbitals are of close values in these systems. Derivatives of
alkanes containing slightly more electronegative heteroatoms also are
embraced by the Hamiltonian matrix concerned.

Let us now turn to the overview of the relevant results. First, substituting
Eq.(3.3.1) into Eq.(3.1.1) yields the following simple transformation matrix%
\begin{equation}
\mathbf{U}^{\prime }=\frac{1}{\sqrt{2}}\left\vert 
\begin{array}{cc}
\mathbf{I} & \mathbf{I} \\ 
\mathbf{I} & \mathbf{-I}%
\end{array}%
\right\vert ,  \tag{3.3.2}
\end{equation}%
which represents passing to bonding and antibonding combinations of orbitals
of individual bonds, usually referred to as bond orbitals (BOs) [The
superscript $\prime $ is used here and below to distinguish the case
concerned from that of subsection 3.1]. Analogously, employment of
Eq.(3.3.1) within Eq.(3.1.2) ensures coincidence of matrices both $\mathbf{K}
$ and $\mathbf{L}$ with the unit matrix ($\mathbf{I}$). Moreover, the same
refers to the eigenblocks $\mathbf{E}_{(+)}$ and $\mathbf{E}_{(-)}$ defined
by Eq.(3.1.3), i.e. $\mathbf{E}_{(+)}=\mathbf{E}_{(-)}=\mathbf{I}.$ The
latter result, in turn, is nothing more than the assumption of Eq.(2.4.1).
Consequently, the explicit matrix form of the PNCMO theory of subsection 2.4
proves to be applicable in the present case along with the relevant
implications exhibited in Eqs.(2.4.2)-(2.4.9), where%
\begin{align}
\mathbf{T} =&\frac{1}{2}[(\mathbf{A+F)+(M}^{+}\mathbf{+M})],\quad \mathbf{Q}%
=\frac{1}{2}[(\mathbf{A+F)-(M}^{+}\mathbf{+M})],  \nonumber \\
\mathbf{R} =&\frac{1}{2}[(\mathbf{A-F)+(M}^{+}\mathbf{-M})].  \tag{3.3.3}
\end{align}%
[The above relations also easily result from those of Eq.(3.1.4) after
employment of Eq.(3.3.1)]. It is evident that the afore-defined BOs take the
role of our principal basis functions $\{\varphi \}$\ (Sections 1 and 2).

Let us now consider members of the power series for the relevant
retransformed CBO matrix $\widetilde{\mathbf{P}}^{\prime }.$\ As is seen
from Eqs.(3.1.6) and (3.1.12)-(3.1.20), substituting the unit matrix ($%
\mathbf{I}$) for matrices $\mathbf{B}$, $\mathbf{K}$ and $\mathbf{L}$\
yields significant simplifications of the relevant expressions: First, the
zero order term of the expansion concerned ($\widetilde{\mathbf{P}}%
_{(0)}^{\prime }$) now consists of unit matrices as it was the case with the
transformation matrix of Eq.(3.3.2), viz.%
\begin{equation}
\widetilde{\mathbf{P}}_{(0)}^{\prime }=\left\vert 
\begin{array}{cc}
\mathbf{I} & \mathbf{I} \\ 
\mathbf{I} & \mathbf{I}%
\end{array}%
\right\vert .  \tag{3.3.4}
\end{equation}%
This result indicates uniform values (equal to 1) both of populations of
basis functions $\{\chi \}$\ and of all bond orders inside the
strongly-interacting pairs of orbitals $\chi _{1,i}$\ and $\chi _{2,n+i}$\
within the zero order approximation. Second, a single (sub)matrix (denoted
below by $\mathbf{\Pi }_{(k)})$\ arises instead of the former two different
(sub)matrices $\mathbf{Y}_{(k)}$\ and $\mathbf{Z}_{(k)}$\ of Eq.(3.1.13)
within diagonal positions of the first component ($\widetilde{\mathbf{P}}%
_{(k)}^{\prime (g)}$) of the $k$th order correction ($\widetilde{\mathbf{P}}%
_{(k)}^{\prime }$). \ Moreover, this new (sub)matrix is proportional to the
symmetric part ($\mathbf{G}_{(k)}^{\prime \ast }$) of the relevant principal
matrix $\mathbf{G}_{(k)}^{\prime }.$\ Analogously, the present version ($%
\mathbf{V}_{(k)}^{\prime }$) of the former (sub)matrix $\mathbf{V}_{(k)}^{{}}
$\ of Eqs.(3.1.15) and (3.1.16) is proportional to the skew-symmetric part ($%
\mathbf{G}_{(k)}^{\prime \circ }$) of the same matrix $\mathbf{G}%
_{(k)}^{\prime }$. Further, the former (sub)matrices $\mathbf{S}_{(k)}$\ and 
$\mathbf{N}_{(k)}$\ of the remaining component ($\widetilde{\mathbf{P}}%
_{(k)}^{\prime (x)}$) of the same correction $\widetilde{\mathbf{P}}%
_{(k)}^{\prime }$\ defined by Eqs.(3.1.17) and (3.1.18) now turn into a
single matrix denoted below by $\mathbf{\Lambda }_{(k)}$. Moreover, the
latter becomes expressible via the intrasubset population matrices only
and/or in terms of intersubset delocalization matrices (relations of
Eq.(1.3.16) should be invoked in this case). The same refers also to the
present analogue of the submatrix $\mathbf{J}_{(k)}.$\ We then obtain that 
\begin{equation}
\widetilde{\mathbf{P}}_{(1)}^{\prime }=\left\vert 
\begin{array}{cc}
\mathbf{\Pi }_{(1)} & 2\mathbf{G}_{(1)}^{\prime \circ } \\ 
-2\mathbf{G}_{(k)}^{\prime \circ } & -\mathbf{\Pi }_{(1)}%
\end{array}%
\right\vert   \tag{3.3.5}
\end{equation}%
and 
\begin{equation}
\widetilde{\mathbf{P}}_{(k)}^{\prime }=\left\vert 
\begin{array}{cc}
\mathbf{\Lambda }_{(k)}+\mathbf{\Pi }_{(k)} & \mathbf{J}_{(k)}+2\mathbf{G}%
_{(k)}^{\prime \circ } \\ 
\mathbf{J}_{(k)}^{+}-2\mathbf{G}_{(k)}^{\prime \circ } & \mathbf{\Lambda }%
_{(k)}-\mathbf{\Pi }_{(k)}%
\end{array}%
\right\vert   \tag{3.3.6}
\end{equation}%
for $k=1$ and $k=2,3..,$ respectively, where%
\begin{equation}
\mathbf{\Pi }_{(k)}=-2\mathbf{G}_{(k)}^{\prime \ast },\quad k=1,2,3... 
\tag{3.3.7}
\end{equation}%
\begin{equation}
\mathbf{\Lambda }_{(k)}=\frac{1}{2}(\mathbf{X}_{(+)}^{(k)}+\mathbf{X}%
_{(-)}^{(k)})=-(\mathbf{D}_{(+)}^{(k)}-\mathbf{D}_{(-)}^{(k)}),  \tag{3.3.8}
\end{equation}%
\begin{equation}
\mathbf{J}_{(k)}=\frac{1}{2}(\mathbf{X}_{(+)}^{(k)}-\mathbf{X}%
_{(-)}^{(k)})=-(\mathbf{D}_{(+)}^{(k)}+\mathbf{D}_{(-)}^{(k)})  \tag{3.3.9}
\end{equation}%
and 
\begin{equation}
\mathbf{G}_{(k)}^{\prime \ast }=\frac{1}{2}(\mathbf{G}_{(1)}^{\prime }+%
\mathbf{G}_{(1)}^{\prime +}),\quad \mathbf{G}_{(1)}^{\prime \circ }=\frac{1}{%
2}(\mathbf{G}_{(1)}^{\prime }-\mathbf{G}_{(1)}^{\prime +}).  \tag{3.3.10}
\end{equation}%
Let us now discuss the role of the above-enumerated (sub)matrices in the
formation of charge and bond order redistribution due to perturbation in the
system(s) concerned. As is seen from Eqs.(3.3.5) and (3.3.6), the
(sub)matrices $\mathbf{\Pi }_{(k)},$ $k=1,2,3...$\ yield dipole-like
contributions ($\pm \Pi _{(k)ii}$) to populations of strongly-interacting
orbitals\ $\chi _{1,i}$\ and $\chi _{2,n+i}.$\ In this connection, $\mathbf{%
\Pi }_{(k)},$ $k=1,2,3...$\ have been called the (intrabond) polarization
matrices of the $k$th order [80]. On the basis of Eqs.(2.4.2), (3.3.3) and
(3.3.10), the first two representatives of the series of matrices $\mathbf{%
\Pi }_{(k)}$\ are expressible as follows%
\begin{equation}
\mathbf{\Pi }_{(1)}=\frac{1}{2}(\mathbf{A}-\mathbf{F}),\quad \ \mathbf{\Pi }%
_{(2)}=\frac{1}{4}(\mathbf{FM+M}^{+}\mathbf{F-AM}^{+}\mathbf{-MA})\ . 
\tag{3.3.11}
\end{equation}%
The first relation of Eq.(3.3.11) indicates that diagonal elements \ $\Pi
_{(1)ii}$ of the (sub)matrix $\mathbf{\Pi }_{(1)}$\ are determined by local
differences in Coulomb parameters of orbitals $\chi _{1,i}$\ and $\chi
_{2,n+i}$ in accordance with the expectation. At the same time, the right
relation points to a non-local origin of the second order dipoles $\pm \Pi
_{(2)ii}$. Finally, an evident connection deserves mention between intrabond
polarization and the rule of the alternating polarity. Indeed, (sub)matrices 
$\mathbf{Y}_{(1)}^{(bd)}$\ and $\mathbf{Z}_{(1)}^{(bd)}$\ of Eq.(3.2.4)
yield the above-mentioned rule as discussed in the subsection 3.2. Again,
the same (sub)matrices are the prototypes of the polarization matrix $%
\mathbf{\Pi }_{(1)}.$\ 

By contrast, (sub)matrices $\mathbf{\Lambda }_{(k)}$\ of Eq.(3.3.8) yield
uniform contributions to populations of the same orbitals $\chi _{1,i}$\ and 
$\chi _{2,n+i}.$\ Moreover, the sum $X_{(+)ii}^{(k)}+X_{(-)ii}^{(k)}$\
contained within the definition of the diagonal element $\Lambda _{(k)ii}$\
coincides with the total population of the $k$th order lost (acquired) by
the $I$th bond due to perturbation that, in turn, is devided equally among
orbitals $\chi _{1,i}$\ and $\chi _{2,n+i}.$\ Thus, (sub)matrices $\mathbf{%
\Lambda }_{(k)}$\ describe redistribution (transfer) of population among our
bonds and these have been referred to [80] as the charge transfer matrices.

Let us now turn to alterations in bond orders inside pairs of
strongly-interacting orbitals $\chi _{1,i}$\ and $\chi _{2,n+i}$\
(alternatively called below the "internal" bond orders) resulting from
diagonal elements of the off-diagonal blocks of corrections $\widetilde{%
\mathbf{P}}_{(k)}^{\prime }$\ of Eqs.(3.3.5) and (3.3.6). It is evident that
skew-symmetric (sub)matrices $2\mathbf{G}_{(k)}^{\prime \circ }$ yield zero
contributions to these bond orders in contrast to the former matrices $%
\mathbf{V}_{(k)}$\ (Subsect. 3.2). Consequently, diagonal elements ($%
J_{(k)ii}$) of (sub)matrices \ $\mathbf{J}_{(k)},k=2,3,..$ are now the only
contributors to the "internal" bond orders. Moreover, these decisive
elements are expressible either in terms of population alterations of the
same (i.e. $k$th) order referring to the BOs of the given bond ($\varphi
_{(+)i}$ and $\varphi _{(-)i}$) or via the respective corrections to
delocalization coefficients of NCMOs $\psi _{(+)i}$ and $\psi _{(-)i}$\ as
Eq.(3.3.9) shows, viz.%
\begin{equation}
J_{(k)ii}=\frac{1}{2}%
(X_{(+)ii}^{(k)}-X_{(-)ii}^{(k)})=-(D_{(+)ii}^{(k)}+D_{(-)ii}^{(k)}). 
\tag{3.3.12}
\end{equation}%
As is seen from the above formula, a negative sign of the element $J_{(k)ii}$%
\ and thereby lowering of the order of the $I$th bond unambiguously follows
if the relevant increments to delocalization coefficients of both NCMOs
attached to the given bond ($\psi _{(+)i}$ and $\psi _{(-)i}$) are positive
quantities and vice versa. In other words, we actually arrive at a local
relation between the alteration in the order of the $I$th bond and
additional (de)localization of respective two NCMOs. Alternatively, a
negative sign of the element $J_{(k)ii}$\ is ensured if the BBO $\varphi
_{(+)i}$ of the given bond looses its population whilst the relevant ABO $%
\varphi _{(-)i}$\ acquires it so that $X_{(+)ii}^{(k)}<0$ and $%
X_{(-)ii}^{(k)}>0.$\ This conclusion evidently is not surprising.

Let us recall now that intrasubset population matrices of any order $k$
(i.e. $\mathbf{X}_{(+)}^{(k)}$ and $\mathbf{X}_{(-)}^{(k)})$\ are
representable via the intersubset coupling matrices of lower orders, viz. $%
\mathbf{G}_{(k-1)},$ $\mathbf{G}_{(k-2)}^{{}},$\ etc. (see Eq.(1.1.10)), the
latter determining the relevant corrections to intersubset bond orders
between BBOs and ABOs as Eq.(1.1.5) indicates. This implies that the extent
of the alteration in the "internal" bond order of the $I$th bond depends
upon orders of "external" bonds formed by its BOs $\varphi _{(+)i}$ and $%
\varphi _{(-)i}$\ with other basis orbitals ($\varphi _{(+)j},j\neq i$ and $%
\varphi _{(-)l},l\neq i$) within previous terms of the power series.
Moreover, the stronger these "external" bonds become, the more the
"internal' bond order is altered. Thus, we now arrive at explicit
interdependences between the "internal" and "external" bond orders.

For illustration, let us consider the most important contributors to the
"internal" bond orders, viz. the diagonal elements ($J_{(2)ii})$\ of the
second order matrix \textbf{\ }$\mathbf{J}_{(2)}.$\ After invoking the
expressions for $\mathbf{X}_{(+)}^{(2)}$ and $\mathbf{X}_{(-)}^{(2)}$\ shown
in Eq.(1.1.10) and substituting them into Eq. (3.3.12), we obtain that%
\begin{equation}
J_{(2)ii}=-[\mathop{\displaystyle \sum }\limits_{(-)m}(G_{(1)im})^{2}+%
\mathop{\displaystyle \sum }\limits_{(+)j}(G_{(1)ji})^{2}]<0  \tag{3.3.13}
\end{equation}%
i.e. any element $J_{(2)ii}$\ is an \textit{a priori }negative quantity.
Thus, bond orders inside pairs of strongly-interacting orbitals $\chi _{1,i}$%
\ and $\chi _{2,n+i}$\ are unambiguously predicted to be reduced to within
the second order terms inclusive. It is also seen that lowering of the order
of the $I$th bond is proportional to the sum of squares of orders of
"external" bonds that are formed by its orbitals ($\varphi _{(+)i}$ and $%
\varphi _{(-)i}$) with BOs of other bonds. The negative sign of $J_{(2)ii}$
may be also alternatively traced back to the fact that BBOs always loose
their population, whereas ABOs acquire it for $k=2$ (Subsect. 1.1).
Therefore, the rebonding effect (introduced in the subsection 3.2) now
acquires a more rigorous form.

Finally, the relevant total energy is worth some discussion. Thus,
substituting Eq.(3.3.1) into Eq.(3.1.22) shows that the zero order term $%
\mathcal{E}_{(0)}$\ now coincides with $2n$, i.e. with the total energy of $n
$ isolated homopolar bonds. This result evidently causes no surprise.
Similarly, from Eq.(3.1.23), we obtain%
\begin{equation}
\mathcal{E}_{(1)}=Tr[\mathbf{A}+\mathbf{F}+2\mathbf{M}].  \tag{3.3.14}
\end{equation}%
Thus, the first order correction $\mathcal{E}_{(1)}$\ represents the
energetic effect of deviation of the actual bonds of our system from uniform
homopolar ones. The present analogue of Eq.(3.1.24), in turn, takes the
following simple form 
\begin{equation}
\mathcal{E}_{(k)}=-\frac{2}{k-1}Tr\mathbf{J}_{(k)},\quad k=2,3.. 
\tag{3.3.15}
\end{equation}%
Thus, the energy correction of any order ($k$) is now proportional to the
sum of alterations in the "internal" bond orders. The same evidently refers
also to the relevant $\alpha -$component ($\mathcal{E}_{(k)}^{(\alpha )}$).
Hence, we actually arrive at a new interpretation of the principal relation
of Eq.(1.4.5), wherein\ alterations in the "internal" bond orders play the
role of the former charge transfer (Subsection 2.3). Emergence of such an
alternative is not surprising if we recall the direct relation between
elements $J_{(k)ii}$ and changes in populations of BOs $\varphi _{(+)i}$ and 
$\varphi _{(-)i}$ seen from Eq.(3.3.12). It should be kept in mind, however,
that the $\beta -$component of the same energy correction also generally
embraces populations of basis orbitals along with the "external" bond
orders. For example, the present form of Eq.(3.1.25) is as follows%
\begin{equation}
\mathcal{E}_{(2)}^{(\beta )}=Tr[\mathbf{\Pi }_{(1)}(\mathbf{A-F)]+}2Tr%
\mathbf{[G}_{(1)}^{\prime \circ }(\mathbf{M}^{+}\mathbf{-M)]}  \tag{3.3.16}
\end{equation}%
and contains the intrabond dipoles\ $\pm \Pi _{(1)ii}.$

\begin{center}
\textbf{3.4. Systems consisting of uniform homopolar bonds. An analogue of
the classical concept of conjugation}
\end{center}

The last and the simplest case of quasi-degenerate systems (discussed below)
corresponds to combination of the principal assumption of the above
subsection shown in Eq.(3.3.1) and of the anti-block-diagonal first order
matrix $\widetilde{\mathbf{H}}_{(1)}^{(ab)}$\ of Eq.(3.2.2). In other words,
we will now consider alternant systems representable by the following total
Hamiltonian matrix 
\begin{equation}
\widetilde{\mathbf{H}}_{(c)}=\widetilde{\mathbf{H}}_{(0)}^{\prime }+%
\widetilde{\mathbf{H}}_{(1)}^{(ab)}=\left\vert 
\begin{array}{cc}
\mathbf{0} & \mathbf{I} \\ 
\mathbf{I} & \mathbf{0}%
\end{array}%
\right\vert +\left\vert 
\begin{array}{cc}
\mathbf{0} & \mathbf{M} \\ 
\mathbf{M}^{+} & \mathbf{0}%
\end{array}%
\right\vert   \tag{3.4.1}
\end{equation}%
resulting from that of the above subsection under an assumption of vanishing
submatrices $\mathbf{A}$ and $\mathbf{F}$. The results under present
interest also are easily obtainable from those of subsection 3.3 under
employment of the equality $\mathbf{A=F=0.}$\ The reason why this specific
case is distinguished and discussed separately lies in its relevance to an
important and wide class of chemical compounds, viz. to acyclic conjugated
hydrocarbons [The superscript $(c)$ of the left-hand side of Eq.(3.4.1) is
used just in this connection]. Thus, let us start with the perturbational
perspective on these popular compounds in the framework of the H\"{u}ckel
type approximation.

Suppose that a certain conjugated hydrocarbon contains two types of uniform
bonds, namely strong double (C$=$C) bonds and weak single (C$-$C) ones so
that resonance parameters of the former exceed significantly those of the
latter in the basis of 2p$_{z}$ AOs of carbon atoms. Moreover, let our
hydrocarbon have an even number of carbon atoms that will be denoted by $2n$
as previously. The 2p$_{z}$ AOs of these atoms will then compose our $%
2n\times 2n-$dimensional basis set $\{\chi \}.$\ It is also assumed that our
hydrocarbon contains no adjacent double bonds and no cycles having an odd
number of carbon atoms. Evidently, this model primarily refers to acyclic
polyenes [40, 58, 73, 77]. Nevertheless, it has been applied recently also
to individual Kekul\'{e} valence structures of related (poly)cyclic
compounds [38, 61].

Further, let our hydrocarbon contain exactly $n$ double (C=C) bonds. Given
that the basis orbitals $\{\chi \}$ are enumerated as described in the above
subsection (i.e. the AOs belonging to the $I$th C=C bond acquire the numbers 
$i$ and $n+i$), the system concerned may be considered as a set of $n$
weakly-interacting C=C bonds, where the single (C$-$C) bonds serve as a
means of this interaction.

Let us assume finally that Coulomb parameters of all AOs $\{\chi \}$ are
uniform and coincide with our energy reference point. This evidently implies
zero diagonal elements of the relevant Hamiltonian matrix. Similarly,
uniform values of resonance parameters of all C=C bonds will serve as the
energy unit. These parameters will be entirely included into the zero order
member of our Hamiltonian matrix. Meanwhile, resonance parameters of single
(C$-$C) bonds will be incorporated into the relevant first order member. We
then immediately arrive at the Hamiltonian matrix of Eq.(3.4.1), where $%
n\times n-$dimensional (sub)matrices $\mathbf{I}$ and $\mathbf{M}$\
correspondingly embrace resonance parameters of C=C and C$-$C bonds. It is
also evident that diagonal elements of the (sub)matrix $\mathbf{M}$ take
zero values, i.e. $M_{ii}=0$\ for any $i$. At the same time, non-zero
off-diagonal elements of the same matrix $M_{ij}\neq 0$ $(i\neq j)$ refer to
C$-$C bonds. Otherwise, these elements vanish.

Let us now dwell on some general properties of hydrocarbons represented by
the Hamiltonian matrix $\widetilde{\mathbf{H}}^{(c)}$\ of Eq.(3.4.1). To
this end, let us employ the equality $\mathbf{A=F=0}$\ within the principal
relations of the above subsection. In particular, from both the last
relation of Eq.(3.3.3) and the first formula of Eq.(2.4.2) it follows that
the first order matrix $\mathbf{G}_{(1)}^{(c)}$ is of the skew-symmetric
nature. Moreover, the same may be easily shown to refer also to the relevant
principal matrices ($\mathbf{G}_{(k)}^{(c)}$) of higher orders ($k=2,3...$).
In summary, we obtain that 
\begin{equation}
\mathbf{G}_{(k)}^{(c)+}=-\mathbf{G}_{(k)}^{(c)},\quad k=1,2,3...  \tag{3.4.2}
\end{equation}%
Let us now turn to implications of the above result. Thus, employment of
Eq.(3.4.2) within Eq.(1.1.10) followed by invoking Eq.(1.3.16) yields the
result exhibited below 
\begin{equation}
\mathbf{X}_{(+)}^{(k,c)}=-\mathbf{X}_{(-)}^{(k,c)},\qquad \mathbf{D}%
_{(+)}^{(k,c)}=\mathbf{D}_{(-)}^{(k,c)}  \tag{3.4.3}
\end{equation}%
instead of the former coincidences of traces of analogous matrices (see
Eqs.(1.1.11) and (1.3.13)). On the basis of Eq.(3.4.3) one can immediately
conclude that%
\begin{equation}
X_{(+)ii}^{(k,c)}=-X_{(-)ii}^{(k,c)},\qquad
D_{(+)ii}^{(k,c)}=D_{(-)ii}^{(k,c)}  \tag{3.4.4}
\end{equation}%
for any $k$ and for any $i$. The first relation of Eq.(3.4.4) shows that the
population of the $k$th order lost by the BBO $\varphi _{(+)i}$\ of the $I$%
th C=C bond coincides with that acquired by the respective ABO $\varphi
_{(-)i}$\ or vice versa (the latter case is allowed for $k=3,4,.$. only
(Subsection 1.1)). Consequently, the double (C=C) bonds of conjugated
hydrocarbons ultimately take (loose) no population and this result causes
little surprise. Analogously, the second relation of Eq.(3.4.4) indicates
that total delocalization coefficients of NCMOs $\psi _{(+)i}$ and $\psi
_{(-)i}$\ of the same C=C bond are uniform.

Let us now turn to the relevant CBO matrix $\widetilde{\mathbf{P}}%
_{(c)}^{{}}.$\ It is evident that the zero order member $\widetilde{\mathbf{P%
}}_{(0)}^{(c)}$\ of the power series concerned coincides with that of the
above subsection, i.e. with $\widetilde{\mathbf{P}}_{(0)}^{\prime }$\ of
Eq.(3.3.4). As a result, interpretation of the latter (subsection 3.3) may
be straightforwardly transferred to $\widetilde{\mathbf{P}}_{(0)}^{(c)}$.
Further, from Eqs.(3.3.8) and (3.4.3) it follows that (sub)matrices $\mathbf{%
\Lambda }_{(k)}^{(c)},$\ $k=2,3..$ vanish within any subsequent correction $%
\widetilde{\mathbf{P}}_{(k)}^{(c)}$\ in accordance with the above-mentioned
zero charge redistribution among C=C bonds. Moreover, the same refers also
to the symmetric parts ($\mathbf{G}_{(k)}^{(c)\ast }$) of the principal
matrices $\mathbf{G}_{(k)}^{(c)}$\ and thereby to polarization matrices $%
\mathbf{\Pi }_{(k)}^{(c)}$\ defined by Eq.(3.3.7). The overall result is
then as follows%
\begin{equation}
\mathbf{\Lambda }_{(k)}^{(c)}=\mathbf{\Pi }_{(k)}^{(c)}=\mathbf{0,\qquad }%
k=1,2,3\mathbf{..}  \tag{3.4.5}
\end{equation}%
and ensures coincidence of the ultimate occupation numbers of AOs $\{\chi \}$%
\ with the relevant zero order increments (equal to 1). At the same time,
the result of Eq.(3.4.5) indicates the anti-block-diagonal constitution of
any correction $\widetilde{\mathbf{P}}_{(k)}^{(c)}$. The principal
(sub)matrix $\mathbf{J}_{(k)}^{(c)}$\ contained within the off-diagonal
blocks of this correction, in turn, results from Eq.(3.3.9) after employment
of Eqs.(3.4.3) and (3.4.4). We then obtain that%
\begin{equation}
\mathbf{J}_{(k)}^{(c)}=\mathbf{X}_{(+)}^{(k,c)}=-2\mathbf{D}%
_{(+)}^{(k,c)},\qquad J_{(k)ii}^{(c)}=X_{(+)ii}^{(k,c)}=-2D_{(+)ii}^{(k,c)}.
\tag{3.4.6}
\end{equation}%
It also deserves mention that the former skew-symmetric part ($\mathbf{G}%
_{(k)}^{\circ })$\ of the matrix $\mathbf{G}_{(k)}^{{}}$\ of Eq.(3.3.5) now
coincides with the total matrix $\mathbf{G}_{(k)}^{(c)}.$\ Collecting the
above-overviewed results followed by invoking Eq.(1.1.10) then yields the
final expression for the correction $\widetilde{\mathbf{P}}_{(k)}^{(c)},$
viz. 
\begin{equation}
\widetilde{\mathbf{P}}_{(k)}^{(c)}=\left\vert 
\begin{array}{cc}
\mathbf{0} & \mathbf{\Omega }_{(k)}^{{}} \\ 
\mathbf{\Omega }_{(k)}^{+} & \mathbf{0}%
\end{array}%
\right\vert ,\qquad k=1,2,3..  \tag{3.4.7}
\end{equation}%
where%
\begin{equation}
\mathbf{\Omega }_{(k)}^{{}}=\mathbf{J}_{(k)}^{(c)}+2\mathbf{G}_{(k)}^{(c)} 
\tag{3.4.8}
\end{equation}%
and 
\begin{align}
\mathbf{J}_{(1)}^{(c)} =&\mathbf{0,\quad J}_{(2)}^{(c)}=2(\mathbf{G}%
_{(1)}^{(c)})^{2},\quad \mathbf{J}_{(3)}^{(c)}=2(\mathbf{G}_{(1)}^{(c)}%
\mathbf{G}_{(2)}^{(c)}+\mathbf{G}_{(2)}^{(c)}\mathbf{G}_{(1)}^{(c)}), 
\nonumber \\
\mathbf{J}_{(4)}^{(c)} =&2(\mathbf{G}_{(1)}^{(c)}\mathbf{G}_{(3)}^{(c)}+%
\mathbf{G}_{(3)}^{(c)}\mathbf{G}_{(1)}^{(c)})+2(\mathbf{G}%
_{(2)}^{(c)})^{2}-2(\mathbf{G}_{(1)}^{(c)})^{4},etc.  \tag{3.4.9}
\end{align}%
It is evident that most of comments of the above subsection (3.3) concerning
the former matrices $\mathbf{J}_{(k)}^{{}}$\ of Eq.(3.3.9) are directly
transferable to the present matrices $\mathbf{J}_{(k)}^{(c)}.$\ In this
connection, we will concentrate our attention here mostly on specific points
referring to the case of conjugated htdrocarbons:

1) The orders of C=C bonds in a certain conjugated hydrocarbon are
determined by diagonal elements $\Omega _{(k)ii}^{{}}$\ of matrices $\mathbf{%
\Omega }_{(k)}^{{}},$\ whereas those of C$-$C bonds follow from the relevant
off-diagonal elements ($\Omega _{(k)ij}^{{}},i\neq j$). Thus, matrices $%
\mathbf{\Omega }_{(k)}^{{}}$\ actually represent redistribution of bond
orders (rebonding) due to conjugation of C=C bonds. In this connection, $%
\mathbf{\Omega }_{(k)}^{{}}$\ has been called the rebonding matrix of the $k$%
th order. It also deserves mention that $\mathbf{\Omega }_{(k)}^{{}}$\ are
now the only non-zero submatrices inside members of the power series for the
CBO matrix $\widetilde{\mathbf{P}}_{(c)}.$\ 

2) The rebonding effect in conjugated hydrocarbons manifests itself as an
interdependence between orders of the newly-formed C$-$C bonds and
alterations in the orders of the "initial" C=C bonds. In particular, the
(negative) second order correction to the order of the $I$th C=C bond
(determined by the element $J_{(2)ii}^{(c)}$) was shown to depend upon the
sum of squares of first order increments to orders of the adjacent C$-$C
bonds. [This relation is based on the one-to-one correspondence between
non-zero elements of the matrix $\mathbf{G}_{(1)}^{(c)}$ and C$-$C bonds
[73]].

3) An alteration in the order of the $I$th C=C bond proves to be directly
related to the respective change of population of its bonding orbital $%
\varphi _{(+)i}$\ and this relation is valid for each $k$ separately as
Eq.(3.4.6) indicates. Moreover, the same alteration is proportional to the
relevant increment to the delocalization coefficient of the occupied NCMO of
the same C=C bond ($\psi _{(+)i}$). Thus, weakening of a certain C=C bond
necessarily is accompanied by an additional delocalization of the respective
single pair of electrons and vice versa.

Let us now turn to the relevant energy expansion. The zero order member of
the latter ($\mathcal{E}_{(0)}^{(c)}$) coincides with 2$n$ as previously
(Subsection 3.3), whereas $\mathcal{E}_{(1)}^{(c)}$\ takes a zero value due
to vanishing diagonal elements of matrices $\mathbf{A,F}$ and $\mathbf{M.}$\
Furthermore, components of terms of higher orders of the energy expansion
are easily obtainable by substituting Eqs.(3.4.1) and (3.4.7) into
Eq.(1.5.7), and these are expressible as follows 
\begin{align}
\mathcal{E}_{(k)}^{(\alpha ,c)} =&2Tr\mathbf{\Omega }_{(k)}=2Tr\mathbf{J}%
_{(k)}^{(c)},  \tag{3.4.10} \\
\quad \mathcal{E}_{(2)}^{(\beta ,c)} =&2Tr[\mathbf{\Omega }_{(k-1)}\mathbf{M%
}^{+}]=2Tr\mathbf{[(\mathbf{J}}_{(k-1)}^{(c)}+2\mathbf{G}_{(k-1)}^{(c)})%
\mathbf{M}^{+}\mathbf{].}  \nonumber
\end{align}%
It is seen that the $\alpha -$component of the $k$th\ order energy ($%
\mathcal{E}_{(k)}^{(\alpha ,c)}$) is proportional to the sum of respective
corrections to orders of C=C bonds in a full aggreement with the results of
the above subsection. In contrast to the previous relation of Eq.(3.3.16),
however, the $\beta -$component $\mathcal{E}_{(2)}^{(\beta ,c)}$ now
contains only the sum of ($k-1$)th order increments to orders of C$-$C bonds
[One-to-one correspondence between non-zero elements of the submatrix $%
\mathbf{M}$\ and C$-$C bonds should be recalled here]. Hence, the components 
$\mathcal{E}_{(k)}^{(\beta ,c)}$\ and $\mathcal{E}_{(k)}^{(\alpha ,c)}$\
represent the energetic increments of two principal aspects of the rebonding
effect, namely of formation of C$-$C bonds and of the consequent alterations
in the orders of C=C bonds, respectively. The relation of Eq.(1.4.5) then
indicates an interdependence between absolute values of these energetic
increments and their opposite signs. In other words, new (i.e. C$-$C) bonds
may be formed only at the expense of the already-existing bonds. This
principal result may be regarded as a quantum-chemical justification of the
classical concept of the limited valence for conjugated hydrocarbons.
Finally, the total correction $\mathcal{E}_{(k)}^{(c)}$\ now represents the
ultimate energetic yield of the above-mentioned two interdependent aspects
of bond order redistribution and has been accordingly called the $k$th order
rebonding energy.

Proportionality of the total correction $\mathcal{E}_{(k)}^{(c)}$\ to $Tr%
\mathbf{J}_{(k)}^{(c)}$ and thereby to the sum of alterations of the $k$th
order in the orders of C=C bonds follows directly from Eq.(3.3.15).
Moreover, the same correction $\mathcal{E}_{(k)}^{(c)}$\ is now
representable as a sum of the relevant energetic increments associated with
formation of C$-$C bonds. Analogous two alternative forms of the total
rebonding energy $\mathcal{E}_{(c)}$ also are possible, viz.%
\begin{equation}
\mathcal{E}_{(c)}=\mathop{\displaystyle \sum }\limits_{k=2}^{\infty _{{}}}%
\frac{2}{1-k}Tr\mathbf{J}_{(k)}^{(c)},\quad \mathcal{E}_{(c)}=%
\mathop{\displaystyle \sum }\limits_{k=2}^{\infty _{{}}}\frac{2}{k}Tr(%
\mathbf{\Omega }_{(k-1)}\mathbf{M}^{+})  \tag{3.4.11}
\end{equation}%
and exhibit additivity with respect to individual contributors (i.e. C=C and
C$-$C bonds, respectively). Besides, employment of the expression for $%
\mathbf{J}_{(k)}^{(c)}$\ in terms of the matrix $\mathbf{D}_{(+)}^{(k)}$
shown in Eq.(3.4.6) yields the earlier-discussed formula for $\mathcal{E}%
_{(k)}$\ of Eq.(2.4.5) in accordance with the expectation. Thus, the
rebonding energy actually coincides with the total delocalization energy
inside the subset $\{\varphi _{(+)}\}$\ of initially-occupied orbitals of
C=C bonds.

It is seen, therefore, that the rebonding effect is the only factor
determining both the CBO matrix and the total energy of conjugated
hydrocarbons consisting of uniform C=C bonds connected by uniform C$-$C
bonds. Moreover, the above-established weakening of the former bonds due to
formation of the latter is in line with the observed trends in bond lenghts
of acyclic conjugated hydrocarbons [81]. These results allow us to conclude
that the rebonding effect is the quantum-chemical analogue of the classical
concept of conjugation. The dual nature of the effect also is among the
conclusions, i.e. it may be considered either as an interaction between C=C
bonds or as a sum of the relavant intrabond effects.

\begin{center}
\textbf{REFERENCES}
\end{center}

1. M.J.S. Dewar, and R.C. Dougherty, \textit{The PMO Theory of Organic
Chemistry} (Plenum Press, New York, 1975).

2. D.-K. Seo, G. Papoian and R. Hoffmann, Int. J. Quant Chem. \textbf{77},
408 (2000).

3. L. Rincon, J. Mol. Struct. (Theochem) \textbf{731}, 213 (2005).

4. L. Z\"{u}licke, \textit{Quantenchemie, Bd.1, Grundlagen und Algemeine
Methoden} (Deut- scher Verlag der Wissenschaften, Berlin, 1973).

5. L.D. Landau, E.M. Lifshits, \textit{Quantum Mechanics. The
Non-relativistic Theory }(Nauka, Moscow, 1974).

6. E. H\"{u}ckel, Z. Phys. \textbf{70}, 204 (1931).

7. E. H\"{u}ckel, Z. Phys. \textbf{76}, 628 (1932).

8. A. Streitwieser, Jr., \textit{Molecular Orbital Theory} (John Wiley
\&Sons, New York, 1961).

9. W. Kutzelnigg, J. Comput. Chem. \textbf{28}, 25 (2007).

10. C.A. Coulson, B.O'Leary and R.B. Mallion, \textit{H\"{u}ckel Theory for
Organic Chemistry} (Academic Press, London, 1978).

11. S. Huzinaga, \textit{The MO Method} (Mir, Moscow, 1983) [in Russian].

12. C.A.Coulson and H.C. Longuet-Higgins, Proc. Roy. Soc. (London) \textbf{%
A192}, 16 (1947); \textbf{A193}, 447 (1948).

13. M.J.S. Dewar, \textit{The Molecular Orbital Theory of Organic Chemistry}
(McGraw-Hill, New York, 1969).

14. M.M. Mestechkin, \textit{The Density Matrix Method in Quantum Chemistry}
(Naukova Dumka, Kiev, 1977) [in Russian].

15. I. Gutman, Z. Naturforsch 36A, 1112 (1981).

16. J. March, \textit{Advanced Organic Chemistry, Reactions, Mechanisms and
Structure} (Wiley Interscience, New York, 1985).

17. F.A. Carroll, \textit{Perspectives on Structure and Mechanism in Organic
Chemistry} (Brooks/Cole, Pacific Grove, CA, 1998).

18. M. Edenborough, \textit{Organic Reaction Mechanisms. A Step by Step
Approach} (Taylor and Francies, London, 1999).

19. A.N. Vereshtchagin, \textit{The Inductive Effect} (Nauka, Moscow, 1987)
[in Russian].

20. O. Chalvet (Ed.), \textit{Localization and Delocalization in Quantum
Chemistry, Atoms and Molecules in the Ground State}, Vol.1 (Reidel,
Dordrecht, 1975).

21. I. Mayer, Chem. Phys. Lett. \textbf{89}, 390 (1982).

22. P.R. Surjan, I. Mayer, and M. Ketesz, J. Chem. Phys.\textbf{\ 77}, 2454
(1982).

23. I. Mayer, and P.R Surjan, J. Chem. Phys. \textbf{80}, 5649 (1984).

24. J.P. Daudey, Chem. Phys. Lett. \textbf{24}, 574 (1974).

25. V. Gineityte, Int. J. Quant. Chem. \textbf{68}, 119 (1998).

26. V. Gineityte, Lith. J. Phys. \textbf{44}, 219 (2004).

27. V. Gineityte, Int. J. Quant. Chem. \textbf{72}, 559 (1999).

28. V. Gineityte, J. Mol. Struct. (Theochem), \textbf{333}, 297 (1995).

29. R. McWeeny, \textit{Methods in Molecular Quantum Mechanics,} 2nd. ed.
(Academic Press, London, 1992).

30. M.M. Mestechkin, Teoretitch. Eksperim. Khim. \textbf{4}, 154 (1968).

31. M. Mestechkin, Int. J. Quant. Chem. \textbf{15}, 601 (1979).

32. M.M. Mestechkin, Teoretitch. Eksperim. Khim. \textbf{12}, 739 (1976).

33. L. Cohen and C. Frischberg, J. Chem. Phys. \textbf{65}, 4234 (1976).

34. M. Simonetta and E. Gianinetti, In: \textit{Molecular Orbitals in
Chemistry, Physics and Biology}, Eds. P.-O. L\"{o}wdin, B. Pullman
(Academic, New York, London, 1964), pp. 83-112.

35. R. McWeeny, Phys. Rev. \textbf{126}, 1028 (1962).

36. V. Gineityte, J. Mol. Struct. (Theochem) \textbf{343}, 183 (1995).

37. V. Gineityte, J. Mol. Struct. (Theochem) \textbf{585}, 15 (2002).

38. V. Gineityte, MATCH Commun. Math. Comput. Chem. \textbf{72}, 39 (2014).

39. V. Gineityte, Int. J. Quant. Chem. \textbf{94}, 302 (2003).

40. V. Gineityte, \textit{On Relative Stabilities of Distinct Polyenes. An
Extension of the Concept of Conjugated Paths }ArXiv (2015) \
https://arxiv.org./abs/1501.04734.

41. V. Gineityte, Z. Naturforsch. \textbf{64A}, 132 (2009).

42. V. Gineityte, Int. J. Chem. Model. \textbf{4}, 189 (2012).

43. P. Lankaster, \textit{Theory of Matrices }(Academic Press, New York,
1969).

44. F. Jensen, \textit{Introduction to Computational Chemistry}, 2nd. ed.
(John Wiley \& Sons, 2007).

45. S. Liu, J.M. Perez-Jorda and W. Yang, J. Chem. Phys. \textbf{112}, 1634
(2000).

46. Z. Szekeres and P.R. Surjan, Chem. Phys. Lett. \textbf{369}, 125 (2003).

47. S.F. Boys, Rev. Mod. Phys. \textbf{32}, 296 (1960).

48. C. Edmiston and K. Ruedenberg, Rev. Mod. Phys. \textbf{34}, 457 (1963).

49. C. Edmiston and K. Ruedenberg, J. Chem. Phys. \textbf{43}, s97 (1965).

50. J. Pipek and P.G. Mezey, J. Chem. Phys. \textbf{90}, 4916 (1989).

51. V. Gineityte, J. Mol. Struct. (Theochem), \textbf{288}, 111 (1993).

52. V. Gineityte, Lith. J. Phys. \textbf{51}, 107 (2011).

53. V. Gineityte, Int. J. Quant. Chem. \textbf{101}, 274 (2005).

54. V. Gineityte, Int. J. Quant. Chem. \textbf{105}, 232 (2005).

55. V. Gineityte, Lith. J. Phys. \textbf{42}, 397 (2002).

56. G.N. Lewis, J. Amer. Chem. Soc. \textbf{38}, 762 (1916).

57. T.C. Koopmans, Physica, \textbf{1}, 104 (1934).

58. V. Gineityte, J. Mol. Struct. (Theochem) \textbf{487}, 231 (1999).

59. V. Gineityte, J. Mol. Struct. (Theochem) \textbf{680}, 199 (2004).

60. V. Gineityte, Int. J. Quant. Chem. \textbf{108}, 1141 (2008).

61. V. Gineityte, Monatsh. Chem., (2018).
https://doi.org/10.1007/s00706-017-2133-3

62. R. Zahradnik, R. Polak, \textit{Elements of Quantum Chemistry} (Plenum
Press, New York, 1980).

63. V. Gineityte, J. Mol. Struct. (Theochem) \textbf{364}, 85 (1996).

64. V. Gineityte, J. Mol. Struct. (Theochem) \textbf{532}, 257 (2000).

65. V. Gineityte, J. Mol. Struct. (Theochem) \textbf{546}, 107 (2001).

66. V. Gineityte, \textit{Quasi-classical Alternatives in Quantum Chemistry, 
}ArXiv\textit{\ }(2014)\textit{\ \ }

https://arxiv.org/abs/1402.6268.

67. \ M.J.S. Dewar, R. Pettit, J. Chem. Soc. 1625 (1954).

68. M.J.S. Dewar, J. Amer. Chem. Soc. \textbf{106}, 669 (1984).

69. V.F.Traven, \textit{Electronic Structure and Properties of Organic
Molecules} (Khimia, Moscow, 1989) [in Russian].

70. V. Gineityte, Int. J. Quant. Chem. \textbf{77}, 534 (2000).

71. V. Gineityte, Int. J. Chem. Model. \textbf{5}, 99 (2013).

72. R. A. Horn and C.R. Johnson, \textit{Matrix Analysis }(Cambridge Univ.
Press, Cambridge, 1986).

73. V. Gineityte, Monatsh. Chem. \textbf{147}, 1303 (2016).

74. G.G. Hall, Proc. Roy. Soc. (London), \textbf{A229}, 251 (1955).

75. A.V. Luzanov, J. Structural Chemistry, \textbf{21}, 703 (1981).

76. V. Gineityte, Int. J. Quant. Chem. \textbf{106}, 2145 (2006).

77. V. Gineityte, J. Mol. Struct. (Theochem) \textbf{497}, 83 (2000).

78. V. Gineityte, Croat. Chem. Acta \textbf{81}, 487 (2008).

79. V. Gineityte, Croat. Chem. Acta \textbf{86}, 171 (2013).

80. V. Gineityte, J. Mol. Struct. (Theochem) \textbf{430}, 97 (1998).

81. B. Milian-Medina, J. Gierschner, WIREs Comput. Mol. Sci. \textbf{2}, 513
(2012).

\end{document}